\documentclass[trackchanges, twocolumn,twocolappendix]{aastex7}

\begin{document}

\title{The YREC Stellar Evolution Code: Public Data Release}

\author[0000-0002-7549-7766]{Marc H. Pinsonneault}
\affiliation{Department of Astronomy, The Ohio State University, Columbus, OH 43210, USA}
\affiliation{Center for Cosmology and AstroParticle Physics (CCAPP), The Ohio State University, 191 West Woodruff Ave, Columbus, OH 43210, USA}
\email[show]{pinsonneault.1@osu.edu}

\author[0000-0002-4284-8638]{Jennifer L.~van~Saders} 
\affiliation{Institute for Astronomy, University of Hawai`i, 2680 Woodlawn Drive, Honolulu, HI 96822, USA}
\email{jlvs@hawaii.edu}

\author[0000-0002-8849-9816]{Lyra Cao}
\affiliation{Department of Physics and Astronomy, Vanderbilt University, Nashville, TN 37212, USA}
\email{lyra.cao@Vanderbilt.Edu}

\author[0000-0002-4818-7885]{Jamie Tayar}
\affiliation{Department of Astronomy, University of Florida, Gainesville, FL 32611, USA}
\email{jtayar@ufl.edu}

\author[0000-0002-8004-0795]{Franck Delahaye}
\affiliation{LUX, Observatoire de Paris, Université PSL, Sorbonne Université, CNRS, 75005 Paris, France}
\email{franck.delahaye@observatoiredeparis.psl.eu}

\author[0000-0002-1333-8866]{Leslie M. Morales}
\affiliation{Department of Astronomy, University of Florida, Gainesville, FL 32611, USA}
\email{l.morales@ufl.edu}

\author[0009-0006-0035-7431]{Rachel A. Patton}
\affiliation{Department of Physics and Astronomy and Pittsburgh Particle Physics, Astrophysics and Cosmology Center (PITT PACC), University
of Pittsburgh, 3941 O’Hara Street, Pittsburgh, PA 15260, USA}
\email{rap342@pitt.edu}

\author [0009-0008-8004-851X]{Matthew C. Rendina}
\affiliation{Department of Astronomy, The Ohio State University, Columbus, OH 43210, USA}
\email{rendina.8@osu.edu}

\author[0000-0002-7550-7151]{Joel C. Zinn}
\affiliation{Department of Physics and Astronomy, California State University, Long Beach, Long Beach, CA 90840, USA}
\email{joel.zinn@csulb.edu}

\author[0000-0002-9879-3904]{Zachary R. Claytor}
\affiliation{Space Telescope Science Institute, 3700 San Martin Drive, Baltimore, MD 21218, USA}
\email{zclaytor@stsci.edu}

\author[0000-0002-8674-9922]{Amanda L. Ash}
\affiliation{Department of Astronomy, The Ohio State University, Columbus, OH 43210, USA}
\email{ash.172@osu.edu}

\author[0009-0001-4446-833X]{Susan Byrom}
\affiliation{Department of Astronomy, University of Florida, Gainesville, FL 32611, USA}
\email{s.byrom@ufl.edu}

\author[0000-0002-1699-6944]{Kaili Cao}
\affiliation{Center for Cosmology and AstroParticle Physics (CCAPP), The Ohio State University, 191 West Woodruff Ave, Columbus, OH 43210, USA}
\affiliation{Department of Physics, The Ohio State University, 191 West Woodruff Ave, Columbus, OH 43210, USA}
\email{cao.1191@osu.edu}

\author[0009-0008-8819-0481]{Vincent A. Smedile}
\affiliation{Department of Astronomy, The Ohio State University, Columbus, OH 43210, USA}
\affiliation{Center for Cosmology and AstroParticle Physics (CCAPP), The Ohio State University, 191 West Woodruff Ave, Columbus, OH 43210, USA}
\email{smedile.1@osu.edu}

\begin{abstract}

In this paper we present the public release of the Yale Rotating Evolution Code (YREC). YREC is a stellar evolution code that covers brown dwarfs and stars across a wide range of masses, and evolutionary states from the pre-MS through helium burning. We summarize the key ingredients of the code, document the code performance, and discuss its strengths and limitations. We present libraries of input files, documentation, sample use cases, and scripts. In addition to usage as a research tool, we highlight the utility of the code for educational purposes. 

\end{abstract}


\section{Introduction}
\label{section:Introduction}
The theory of stellar structure and evolution is deep and beautiful. Its bedrock lies in fundamental physical principles. The equations of stellar structure are expressions of the conservation of mass, energy, and momentum, complemented by the second law of thermodynamics.  The remarkably short free-fall timescale ensures that hydrostatic balance is a superb approximation for almost all stars in almost all circumstances, greatly simplifying the general problem. The transport and generation of energy are also closely matched in stars, so stars are near-equilibrium structures. As a consequence, stellar structure, or the solution of the equations at an instant in time, is a tractable problem. It is only over oceans of time that the graceful arc of stellar evolution proceeds, driven by the slow but irreversible transformation of light species into heavy ones by nuclear reactions. 

What might seem at first to be an impossible problem can then be cast in a straightforward form. It simply requires tools from the full arsenal of modern physics---quantum mechanics, thermodynamics, statistical mechanics, electricity and magnetism, nuclear physics, and fluid mechanics. The transport of radiation through matter can be captured by an effective mean free path of a photon, expressed as an opacity. Nuclear reaction rates and the equation of state can similarly be defined as functions of the local temperature, density and composition. Even difficult physical problems, such as the transport of energy by turbulent convective motions, can be collapsed into a local prescription. Solving these equations, of course, is not the same as setting them up.

Analytic work has a distinguished history in stellar theory, but the limitations of these tools became very clear at an early stage. The microphysics are clearly posed questions, but not ones with answers in convenient analytical forms. The correct solution to the equations need not be easy to compute. The dawn of the computer age in the 1950s and 1960s led directly to the first generation of numerical stellar structure codes. Astrophysicists were able to use compact codes and kilobytes of memory, deployed with punch cards, to decode the life cycle of stars. A number of independent codes, still in use today, were developed. In this paper we present the public release of the Yale Rotating Stellar Evolution Code (YREC), which had its origin in this era. 

Despite the variety of stellar codes that were developed, only a few major codes are public: the MESA code \citep{paxton+2011, paxton+2013, paxton+2015, Paxton2018, Paxton2019, Jermyn2023}, CESAM \citep{Morel97, Morel08, Manchon25}, the Eggleton STARS code \citep{Eggleton1972,Eggleton2011}, and 
the ESTER code \citep{Espinosa2013,Rieutord2016J,White2025}. MESA has been particularly high in impact and is widely used. This is not because other authors were unwilling to share and compare results:  a good example of detailed code comparisons within the community can be found in \citet{Bahcall1992} for solar neutrinos. Instead, the scarcity of public tools reflects the mismatch between the era in which those codes were developed and the time frame when the open source movement blossomed. Proprietary codes are also typically not built with the needs of a large and diverse user base in mind. Some are carefully curated gems. But walled gardens can also be filled with weeds. Releasing such codes therefore requires careful boundaries, which we lay out below.

Users are free to modify the YREC source code, but the code as written was not designed with modular user inputs in mind. In future releases, we will prioritize making the code easier to alter. YREC will also do what you tell it to do, without second-guessing. This means that the code can and will crash if you assign, for example, a super-critical rotation rate. The documentation does discuss known failure modes and how to diagnose them. Many of the inputs for YREC were designed for specific science cases, and the code is not guaranteed to perform well outside of that domain, where the assumptions break down. Finally, YREC is written in glorious Fortran, F77 to be specific, but it is F90 compatible.
With these cautions in mind, the modern open source movement has proven to be fruitful and powerful, and we believe that YREC will be a useful tool for research and education.

\subsection{Why YREC?}

YREC is a stellar evolution code optimized for the study of hydrostatic stars from the pre-main sequence through to the helium burning phase of evolution. YREC can also be used to study giant planets and white dwarfs. It is fast, precise, and accurate in the domains where it functions. YREC has been extensively used for both stellar population and stellar physics studies, including high-precision solar models \citep{Basinger2024}. 

YREC uses modern input microphysics (equation of state, energy transport, energy generation, surface boundary conditions) with a variety of user-configurable options. YREC is a relaxation code, and as written, it requires a starting model to run. These models can then be relaxed to a different set of initial conditions. 
It has the option to include important physical processes not considered in classical stellar models. For example, YREC includes the structural effects of rotation using a pseudo-2D approximation. The treatment of angular momentum evolution is flexible, including star-disk coupling, rotationally induced mixing, internal angular momentum transport, angular momentum loss from magnetized winds, and the ability to model different rotation profiles in convective regions. It also includes gravitational settling and the structural effects of starspots and their associated magnetic fields.

\subsection{Outline of the Paper}

We describe the input physics and numerical approach of the YREC code in Section \ref{section:Codephysics}. We also note similarities and differences with the public MESA code. 

Stellar evolution codes adopt a set of input physics and run a single initial condition (mass, composition, rotation rate, and age) to a prescribed end state. A test suite of runs highlights the various capabilities of the code. The test suite is also an essential tool for code development, as it allows users to check that code changes have not caused unanticipated problems. In Section \ref{section:Models}, we present our test suite, sample applications, and comparisons with other codes in the literature. Scripts and helper codes are discussed in Section \ref{section:Scripts}. In the discussion and summary, we lay out the current domains where the code is recommended for usage and future plans for upgrades. Details of how to access the code repository, input files, documentation, and other resources are given in the appendices.

\section{Physics and Numerics of the YREC Code}

\label{section:Codephysics}

YREC has been in use as a stellar structure and evolution code for more than 60 years. The code was first described in \citet{Larson1964}, and subsequently updated and expanded as described by \citet{Prather1976}. The most recent summaries of the standard model physics are \citet{Demarque2008}, which included the addition of output for computing pulsations; and \citet{vansaders2012}, which includes a series of updates to the input physics. Microscopic diffusion was added by \citet{Bahcall1992,Bahcall1995}, along with a complete overhaul of the treatment of nuclear reaction rates. We also note that the DSEP code, used for a popular set of isochrones \citep{Dotter2008}, was based on YREC and shares much of its structure.

\citet{Pinsonneault1989} added rotation and rotationally induced mixing. There were then a series of updates to the angular momentum evolution model. This included updates to the angular momentum loss model \citep{Krishnamurthi1997,vansaders2013}, updates to the rotationally induced mixing model \citep{Pinsonneault2002,Somers2016}, and the inclusion of the structural effects of starspots \citep{Somers2015,Somers2020}.

\subsection{Overall Code Structure and Repositories}

YREC is a relaxation code that can be run in 2 modes. An evolving run takes an input model and runs to a specified end state. A rescaling model uses a prior solution that is close to the desired conditions and perturbs it. Rescaling can be done in the fully convective pre-main sequence (pre-MS), the zero-age main sequence (ZAMS), or the zero-age core He burning state (ZAHB). For the pre-MS, there is no time independent solution of the structure equations. A pre-MS model must therefore evolve in time when rescaling. On the ZAMS and the ZAHB, the code rescales with a zero time step. We note that many useful stop conditions are not enabled for rescaling runs of any form. If rescaling is desired, it is therefore recommended to do runs in two stages: an initial rescaling run followed by an evolving run with a stop condition.

Running YREC requires the user to download the source code, and then to use a fortran compiler to compile the subroutines and link them. The code, documentation, and instructions on how to run the code can be found at \url{https://github.com/yreclab/yrec}. \footnote{The release paper version of the code can be found in the \url{https://github.com/yreclab/yrec/tree/yrec_2026_release} branch.} YREC also requires libraries of input tables and starting models, which are available on the repository under the \texttt{input} and \texttt{startmodels} directories, respectively. These data are also available in the YREC Zenodo repository (https://zenodo.org/communities/yrecmodels.)

The code uses two namelists that include global run instructions, file input and output paths, numerical controls, and an extensive list of user-controllable options for the input physics. Running the code also requires a converged starting model and tabulated input data for the microphysics. Suites of models can be run using scripts, and the output can be adopted to investigate a range of astrophysical problems.

The github repository contains a \texttt{testsuite} directory that includes the namelists and output for the test suite, which can also be found on the Zenodo repository. The Zenodo repository also has sections for scripts, classroom exercises, a base model grid covering a range of masses and metallicities, and sample cases used to generate all of the figures and tables in this paper.

\subsection{Numerical Methods}

YREC solves the equations of stellar structure using the Henyey technique. In this method, the four coupled equations of stellar structure are converted into finite difference equations. A valid solution of the structure equations at a given time is then perturbed to account for composition changes and gravitational potential energy release. The start-of-timestep structure determines the locations of convective regions. It is also used to compute nuclear reaction rates, which are then used to infer composition changes in an implicit scheme. Composition changes drive changes in the structure variables, which are inferred, along with their derivatives, in microphysics routines. As an example, the conversion of hydrogen to helium lowers the number of free particles per gram, reducing pressure at fixed temperature and density. The model will therefore require compensating changes in temperature and density to restore hydrostatic equilibrium, which in turn cause feedback for other state variables. The code then solves iteratively for the required changes in the structure variables until it reaches a user-specified set of numerical precision criteria.

Stellar evolution is a two point boundary value problem. At the center, YREC uses a Taylor series expansion to map the inner shell to the central values. This approach is widely used in stellar codes, as it avoids the divergence of logarithmic structure variables at zero mass and radius. Close to the surface the assumption of a short photon mean free path becomes invalid. All stellar evolution codes therefore impose a solution of the radiative transfer equations as an atmospheric boundary condition. This can either be a direct integration, such as a gray atmosphere, or a table look-up from a grid of full model atmosphere calculations.

An important difference between YREC and MESA is the treatment of the outer boundary condition below the atmosphere. For a two point boundary value problem, the governing equations must be integrated both inwards from the surface and outwards from the center. The two solutions and their derivatives are matched at a fitting point, and the location of the fitting point is a numerical choice that reflects tradeoffs. The base of the atmosphere is a possible location for the fitting point, but this is not required. A deeper fitting point is usually more stable, but the inwards integration usually makes simplifying assumptions. For example, in YREC the luminosity is constant in the envelope integration. MESA models typically adopt a fitting point at the base of the atmosphere.\footnote{For some tabulated model atmospheres, there is a choice in MESA to adopt a deeper fitting point, up to $\tau = 100$, but this is still relatively close to the surface.}

YREC adopts an outer fitting point well below the photosphere. At this fitting point, the structure variables $P$, $T$, $R$ and $L$ from the interior solution and their derivatives must be matched to the corresponding values from a downwards envelope integration. The code also needs information on how these envelope integration properties change as a function of $L$ and $T_{\rm eff}$. YREC therefore integrates three solutions close to the model $L$ and $T_{\rm eff}$, and interpolates between them in this envelope triangle as the model solution evolves. If the solution moves outside of the triangle the code will flip it to re-center the triangle around the current model location.

Although very little mass is usually in this envelope, typically one part in $10^4$ or less, this is a regime where the pressure and density are dropping very rapidly in stars. As a result, not retaining the outer layer structure yields a significant increase in computing speed. The structure equations are also more stable with a deeper fitting point. The cost is that information on the detailed outer layer physics is not stored in the interiors model. YREC therefore has options to post-process an envelope solution, for both informational purposes and for practical applications such as computing oscillation frequencies.

\begin{figure}
    
    \centering
    \includegraphics[width=8 cm]{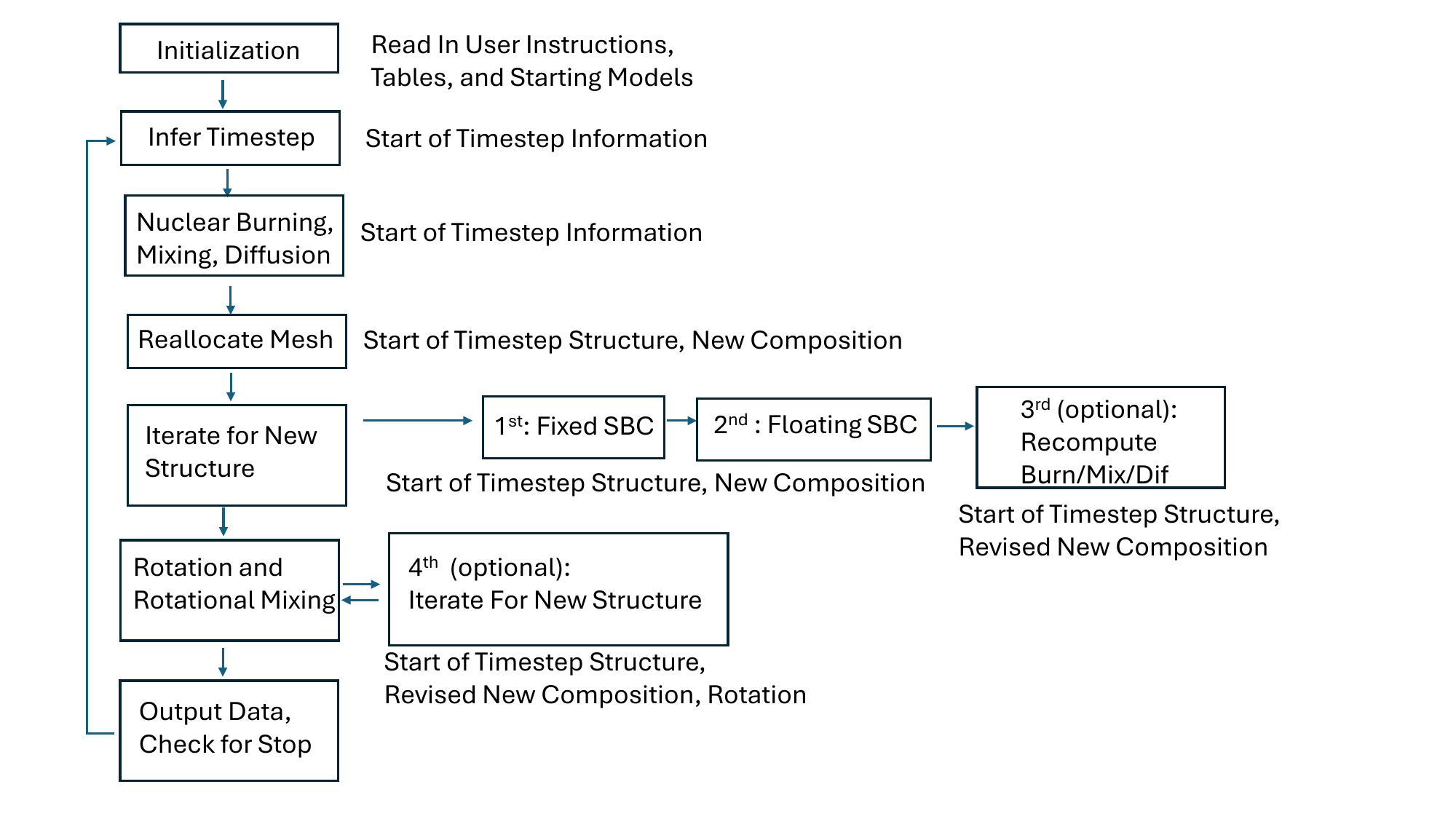}
    \caption{A schematic illustration of the steps in the YREC code.}

\label{fig:flowchart}

\end{figure}

YREC solves the structure equations in up to four iterative loops, two of which are mandatory. In the first loop, the $L$ and $T_{\rm eff}$ at which the surface boundary conditions (SBCs) are evaluated is held fixed. In the second, main loop, the location of the ``envelope triangle" is allowed to change dynamically. The third, optional, level of iteration allows the code to use the end-of-timestep structure to compute the composition change, rather than the start-of-timestep structure. Microscopic diffusion and mixing from rotation are computed after the model structure has converged in a separate step, so the converged structure in the first through third level of iteration does not account for all of the changes in the composition profile or for changes in the structural effects of rotation during a timestep.  The fourth level of iteration allows the code to iterate between the rotation, diffusion, and structure solutions. The overall program flow is illustrated in Figure \ref{fig:flowchart}.

For each model, the code begins by reallocating the spatial mesh points. Mesh points are allocated based on the requirement that the state variables not change from one point to the next by more than a critical value. In the deep core, mass is usually the limiting factor. Luminosity governs the spacings in nuclear burning regions, while pressure is the governing ingredient for envelopes. Points with composition or rotation rate discontinuities can be flagged to prevent rezoning across them.

At the end of the timestep, the code computes a series of limits for the next timestep; the most stringent relevant criterion is used. The code uses the instantaneous rate of nuclear reactions to set a limit on the timestep for the next model. It also computes the rate of change of the structure variables, from start to end of timestep, to set another limit on the timestep; this is adopted if shorter than the nuclear burning timestep. This second criterion is essential for pre-MS models, for which the energy source is gravitational potential energy. For the main sequence and beyond, this alternative timestep criterion is disabled; otherwise it induces very small time steps in the shell hydrogen burning stage.

The code then writes data to output files as requested, and checks whether a stop condition has been achieved. If the run continues, the code then returns to the start of the loop described above.

\subsection{Classical Stellar Model Physics}

Stellar evolution codes are designed for exploration. It is therefore natural that YREC has many options for how to treat important physical processes. The YREC code has evolved over time, and it has retained backwards compatibility with historical choices for the input physics. In this paper we do not dwell on these charming relics, turning our attention instead to the options recommended for use today.

\subsubsection{The Equation of State}

The equation of state relates pressure to temperature, density, and composition, and it also encodes extensive thermodynamic information. As such, it is used for a wide range of purposes in stellar evolution codes. The current version of YREC has three different choices for an equation of state.

The recommended equation of state is the 2005 update of the OPAL EoS \citep{Rogers2002}. We note that this EoS has a fixed heavy element content. Users can generate and use input tables with different metal contents, but the metallicity is fixed for a given run of the code. In the high-density and cool domain, the OPAL table does not cover some of the phase space of interest. The \citet{Saumon1995}, or SCV, equation of state is a good alternative here.

The default equation of state, which we refer to as the Yale EoS \citep{Prather1976}, is available across the full phase space. It includes radiation pressure and degeneracy pressure. The latter is solved for using tabulated solutions to the Fermi integrals, and it includes relativistic corrections. For the gas pressure, the code solves the Saha equations for hydrogen, helium, and metals at low temperature. Above a user-specified threshold, it assumes a fully ionized plasma. The Debye-Huckle correction is also included, to account for electrostatic effects. This EoS is extremely fast, with smooth analytic derivatives, and we recommend its usage for educational applications. However, the code assumptions break down for cool dense plasmas, especially for lower main sequence stars. Most published applications of YREC therefore use tabulated EoS data that includes a more sophisticated treatment, particularly for collective plasma effects.

YREC has a tie-breaker algorithm. It will use the OPAL EoS when possible if it is selected as an option. Outside the table domain, it then checks if the SCV EoS is available (if specified); the Yale EoS is the backup if neither is available. In all cases the code ramps between solutions close to table edges. We caution that thermodynamic consistency is not in general preserved in these cases, which can cause numerical issues in model domains close to table edges. The implementation is different in detail than that in MESA, but both codes use a broadly similar approach.

\subsubsection{Radiative and Conductive Energy Transport}

The opacity of matter to light depends, in complicated ways, on the composition, density, and temperature.  

The calculation of opacities requires several steps. It is necessary i)
to determine the ionic state of the absorbing medium as well as the
population of the atomic levels of its constituents, ii) to calculate
the absorbing coefficients for different processes (photo-excitation,
photoionization), and iii) to include plasma effects (the ions cannot
be considered as isolated). Accuracy versus completeness is still at play nowadays as it was 30 years ago. The atomic systems considered today may be much larger, thanks to the computer power available, but the calculations still require approximations and compromises since they cannot be infinite.

In stellar interiors, local thermal equilibrium (LTE) is a very good approximation since the mean free path of a photons is very small. In this case, we can use a diffusion approximation for energy transport, with a diffusion coefficient that is the product of c/3 and an effective photon mean free path. The latter can be well approximated by a weighted average of the monochromatic opacity of the mixture of elements present in the stellar plasma. This Rosseland mean opacity is defined, per gram, by 
\begin{equation}
    \frac{1}{\kappa}=\int\frac{1}{\kappa_\nu} \frac{\partial B_\nu}{\partial T} d\nu
\end{equation}
with $B_\nu$ being the Planck function characterizing the energy distribution of blackbody radiation and $\kappa_\nu$ being the monochromatic opacity of the mixture.
\begin{equation}
    \kappa_\nu = \sum_{s=species}f_\nu(s)(\kappa^s_{BB}(s) + \kappa^s_{BF}(s)) + \kappa_{FF}
\end{equation}
with $f_s$ being the fractions of species s, and where BB stands for bound-bound transitions (photo excitation processes), BF stands for bound-free transitions (photo ionization processes) and FF stands for free-free transitions (inverse Bremsstrahlung processes).
For the atomic opacities, the main providers are the OPAL project \citep{iglesias_rogers1996} and the Opacity Project \citep{Badnell2005}, hereafter OP. For these, each of the processes (BB and BF) have to be calculated for each ion of each species and combined with a detailed equation of state. If we consider the opacities the OP distributes over its website (\url{https://cdsweb.u-strasbg.fr/topbase/testop/home.html}) this corresponds to 241 ions for the 17 species (H to Ni). Since the Rosseland mean opacity is a harmonic mean, the details of the lines and resonances as well as the plasma effects (mainly their broadening) are crucial and can lead to large differences depending on the resolution of the photo-ionization cross sections, line transitions, and the broadening theory. 

Molecular opacities require a molecular EoS, which is far more complex than the atomic case. The same holds true for the molecular energy levels and transitions relative to the atomic ones. However, for astrophysical conditions molecules are rare above temperatures of order 6000 K, and hydrogen ionization is the dominant opacity source in any case at temperatures of order $10^4 $~K. YREC, like other stellar evolution codes, therefore uses dedicated molecular opacity tables computed only for low temperatures.

YREC uses three families of opacity tables: conductive, atomic, and molecular. Our conductive opacities are taken from \citet{Cassisi2007}. We use molecular opacities from \citet{Ferguson2005} below log T = 4.0, atomic opacities above log T = 4.1, and a ramp between the two for intermediate temperatures. There are three families of atomic opacities that the code can use---OPAL tables for carbon and oxygen-rich mixtures \citep{Iglesias1993}, OPAL tables for H, He, and metal mixtures \citep{iglesias_rogers1996} and OP tables for H, He, and metal mixtures \citep{Badnell2005}. The first set of opacities are used in helium-burning regions, where the heavy element mixture changes significantly. For both OP and OPAL, the heavy element mixture is held fixed, and the opacity is interpolated as a function of X, Z, $\rho$ and T. These tables, with these assumptions, are widely used in stellar models, and there are web tools available to build tables for user-specified heavy element mixtures (for the Opacity Project, \url{https://opacity-cs.obspm.fr/opacity/index.html}). YREC uses tables constructed for a variety of heavy element mixtures, typically tied to different assumed solar abundance measurements.  Our base heavy element mixture for this paper is \citet{grevesse_sauval1998}, which produces solar models in good agreement with helioseismology. We also include tables based on the published solar mixtures of \citet{Magg2022}. \citet{Asplund2021}, and \citet{Lodders2025}.

\subsubsection{Convection and Convective Boundary Layer Physics}

YREC uses the mixing length theory (MLT) of convection \citep{BohmVitense1958} to infer the actual temperature gradient in regimes that are unstable against convection. This requires knowledge of the radiative and adiabatic temperature gradients. The  code considers the Schwarzschild criterion for instability, not the Ledoux criterion. See \citet{Demarque2008} for a fuller discussion of this point.

The effective convective flux can be modified in models that include starspots (section~\ref{sec:starspots} below). In convective cores the specific heat is quite high and there is an option to skip MLT calculations; in this case the adopted temperature gradient is the adiabatic one.

YREC adopts a step overshoot model, expressed as a number of pressure scale heights at the boundary. Overshoot can be enabled separately for cores, envelopes, and intermediate regimes.

The code has a second, recommended control, which limits the overshoot domain to a fraction of the physical size of the core \citep{Woo2001}. If adopted, the code selects the smaller of these two measures. Checking against the size of the core is important to properly model small convective cores, because the pressure scale height formally diverges at zero radius and becomes very large for small cores.

Overshoot as treated here is more properly overmixing, as the radiative temperature gradient is retained. There is a code option to enforce adiabatic overshoot.

YREC also considers semi-convection, using the \citet{Castellani1971} approach. We are in the process of updating this to a newer approach, but do not yet have a publicly supported alternative.

\subsubsection{Nuclear Reaction Rates}

Nuclear reaction rates are important for both energy generation and composition change. YREC is designed for the hydrogen and helium burning stages of evolution. It also includes light element burning ($^6$Li, $^7$Li, $^9$Be) in the trace element approximation, where the contribution to energy generation is neglected. This is an excellent approximation for these rare species. The code includes full non-equilibrium burning for $^1$H, $^3$He, $^4$He, $^{12}$C, $^{13}$C, $^{14}$N, $^{16}$O, and $^{18}$O.

Deuterium burning is not explicitly traced in the pp chain; however, for young stars non-equilibrium burning of $^2$H into $^3$He is important. The code therefore does track the burning, and associated energy generation, for pre-MS deuterium. It does not, however, treat the pp reaction as a source, so deuterium in the models simply goes to zero, typically early in the pre-MS phase.

For dense nuclear reaction grids, most stellar evolution codes use either analytic fits to reaction rates or a table look-up. The NACRE \citep{Angulo1999} and NACRE-2 \citep{Xu2013} compilations of rates are commonly used. By default, MESA nuclear reaction rates are from JINA REACLIB \citep{Cyburt2010} and NACRE \citep{Angulo1999}. The YREC code uses this approach for He burning, adopting the energy yields and rates of the \citet{Caughlin1988} compilation for triple $\alpha$, $^{12}$C$(\alpha,\gamma)^{16}$O, $^{13}$C$(\alpha,n)^{16}$O, and $^{14}$N$(\alpha,\gamma)^{18}$O.

From the 1990s onwards, YREC has adopted the  \citet{Bahcall1992} approach for hydrogen burning, appropriate for reactions where resonances do not play a strong role. In this approach, we use the charges, masses, the cross section extrapolated to zero energy, $S_0$, and its first and second derivatives, to directly compute reaction rates. YREC includes electron screening of reactions, which includes the intermediate and strong screening cases. However, for general use the default code settings recommend the usage of weak screening \citep{Salpeter1954}. The adopted cross sections are taken from \citet{Adelberger2011}, sometimes referred to as Solar Fusion II. The recently published Solar Fusion III rates \citep{Acharya2025} are an alternative, and a case using these rates is included in our test suite.

Solar neutrino fluxes are computed for all pp and CNO neutrino sources, adopting equilibrium abundances for all isotopes not explicitly tracked. Neutrino losses are subtracted from the effective energy yield of nuclear reactions. YREC uses the \citet{Itoh1996} rates for continuum neutrino cooling. These are important processes for evolved stars, and they can lead to off-center He ignition. 

\subsubsection{Mixing and Composition Change}

YREC solves for non-equilibrium abundance changes due to nuclear reactions using an implicit scheme. For convective regions, instantaneous mixing is assumed, and the reaction rates are therefore mass-averaged across the convection zone. The nuclear reaction rates are computed using start-of-timestep values for the structure variables. There are options to refine these estimates using end-of-timestep values instead. Microscopic diffusion and rotationally induced mixing are treated in separate, sequential steps. 

We note that there is a diversity of approaches in the literature in how to treat the interaction between composition change and the evolution of the stellar structure. At one limit, the \citet{Eggleton1972} method simultaneously solves for composition change and the change in the structural variables. The prime author of the MESA code built his first evolution code using Eggleton as a basis \citep{Paxton2004}, and the MESA code is built using the same principle. The YREC sequential approach is more common, and faster, but it is by no means universal. The same comment applies to the choice of where to evaluate the nuclear reaction rates in the timestep, or how to treat a mixed convection zone whose depth varies during the timestep.

In practice, the numerical controls are typically adjusted such that these different approximations yield consistent results within the domains where they are valid. As an example of a case where these differences matter, consider the case of instantaneous mixing in convection zones. This is usually an excellent approximation, but it breaks down in interesting domains, such as AGB shell flashes. 

\subsubsection{Boundary Conditions}

The boundary conditions are evaluated at three pairs of (L, $T_{\rm eff}$) close to the model values (see section 2.2). The structure variables (P, T, R, L) of the interior solution and their derivatives are matched to the envelope solution, and the three different solutions in turn provides derivatives as a function of (L, $T_{\rm eff}$). 

For the atmosphere, the code assumes constant M, R, and L, and integrates P($T(\tau)$) as a function of optical depth $\tau$ to a fitting point. For the default option, a gray atmosphere, this is $\tau = 2/3$. Other fitting points can be chosen. The integration starts at an arbitrary low density and requires a $T(\tau)$ relationship. Users can also use a table look-up from libraries of model atmospheres. For the existing tables, an optical depth of 2/3 is also adopted. The available options are \citet{1993Kurucz}, \citet{Castelli2003}, and \citet{Allard1995}. The Allard tables are available for solar metallicity only, while the other two are available for a range of metallicities. If the models go outside of the table grid on the hot edge, the table look-up is disabled and the code defaults to an Eddington gray atmosphere. The run will terminate if the code falls outside of the table at high L or low $T_{\rm eff}$.

The atmosphere is then used as a starting point for an envelope integration, which assumes a constant L. The remaining three structure equations are integrated down to a fitting point using a Bulirsch-Stoer integration scheme with an adaptive step size. The physics used in the envelope integration are the same as those used in the interiors model. Only the lower boundary condition of the envelope integration is retained, and these results are output in the last and stored model headers. 

\subsection{Microscopic Diffusion}

In a traditional stellar model, radiative regions are strictly stable. In the real world, this is not true. Heavy species sink relative to light ones due to gravitational settling and thermal diffusion. Large atoms can be driven up relative to nuclei and small atoms by radiation pressure, sometimes referred to as radiative levitation. Collectively, we refer to these mechanisms as microscopic diffusion. The timescale for diffusion is shortest in the outer layers of stars, where the density is low, and it becomes shorter in stars with thin surface convection zones. Mass loss can counteract diffusion, and diffusion therefore becomes less important on the upper main sequence with the onset of strong radiation pressure-driven winds. Microscopic diffusion is therefore particularly important in intermediate mass stars, between 1.3 and 9 $M_{\odot}$ at solar metallicity. Even in the Sun, however, it impacts surface abundances relative to  birth values at the 10\% level.

YREC models can include gravitational settling and thermal diffusion. We follow the approach of \citet{Burgers1969} as implemented by \citet{Thoul1994}. The code includes microscopic diffusion for metals as a group, helium, and the light species $^6$Li, $^7$Li and $^9$Be. Metals are assumed to all settle at the same rate as fully ionized iron. This is consistent with the approximations used in the opacity tables, which assume a fixed heavy element mixture.

In the limit of very efficient microscopic diffusion, stars will develop layer cake structures in their outer layers; pure H at the top, with a layer of pure He below, followed by zones with progressively heavier species. This causes difficulties for codes in thin surface convection zones due to interpolation issues in opacity and equation of state tables. The opacity for zero H, for example, is quite different from that for even a small trace contribution---see \citet{Farag2024} for a discussion of this point. YREC disables microscopic diffusion calculations when the central hydrogen drops below a user-specified threshold (default 0.001), or if the outer point in the model is radiative. In practice, this means that one can choose an outer fitting point deep enough to avoid pathological solutions on the main sequence, which manifest as the code failing due to interpolation errors. 

\subsection{The Structural Effects of Star Spots} \label{sec:starspots}

Magnetic fields are remarkably hard to model, and thus usually neglected in stellar models. At some level, this omission is reasonable; it is unlikely, for example, that there are core magnetic fields strong enough to compete with gas or radiation pressure, at least in sensible stars. However, there are contexts where magnetic fields cannot be ignored, despite the fondest wishes of theorists. Angular momentum loss by magnetized winds, and magnetic instabilities for rotating stars, can play important roles in models of rotating stars. We discuss magnetic fields in these contexts in Section 2.6, where we address rotation.

Another regime, included in YREC, is the direct impact of star spots on the structure and evolution of cool stars. Stars with modest rotation rates and deep surface convection zones have dynamo-generated magnetic fields. The dynamo produces cool star spots with equipartition-strength magnetic fields. As a consequence, the ambient temperature in un-spotted regions is not the same at the surface averaged effective temperature. Convection is also observed to be inhibited in sunspots, and the same should apply to star spots. As a consequence, star spots can be consequential for stellar structure, as originally pointed out in \citet{Spruit1982}.

Although the solar sunspot filling factor is small, of order 0.1 percent, there are stars where spots are far more important. In active main sequence stars, the filling factor of spots can reach levels of order 25 percent; much higher filling factors, 70 percent or more, are seen in pre-main sequence stars \citep{cao2025}.

YREC accounts for both flux blocking and a modified surface boundary condition \citep{Somers2015}. Star spots can induce radius inflation and impact pre-MS lithium depletion. In fully convective stars the luminosity can also be substantively altered. The star spot filling factor and temperature contrast is specified at the start of the run and held fixed, which allows users to test for the structural effects of star spots.

\subsection{Rotation}

Rotation is a fundamental stellar property. Stars are born with a wide range of rotation rates. As a consequence stars of the same birth mass, composition, and age can have very different properties. Rotation changes stellar structure, altering the surface appearance and changing fundamental properties such as lifetimes. It induces mixing in otherwise stable regions, which can be consequential for chemical evolution. Rotation can also be an age indicator, making it a key diagnostic tool for stellar populations.

Rotation is also challenging to model, as it is intertwined with frontier challenges in stellar theory. The interplay of rotation and magneto-convection generates stellar dynamos. Magnetized winds drive angular momentum loss in some regimes, but not others. In radiative regions magnetic fields, waves, and hydrodynamic mechanisms can transport angular momentum, and induce mixing of chemical species, to varying degrees. In convective regions the overturn timescale is short compared to the evolutionary timescale, so a rotation profile will be enforced by the interplay of rotation and turbulence. The natural timescales of both rotation and turbulence vary by orders of magnitude, however, so there is no reason to expect all stars to have some universal profile. It is thus unsurprising that we lack a consensus model for stellar rotation. YREC includes a variety of options, which includes the freedom to explore quite different underlying physical models for the angular momentum evolution of stars.

There is a global correspondence between the MESA and YREC codes in terms of the treatment of rotation. The MESA method is based on the approach of \citet{Heger2000}, which in turn used \citet{Pinsonneault1989} as a starting point. Both papers, in turn, broadly follow the approach of \citet{Endal1976}.  In detail, however, the different codes have distinctive approaches. 
Due to the contributions of many people, over the years these models have evolved and mutated. We document this natural process of rejuvenation, particularly in the treatment of angular momentum transport, loss, and mixing, here.

\subsubsection{The Initial Conditions for Rotation}

The natural initial condition for rotation is in the star formation phase, which in turn divides into two phases. Stellar modeling typically begins at the end of the hydrodynamic collapse phase. At this phase the star is assumed to have reached its final mass and hydrostatic equilibrium is a good approximation. In YREC, this would correspond to the deuterium burning birthline. Spherical models can be converted into uniformly rotating ones at a user-specified rate for any model, but the birthline is a physically motivated state, at least for low mass stars.

Starting models much further up the Hayashi track exist, but these do not correspond to real stars. This is an important point for initializing rotation. If a user initializes rotation in an artificially large starting model, angular momentum conservation will transform a reasonable sounding guess into an unreasonable outcome.

During the hydrostatic pre-MS, star-disk coupling can occur \citep{Konigl1991,Shu1994}. It is also necessary to explain the evolution of rotation in young stars. As a result, YREC also includes the option to enforce star-disk coupling for a user-specified timescale at a user-specified rate.

For massive stars, it is less clear that the deuterium burning birthline is a true representation of the starting state \citep{Palla1991}. Above a critical mass, of order 5 to 8 solar masses, accreting models develop radiative cores and are born close to the main sequence. For higher mass models, an initial rotation state on the main sequence may therefore be closer to being physically reasonable.

\subsubsection{The Structural Effects of Rotation}

Rotation induces a departure from spherical symmetry which can be significant. The net effect is a reduction in the effective gravitational force, which makes rotating stars behave for most purposes as if they had a lower effective mass. However, in detail the impact depends on the angular momentum distribution. The general case would require a two or three dimensional model. Fortunately, there are reasonable families of models that can be captured using a pseudo-2D angular velocity distribution. In this case, the structural effects of rotation can be computed by measuring the shape of equipotential surfaces as a function of mass. The impact of rotation on the structure can then be included with correction terms applied to the spherically symmetric equations of stellar structure.

YREC uses a modified version of the  \citet{Kippenhahn1970} technique. See \cite{Pinsonneault1989} for a discussion of the differences. This method assumes uniform rotation on equipotential surfaces. The code then solves for the shape of these surfaces, including contributions from the quadrupole term in the potential, and solves a modified set of stellar structure equations.

\subsubsection{Rotation in Convective Regions}

YREC assumes that a rotation law is enforced in convective regions on a short timescale. The code permits rotation profiles of the form $\omega \sim R^{\alpha},-2 \leq \alpha \leq 0$. There are also options to couple the rotation profile to adjacent radiative regions - for example, to match the rate at the base of a convective region to the rate in the radiative zone below it. These options have a significant impact for the rotational evolution of post-main sequence stars, where observational data suggests that core rotation rates can be tens to hundreds of times faster than the envelope rotation rates \citep[e.g.]{Mosser2012rot, Gehan2018}, with important implications for internal angular momentum transport \citep{Tayar2013, Fuller2019}. Both the surface and internal rotation of core He burning stars are important tests of angular momentum loss \citep{Tayar2018}. Rotation and magnetism may also affect mass loss on the red giant branch \citep{Li2025}. 

YREC does not include differential rotation with latitude. For sunlike stars, the average rotation with depth is nearly constant, even if there is latitudinal differential rotation. We therefore recommend that users impose rigid rotation for pre-MS and MS cases. However, they should keep in mind that imposing a uniform rotation rate in convection zones may be a poor approximation for luminous giants. It is difficult to explain the survival of rotation in blue core He-burning stars without strong differential rotation in luminous red giants \citep{Pinsonneault1991, Sills2000}. This differential rotation can also impact the interpretation of red giant core rotation rates \citep{Kissin2015} and core He burning stars \citep{Tayar2018}.

\subsubsection{Angular Momentum Loss From Magnetized Winds}\label{sec:winds}

Users may choose between braking laws that describe the angular momentum loss via magnetized winds. Magnetic braking is the consequence of the interaction of mass loss and magnetic fields. As both are frontier fields in stellar physics, the description of their interaction is unsurprisingly burdened with many assumptions. 

Broadly, the rate of angular momentum (AM) loss is set by a combination of the mass loss rate ($\dot{M}$), the rotation rate ($\Omega$), and the effective magnetic lever arm, parameterized as the Alfv\'{e}n radius ($R_A$). The Alfv\'{e}n radius itself generally depends on $\dot{M}$, stellar magnetic field ($B$), and wind driving physics. The overall field configuration (dipole, quadropole, etc.) affects how strongly the torque scales with the magnetic lever arm. YREC has three built-in options for the braking law that differ in their assumptions about each of these ingredients. 

The first is a modified version of the \citet{Kawaler1988} prescription proposed by \citet{Krishnamurthi1997}. The Kawaler braking law assumes that a dipole magnetic field scales as $B R^2 \sim \Omega$. The torque for slow rotators therefore scales as $\Omega^3$. \citet{Krishnamurthi1997} added a Rossby-scaled magnetic saturation threshold to the standard form of the Kawaler law, beyond which the AM loss scales more weakly with rotation rate. This saturation is a ubiquitous feature of magnetic proxies in stars: they take maximal but constant values at sufficiently rapid rotation and low Rossby number $Ro$ (the rotation period divided by the convective overturn timescale).

The second form is the loss law from \citet{vansaders2013}, which uses a parametrized model of the steady-state MHD wind solutions from \citet{Matt2008} and \citet{Matt2012}. The magnetic field is scaled assuming a linear (Rossby-scaled) dynamo and pressure equipartition at the photosphere: $B\sim Ro^2
P_{phot}^{1/2}$. Mass loss is scaled as $\dot{M} \sim L_{X} \sim (\omega \tau_{cz})^2$ following the observed correlation of $\dot{M}$ with X-ray luminosity \citep{Wood2005} and scaling of the X-ray luminosity with Rossby number \citep{Pizzolato2003}. The liberal use of Rossby scalings causes the torque to be significantly weaker in stars with thin convective envelopes than that in a Kawaler prescription, which better matches the actual rotational behavior of stars near the Kraft break.   

Finally, users can enable a ``custom” mode where they can manually set the exponents that describe the dynamo scalings, Rossby scalings, and effective magnetic index.

Although the braking utilizes an inferred $\dot{M}$, this $\dot{M}$ is not self-consistently incorporated into the structural evolution.  All prescriptions available in YREC allow the user to specify a disk-locking rotation period and duration. Magnetic saturation of the torque is accounted for, with the saturation threshold set by the user. Because the field strengths, mass loss rates, and overall torque magnitudes are not known from first principles, practical use always involves the adjustment of a scale factor to calibrate the magnitude of the torque against benchmark observations. Sample cases of such a calibration are discussed in Section \ref{sec:gyro}.

\subsubsection{Angular Momentum Transport and Rotationally Induced Mixing in Radiative Regions}

The original YREC code included angular momentum transport and the associated mixing via hydrodynamic mechanisms. YREC solves diffusion equations for the transport of angular momentum and the associated rotationally induced mixing. Instantaneous mixing is assumed in convective regions. Angular momentum loss is induced in the surface convection zone as a surface boundary condition.

The secular mechanisms included are meridional circulation, the secular shear instability, and the Goldreich-Schubert-Fricke instability. See \citet{Somers2016} for a recent summary of the assumed input physics. This differs from the \citet{Pinsonneault1989} approach in that the characteristic length scale for the instabilities is the radius, rather than the velocity scale height. The meridional circulation treatment is also modeled on the \citet{zahn1992} approach, neglecting terms of order the second or higher derivative of the angular velocity but including the quadrupole component of the potential. The hyper-diffusion terms in the \citet{zahn1992} method require a very different numerical treatment than the one used here. This expression is broadly similar to others in the literature. We include the reduction in the effective meridional circulation velocity induced by mean molecular weight ($\mu$) gradients following the approach of \citet{Maeder1998}. We also adopt the \citet{Maeder1997} approach for the secular shear instability. The code permits different treatments of the inhibiting effect of mean molecular weight gradients on angular momentum transport and the associated mixing.

A central result from \citet{Pinsonneault1989} was that the efficiency of angular momentum transport must be much higher than that of the associated mixing. This was placed on a physical basis by \citet{Chaboyer1992}, who argued that the turbulence induced by rotation is much more effective in the horizontal direction than in the vertical direction. The former can exchange angular momentum, while mixing is only induced in the vertical direction. YREC treats this as a free parameter, $f_c$, which can be used to adjust the efficiency of rotationally induced mixing. This differs from the \citet{zahn1992} approach, where this is a fixed parameter of the model. In our view, the assumptions embedded in the approach make a free parameter an appropriate choice.

Wave-driven transport is not supported in the public release, but we intend to add in in later releases. Magnetic transport is modeled under two different treatments. Strong magnetic coupling, as that predicted from the Taylor-Spruit mechanism, is effectively equivalent to uniform rotation on the lower main sequence. We therefore include an option to simply assume uniform rotation at all times, which mimics the effects of strong coupling. We also include an option for a diffusive magnetic transport term \citep{Somers2016} that does not induce mixing; this can be adjusted to reproduce, for example, the core-envelope coupling timescale inferred from the spin down of low mass stars, or the core to surface rotation contrast in the Sun. This approach is now used in other codes to mimic magnetic transport \citep{eggenberger2019}. The code also allows users to couple core rotation profiles to differentially rotating convective envelopes.

\section{Sample Applications and Literature Comparisons}

\label{section:Models}

The life cycles of stars are wondrously complex. YREC is a flexible tool, which can simulate many scenarios. 
It is important to understand the precision, accuracy, and domain of validity for the code. We therefore present different physical scenarios as well as different mass, composition, and evolutionary state domains for the code. We begin by laying out the properties of our ``base case'' models, and follow by discussing models with different input physics or initial conditions. We discuss scripts to run sets of models, and tools to extract information from the code, in Section \ref{section:Scripts}.

\subsection{The Base Case and Numerical Precision Settings}\label{basecase}

Our base case adopts the physical assumptions listed in Table \ref{Table:gridphysics}, and---coupled with the test suite cases---represents a reasonable starting point for users developing their own input namelists. 
We use a solar calibration for the convective mixing length $\alpha$ and the birth solar helium $Y_{\odot, init}$. For models with non-solar metal content, we use the difference between the solar and Big Bang nucleosynthetic Y abundance \citep{2020Planck} to define the helium to metal enrichment ratio $\Delta Y/\Delta Z$. We note that other codes and isochrones make different assumptions about this ratio, which also depends on the choice of input physics. Our adopted birth hydrogen and metal abundances (X and Z) as a function of [Fe/H] are also given in Table \ref{Table:gridphysics}, assuming a GS98 abundance pattern.

\begin{deluxetable}{p{3cm}p{3cm}} 
\tabletypesize{\scriptsize}
\tablecaption{Summary of Input Physics for the Base Model Grid \label{Table:gridphysics}}
\tablehead{
    \colhead{Parameter} & \colhead{YREC}}
\startdata
Atmosphere & Gray   \\
Convective Core Overshoot & Lesser of  Step: 0.2H$_{\mathrm{p}}$, 0.15 core size\\
Diffusion & None \\
Equation of State & OPAL\textsuperscript{a,b} {+} SCV\textsuperscript{c} \\
High-Temperature Opacities & OP\textsuperscript{d} \\
Low-Temperature Opacities & \citet{Ferguson2005} \\
Mixing Length & 1.707873\\
Nuclear Reaction Rates & \citet{Adelberger2011} \\
Rotation & None \\
Weak Screening & \citet{Salpeter1954} \\
Mixture and Solar $\mathrm{Z/X}$ & \citet{grevesse_sauval1998}  \\
Solar X & 0.719600 \\
Solar Y & 0.263908 \\
Solar Z & 0.016492 \\
$\Delta\mathrm{Y/}\Delta\mathrm{Z}$ & 0.951249 \\
Surface $\mathrm{(Z/X)}_\odot$ & 0.02292 \\ 
\hline
\multicolumn{2}{c}{Metallicity Map} \\[3pt]
$[\mathrm{Fe/H}]$ & $\quad X \hspace{1.4cm} Z$ \\
$-1.0$ & 0.750950 \quad\quad 0.001721 \\
$-0.5$ & 0.743176 \quad\quad 0.005386 \\
$+0.5$ & 0.654060 \quad\quad 0.047401 \\[6pt]
\enddata
{\scriptsize
\tablenotetext{}{
\textsuperscript{a}~\citet{Rogers1996};
\textsuperscript{b}~\citet{Rogers2002};
\textsuperscript{c}~\citet{Saumon1995};
\textsuperscript{d}~\citet{Badnell2005}
}
}
\end{deluxetable}

The numerical choices in our models are customized to the requirements for different use cases. In some cases, the tolerances for numerical precision had to be relaxed to obtain a solution. We note the domains where the recommended numerical controls differ significantly from the general case in the discussion below.

Our spatial resolution, temporal resolution, and convergence criteria for all cases are defined such that the lifetime of each phase is precise at the 1 percent level. Solar models require $L=L_{\odot}$, $3.827 \times 10^{33}$ erg $s^{-1}$, and $R=R_{\odot}$, $6.958 \times  10^{10}$ cm,  at the solar age, 4.568 Gyr, within one part in $10^{-4}$. The surface $\frac{Z}{X}$ is required to be within 1 percent of the current solar surface $\frac{Z_{\odot}}{X_{\odot}}$, 0.02292 for the \citet{grevesse_sauval1998} mixture. We also require that neutrino fluxes were precise at the 1 percent level. Solar models including rotation also reproduce the solar equatorial rotation rate (25.4 days) and the difference between the meteoritic and the current solar surface $^7$Li abundance (2.3 dex) within 1 percent. These are achieved by varying the scaling constant in the magnetized wind and the efficiency of mixing relative to angular momentum transport, respectively. 

The inner fitting point is set at $10^{-4} M_{\rm tot}$. For typical main sequence conditions, this places the inner fitting point close to $0.01 R_{tot}$. Solar models, and the models showing diffusion and mixing to the terminal-age main sequence (TAMS), have the outer fitting point set at $10^{-7} M_{\odot}$  below the surface. For solar models, this places the outer fitting point close to $0.99 R_{\rm tot}$. General purpose stellar evolution runs have the outer fitting point set at $10^{-4} M_{\rm tot}$  below the surface, which on the main sequence corresponds to roughly $0.95 R_{\rm tot}$. In some cases the fitting point location is adjusted differently to achieve proper numerical results. We also evolved some models in stages, with somewhat different numerical tolerances in different regimes.

We use Hayashi track seed models to generate models at the beginning of the deuterium burning birthline, defined as the location where 1 percent of the birth deuterium has been burned. These models are the starting point for the large majority of the cases that we present here. 
We provide libraries of starting models with a range of masses and metallicities. In addition to the GS98 mix, we also provide models with the \citet{Asplund2021}, \citet{Magg2022}, and \citet{Lodders2025} heavy element mixtures. The default YREC models use the \citet{Lodders2021} isotopic ratios for CNO and the light element abundances D, $^3$He, $^6$Li, $^7$Li and $^9$Be.

\subsubsection{Run-speed comparison with MESA}

We perform rough speed comparisons between MESA and YREC for low mass stars. For YREC, we run a solar mass, solar metallicity model up to the tip of the red giant branch using the same numerical and physics settings as the base model grid in Section \ref{sec:base_case}. For the MESA comparison model, we use MESA version r24.08.1 with most inlist values left at their default settings except those required to match the input physics choices of the YREC model; an example inlist is provided in the sample cases. We adjusted the global mesh variable \texttt{mesh\_delta\_coeff} in MESA for each phase of evolution to roughly match the number of mesh points in the YREC models. The average number of shells across each phase of evolution we consider is within 3\% for both codes. Models were run on a single thread to mimic YREC execution and to make the comparison more hardware agnostic, although in practice users of MESA would utilize all available cores on their machines. From pre-main sequence initialization to the ZAMS, YREC is roughly 8 times faster than MESA at similar mesh resolution; from the ZAMS to TAMS  it is $\sim2\times$ faster; and from the TAMS to TRGB, YREC is $\sim6\times$ faster. ``Default" YREC runs with fewer shells than default MESA: the mesh was made sparser in each of these evolutionary phases to match YREC. In practice, the speed gains may therefore be even greater when models are tuned to achieve the numerical precision required for the science question at hand. Users can use the simplified Yale EoS for another factor of $\sim9$ speed gain over the base case pre-MS to TRGB run, which allows for rapid testing (54 to 72 times faster execution time than MESA) with only modest sacrifices in physical accuracy. 

\subsection{The Test Suite}

A test suite is provided to demonstrate YREC's capabilties and provide for a baseline reference to aid in regression testing during development. It is not yet comprehensive, but is being added to regularly and our aim is to provide full coverage of use cases and model types that YREC supports. Information on setting up and running all or a portion of the pre-defined test suite or adding one's own test cases may be found in the `examples' subdirectory of the project file tree.
 
Our test suite is designed to test physical scenarios in current use, and to include a wide range of physical conditions. We recommend that users who modify the code validate their changes against the standard library of test suite results. Our physical assumptions, the masses, birth compositions, and mixing lengths are listed in Table \ref{Table:testsuite}. We will present the results from the test suite and other models in subsequent sections. The namelists also provide examples of how users can customize the code for different use cases.

\begin{deluxetable}{p{6cm}p{3cm}p{6cm}} 
\tabletypesize{\scriptsize}
\tablecaption{Cases in the YREC test suite. Column 1 contains the name, or names, of the cases; Column 2 is the type, and Column 3 is a brief description. Solar type runs are calibrated solar models, designed to showcase common use cases. Zero-age main sequence (ZAMS) runs, defined as where core H drops 0.005 below the birth value, are for the lower main sequence, and highlight the impact of different choices of surface boundary conditions and equations of state. Brown dwarf runs, taken to 15 Gyr, test code performance below the H burning limit. Terminal-age main sequence (TAMS) runs are taken to core H exhaustion, defined as when the central H drops to 0.0001. TAMS runs show the effects of diffusion, rotation, and rotationally induced mixing below and just above the Kraft break. Evolution runs are longer sets chained together. The 0.3 $M_{\odot}$ case runs first to the ZAMS and next to the TAMS. The 1.0 $M_{\odot}$ case runs first to the TAMS, then to He ignition in a degenerate flash. It then restarts at the zero-age horizontal branch (ZAHB), and ends at core He exhausion (TAHB). The 3.0 and 9.0 $M_{\odot}$ cases are evolved to the TAMS, the ZAHB, and the TAHB in three runs.}
\label{Table:testsuite}
\tablehead{
    \colhead{Case Name} & \colhead{Type}& \colhead{Descriptions}}
\startdata
Test solarGS98 base & Solar & Base case.   \\
Test solarGS98 dif & Solar & Diffusion included.   \\
Test solarAAG21 dif & Solar & AAG21 mix, diffusion included.   \\
Test solarM22M dif & Solar & Magg22 mix, diffusion included.   \\
Test solarGS98 dif rot & Solar & Diffusion, rotation, and rotational mixing included.   \\
Test solarGS98 dif rot fast & Solar & As the dif rot case, but rapid birth rotation.   \\
Test solarGS98 dif rot solid & Solar & As the dif rot case, but solid body rotation.   \\
Test solarGS98 kurucz & Solar & Kurucz atmosphere SBC.   \\
Test solarGS98 SF3 & Solar & Solar fusion III nuclear reaction rates.   \\
Test solarGS98 yaleeos & Solar & Yale EoS. \\  
m0p03feh+0.0GS98 allard 15 Gyr & Brown Dwarf & $0.03 M_{\odot}$, Allard SBC, to 15 Gyr.   \\
m0p05feh+0.0GS98 allard 15 Gyr & Brown Dwarf & $0.05 M_{\odot}$, Allard SBC, to 15 Gyr.   \\
m0p08feh+0.0GS98 allard 15 Gyr & Brown Dwarf & $0.08 M_{\odot}$, Allard SBC, to 15 Gyr.   \\
m0p1feh+0.0GS98 allard ZAMS & ZAMS & $0.1 M_{\odot}$, Allard SBC. \\
m0p1feh+0.0GS98 base ZAMS & ZAMS & $0.1 M_{\odot}$, base case. \\
m0p1feh+0.0GS98 spot25 ZAMS & ZAMS & $0.1 M_{\odot}$, fspot 0.25 case. \\
m0p3feh+0.0GS98 allard ZAMS & ZAMS & $0.3 M_{\odot}$, Allard SBC. \\
m0p3feh+0.0GS98 spot25 ZAMS & ZAMS & $0.3 M_{\odot}$, fspot 0.25 case. \\
m0p3feh+0.0GS98 scveos ZAMS & ZAMS & $0.3 M_{\odot}$, SCV EoS. \\
m0p3feh+0.0GS98 yaleeos ZAMS & ZAMS & $0.3 M_{\odot}$, Yale EoS. \\
m1p0feh+0.0GS98 spot25 ZAMS & ZAMS & $1.0 M_{\odot}$, fspot 0.25 case. \\
m1p0feh+0.0GS98 dif TAMS & TAMS & $1.0 M_{\odot}$, diffusion included, evolved to TAMS. \\
m1p4feh+0.0GS98 dif TAMS & TAMS & $1.4 M_{\odot}$, diffusion included, evolved to TAMS. \\
m0p3feh+0.0GS98 base ZAMS, TAMS & Evolution & $0.3 M_{\odot}$, base case evolved to ZAMS, TAMS. \\
m1p0feh+0.0GS98 base TAMS, HeIgnite, TAHB & Evolution & $1.0 M_{\odot}$, base cases, evolved to the TAMS, He ignition, and from core He burning to core He exhaustion. \\
m3p0feh+0.0GS98 base TAMS, ZAHB, TAHB & Evolution & $3.0 M_{\odot}$, base cases,  evolved to the TAMS, ZAHB and TAHB. \\
m9p0feh+0.0GS98 base TAMS, ZAHB, TAHB & Evolution & $9.0 M_{\odot}$, base cases, evolved to the TAMS, ZAHB and TAHB.\\
m9p0feh+0.0GS98 yaleeos TAMS, ZAHB, TAHB & Evolution & $9.0 M_{\odot}$, base cases with Yale EoS, evolved to the TAMS, ZAHB and TAHB.\\
Solar\_m1p0feh+0p0\_GN93\_TAHB & TAHB & $1.0 M_{\odot}$ Z=0.019 ZAHB evolved to the TAHB, with GN93 opacity tables.\\
Solar\_m0p7feh-0p75\_GN93\_TAHB & TAHB & $0.7 M_{\odot}$ Z=0.003 ZAHB evolved to the TAHB, with GN93 opacity tables.\\
Solar\_m0p95feh+0p0\_GN93\_TAHB & TAHB & $0.95 M_{\odot}$ Z=0.019 ZAHB evolved to the TAHB via ZAHB seed envelope mass re-scaling, with GN93 opacity tables.\\
\enddata

\end{deluxetable}

\subsection{The Base Case Model Grid} \label{sec:base_case}

We adopted the test suite templates to run our base grid of models, using the same physical assumptions described in Section \ref{basecase}. The grid spans masses from 0.3 to 8.0 $M_\odot$ in increments of 0.1 $M_\odot$, and metallicities of [Fe/H] = $-1.0$, $-0.5$, $0.0$, and $+0.5$. Models with initial masses below 2.1 $M_\odot$ were evolved to the tip of the red giant branch, while the remainder were evolved to the onset of core helium burning. We list in Table \ref{Table:gridphysics} the physical assumptions made to generate the grid. 
We use this grid to define the regime where YREC is expected to produce reliable results for stellar evolution and to demonstrate its predictions.

\subsubsection{From the Deuterium Birthline to Core He ignition.}

\begin{figure}[htbp]
    
    \centering
    \includegraphics[width=8 cm]{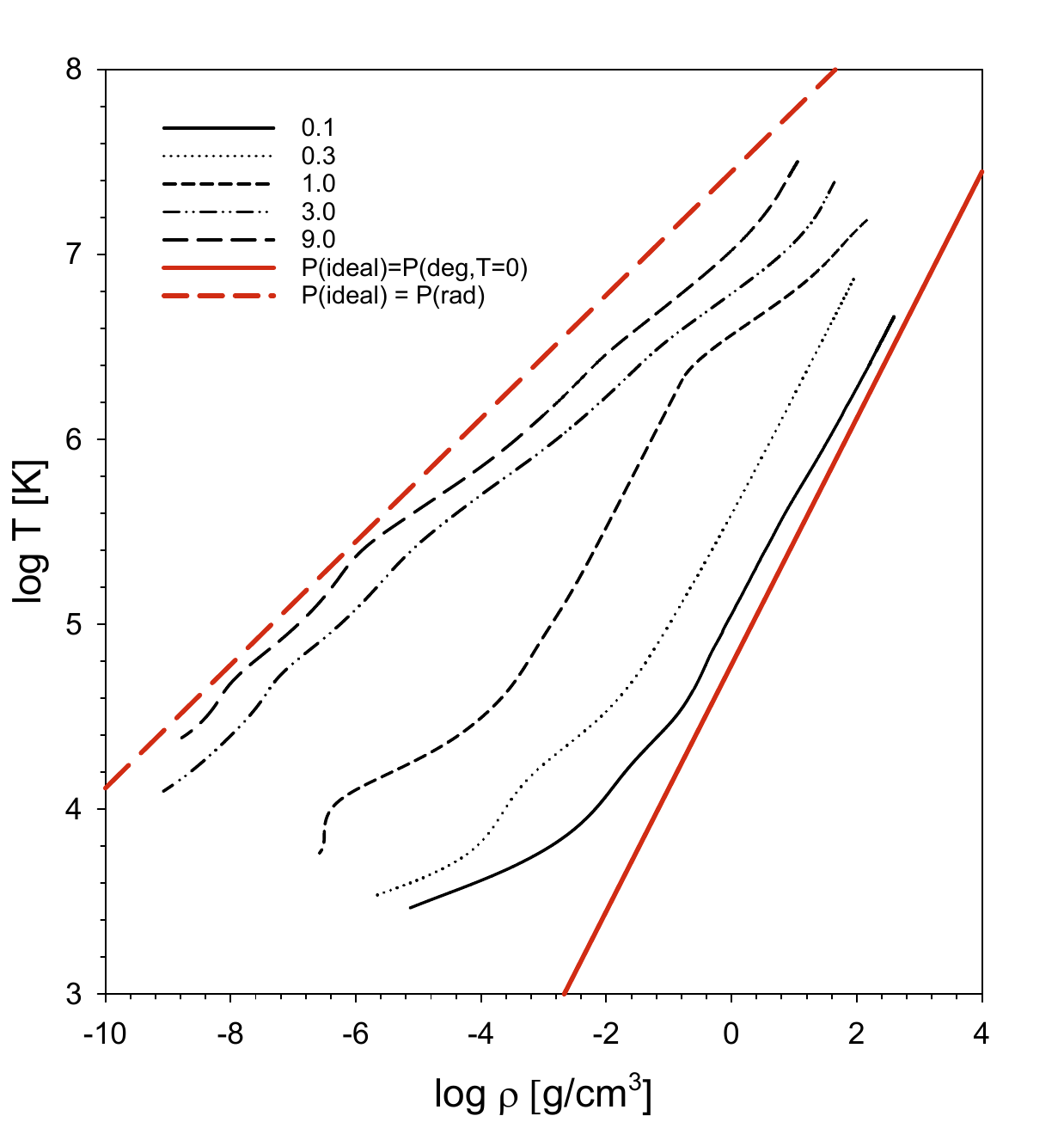}
    \caption{The locus in the $\rho-T$ plane of 0.1(lower right), 0.3, 1, 3 and 9 $M_{\odot}$ models (upper left) at the zero-age main sequence, defined here as the point where the central hydrogen mass fraction is 0.005 lower than the birth value. These are solar metallicity models using the base model physics. The domains where radiation pressure equals gas pressure, and where zero-temperature degeneracy pressure equals gas pressure, are indicated by the respective red lines.}

\label{fig:zamsrhoT}

\end{figure}

Snapshots of the structure of zero-age main sequence models is a logical starting point. 
In Figure \ref{fig:zamsrhoT}, we show the internal structure of representative ZAMS models in the $\rho-T$ plane, an important indicator of the predominant pressure sources from the equation of state.  We also show the boundaries where radiation pressure (upper left) and degeneracy pressure (lower right) dominate.

\begin{figure}[htbp]
\includegraphics[width=8 cm]{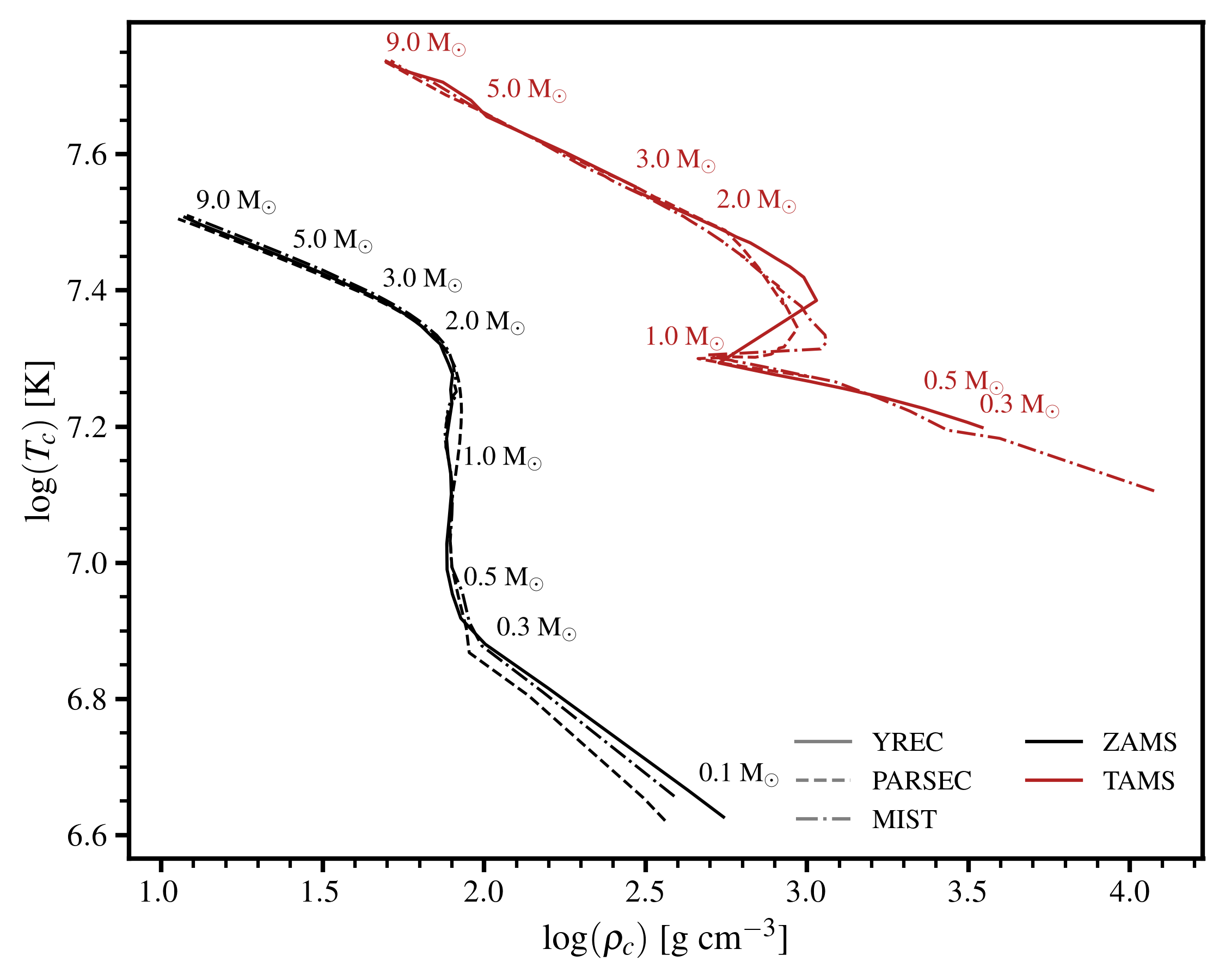}
\caption{Central temperature as a function of central density at ZAMS (black) and TAMS (red) locations. Comparisons are shown between YREC, MIST, and PARSEC evolution tracks for solar metallicity (YREC, $Z = 0.016492$; MIST, $Z = 0.0142$) and $Z = 0.017$ (PARSEC).}
\label{fig:cdensity_ctemp}
\end{figure}

The central temperature and density for models as a function of mass is also an interesting diagnostic of stellar structure. We present our models at the start (ZAMS) and end (TAMS) of the main sequence, compared with PARSEC and MIST, in Figure \ref{fig:cdensity_ctemp}. The upper MS has a shallow slope in central T as a function of mass due to the strong temperature dependence of the CNO cycle; in lower mass stars the pp chain is more important and the slope is steeper there. The steep change in central density close to solar mass reflects the transition from radiative to convective envelopes. Different codes are very close above 0.5 $M_{\odot}$ on the ZAMS, with the differences at lower mass reflecting the importance of the equation of state and surface boundary conditions for these stars. Although both $\rho$ and T increase on the MS, the increases are much larger for $\rho$. Physical conditions can change rapidly after the MS turnoff, which accounts for some of the differences seen between 1 and 2 $M_{\odot}$.

\begin{figure}[htbp]
\includegraphics[width=8 cm]{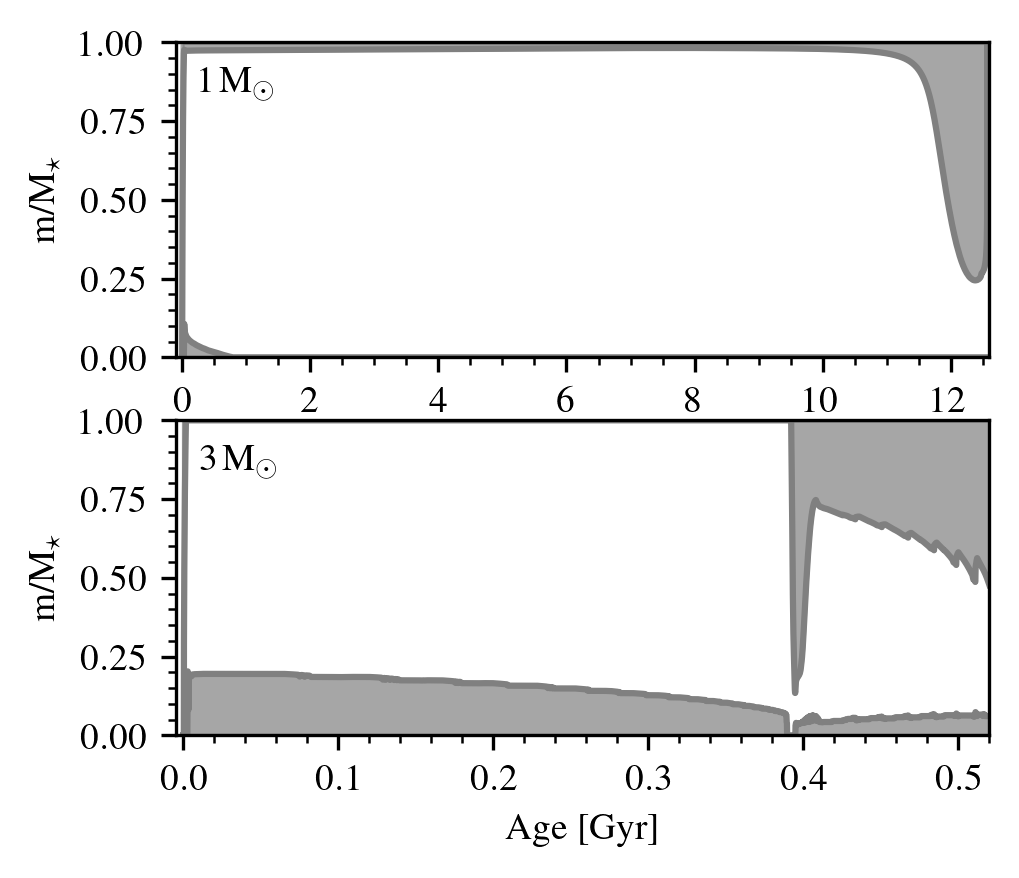}
\caption{Kippenhahn diagram showing the evolution of convective boundaries for the 1M$_\odot$ (top panel) and 3M$_\odot$ (bottom panel) models. Shaded gray regions indicate convective zones shown as a function of age and fractional mass coordinate.}
\label{fig:kipp}
\end{figure}

The depth of core and surface convection zones are another interesting diagnostic of stellar evolution, frequently displayed as a function of time in Kippenhahn diagrams. We show the histories of 1 and 3 $M_{\odot}$ models in Figure \ref{fig:kipp}. The 1 $M_{\odot}$ model has a shallow (in mass) convective envelope on the MS, develops a deep surface convection zone during an extended red giant branch phase, and has a convective core during the core He-burning phase. The 3 $M_{\odot}$ model has a convective core that gradually shrinks during the MS phase, followed by a brief shell H-burning red giant phase. The core He-burning phase is a larger fraction of the lifetime in this model than in the lower mass one.

\begin{figure}[htbp]
\includegraphics[width=8 cm]{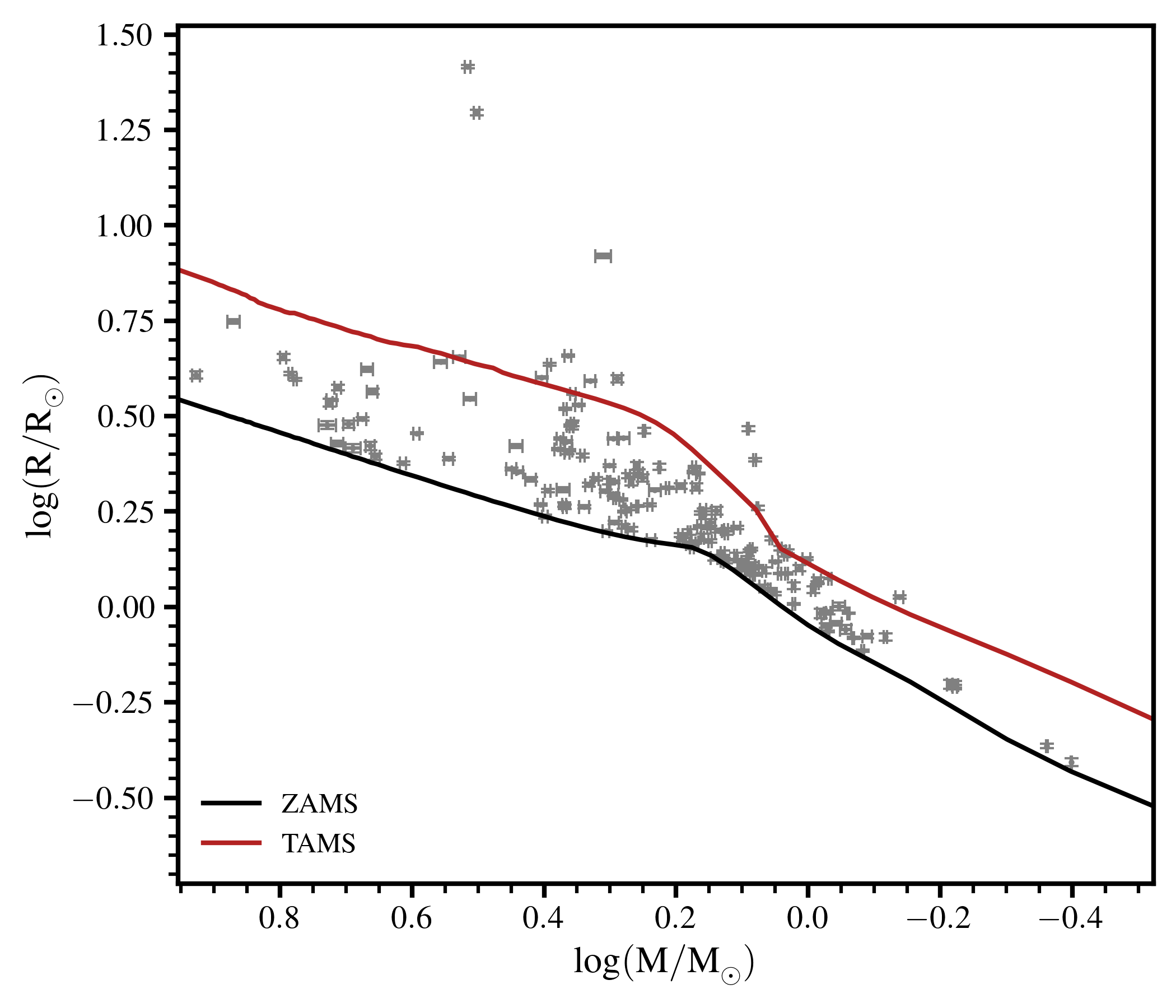}
\caption{Radius as a function of mass for tracks 0.3 - 9.0 M$_\odot$ at ZAMS and TAMS. \citet{2010Torres} masses and radii of eclipsing binaries plotted for reference.}
\label{fig:MR_torres}
\end{figure}

Figure \ref{fig:MR_torres} shows the stellar radius as a function of mass, with the black and red lines corresponding to the ZAMS and TAMS locations. The lower envelope is in excellent agreement with fundamental data from eclipsing binary stars \citep{2010Torres}. The apparent agreement is deceptive for the lower MS, where the TAMS greatly exceeds the age of the Universe. We discuss the ``radius inflation'' problem in Section 3.4.5. 

\begin{figure}[htbp]
\includegraphics[width=8 cm]{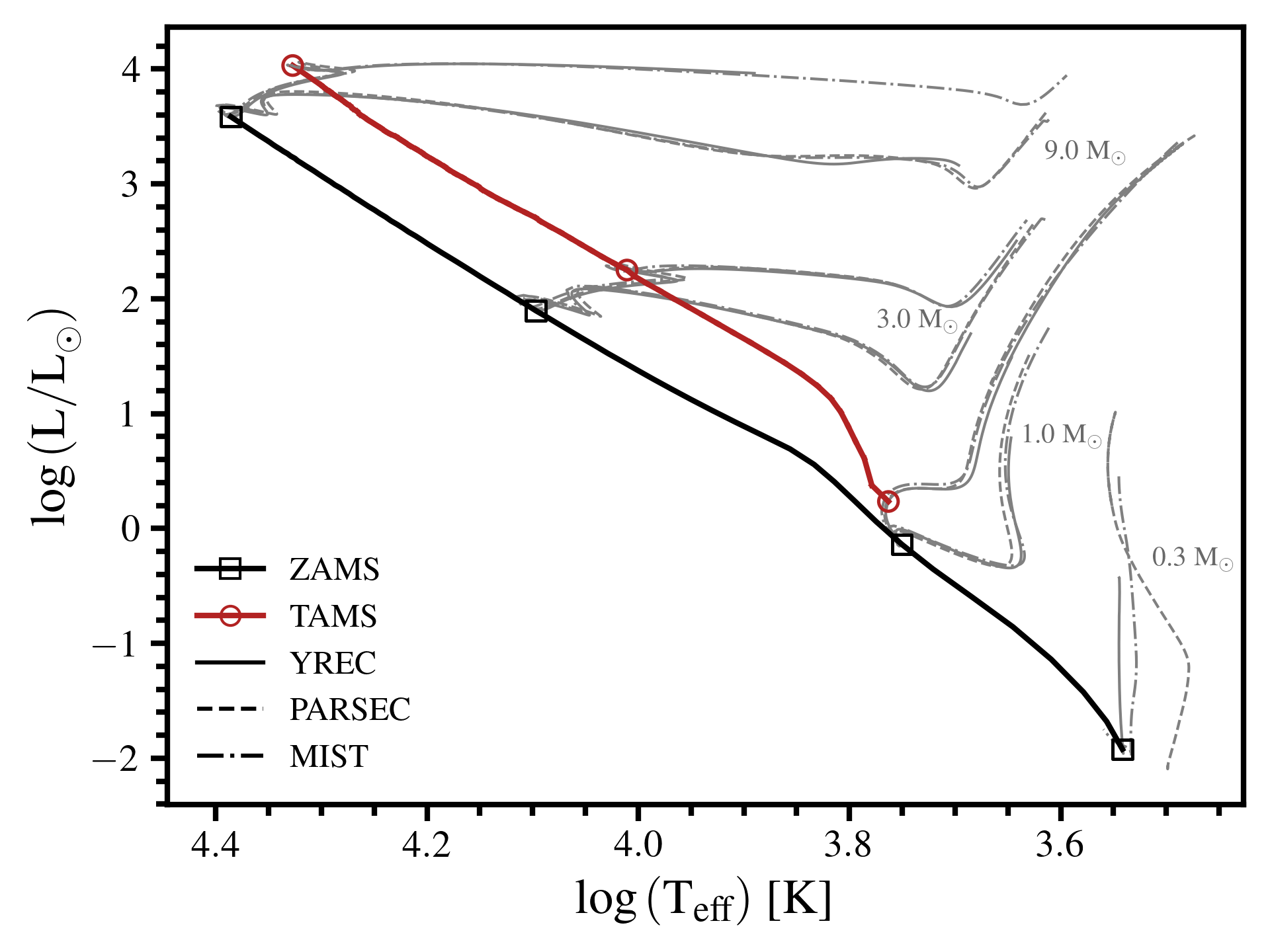}
\caption{Luminosity as a function of temperature for the solar metallicity models. The solid black and red lines show the location of ZAMS and TAMS, respectively. We show a comparison of evolutionary tracks from YREC (solid), PARSEC (dashed), and MIST (dash-dot) model grids for solar metallicity and Z = 0.017 for PARSEC. Tracks are shown for masses of 0.3, 1.0, 3.0, and 9.0 M$_\odot$, as available in each model library. 1.0, 3.0, and 9.0 M$_\odot$ tracks are ended at He core ignition.}
\label{fig:zamsloc}
\end{figure}

In Figure \ref{fig:zamsloc}, we show the evolutionary tracks for solar metallicity models at 0.3, 1.0, 3.0, and 9.0 M$_\odot$. The main sequence band is superposed for reference. The lower branch for each mass track is the pre-MS phase, defined at deuterium ignition. This definition breaks down for the more massive stars, where a physical model would predict a starting point closer to the MS \citep{Palla1991,Yorke2002}. A transient loop above the MS, due to non-equilibrium CN burning, is clearly seen in the higher mass tracks, as is the sudden adjustment at the end of the MS phase from H depletion in a large convective core. The 1 and 3 M$_\odot$ cases are similar in L during the core He-burning phase, while the 9 M$_\odot$ case is significantly brighter due to a larger H-exhausted core on the MS.

The non-rotating models from the MIST \citep{2016Choi} and PARSEC (v1.2 and v2.0, \citealt{2012Bressan}; \citealt{2014Chen}; \citealt{2015Chen}; \citealt{2018Fu} \citealt{2022Nguyen}; \citealt{2025Costa}) libraries are overlaid for comparison with YREC \footnote{The PARSEC library provides only pre-MS to ZAMS for 0.3 M$_\odot$ and does not include tracks at its defined solar metallicity, where Z$_\odot$ = 0.01524 and Z$_{\mathrm{initial}} = 0.01774$ (\citealt{2012Bressan}; \citealt{2014Chen}). For this reason, we use the Z = 0.017 tracks for comparison.}. The tracks for different codes are broadly similar at solar mass and above. Tracks for the lowest mass model (0.3 $M_{\odot}$) are quite different, reflecting the importance of the equation of state and surface boundary conditions. See Section \ref{sec:lowerMS} for a further discussion of this point. 

Table \ref{Table:modelcomparison} provides a summary of stellar parameters at TAMS for the 1M$_\odot$ tracks, including age, luminosity, effective temperature, radius, central density, and central temperature. The TAMS location is defined as the point where the central H abundance drops to 0.0001. Both MIST and PARSEC models show close agreement with our base grid in central temperature, effective temperature, and radius, each agreeing within 3\%. The MIST models exhibit slightly larger offsets in luminosity and age (6 and 10\%, respectively), while the PARSEC models differ primarily in central density and age (6\% and 7\%). These discrepancies can be partially attributed to differences in the precise definition of the TAMS point in each grid and variations in the adopted input physics. For instance, the MIST models use the \citet{1993Kurucz} surface boundary conditions, whereas this grid adopts the Eddington-grey approximation.   

\begin{deluxetable}{cccc} 
\tabletypesize{\scriptsize}
\tablecaption{Comparison of stellar properties between YREC, MIST, and PARSEC for 1M${_{\odot}}$ at TAMS. \label{Table:modelcomparison}}
\tablehead{
    \colhead{Parameter} & \colhead{YREC} & \colhead{MIST} & \colhead{PARSEC}}
\startdata
Age [Gyr] & 9.807 & 8.781 & 9.081 \\
log(L/L$_{\odot}$) & 0.235 & 0.260 & 0.223 \\ 
log(T${_\mathrm{eff}}$) [K] & 3.763 & 3.766 & 3.765 \\ 
log(R/R$_{\odot}$) & 0.114 & 0.120 & 0.106 \\
log($\rho_c$) [g cm$^{-3}$] & 2.812 & 2.798 & 2.784 \\
log($T_c$) [K] & 7.285 & 7.294 & 7.290 
\enddata
\end{deluxetable}

\begin{figure}[htbp]
\includegraphics[width=8 cm]{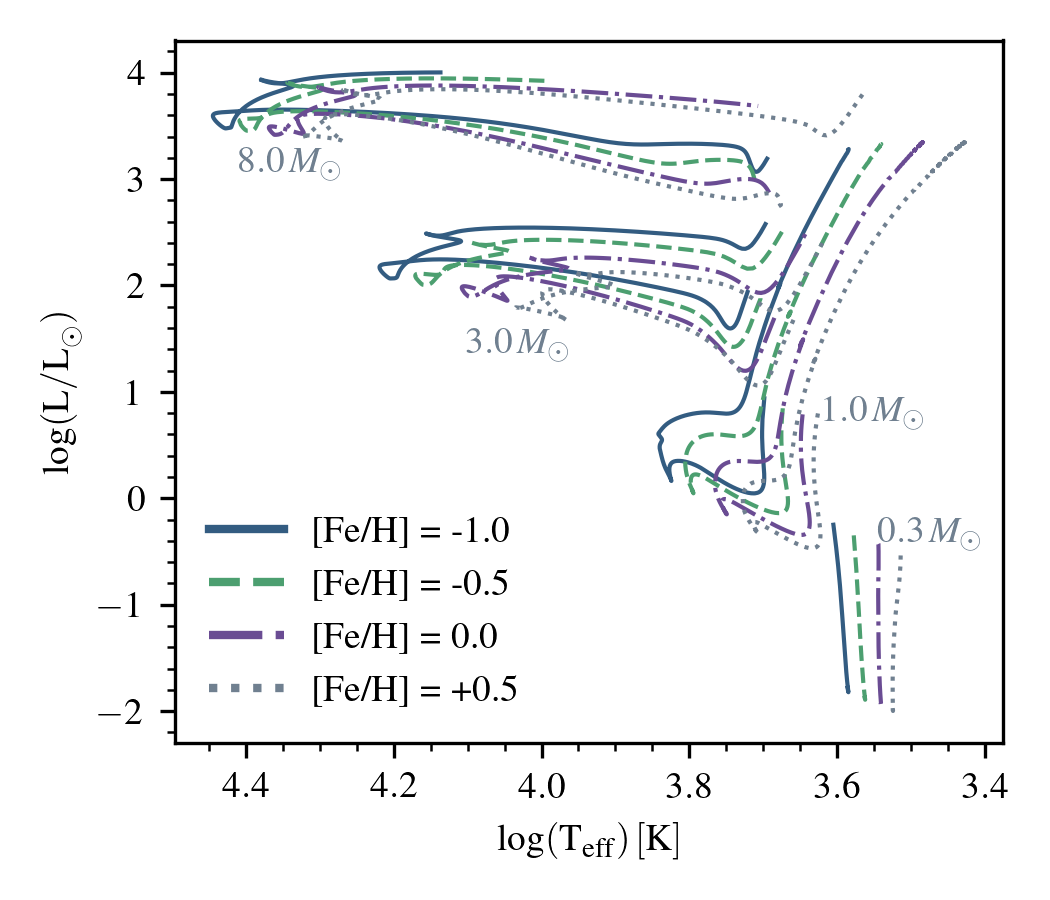}
\caption{H-R diagram showing evolutionary tracks for masses 0.3, 1.0, 3.0, and 8.0 M$_\odot$ from the base grid. We show how each track evolves across metallicities [Fe/H] = -1.0 (solid blue), -0.5 (dashed green), 0.0 (dash-dotted purple), and +0.5 (dotted grey). The 0.3 M$_\odot$ tracks extend to ZAMS, the 1.0 M$_\odot$ tracks to the tip of the RGB, and the 3.0 and 8.0 M$_\odot$ tracks to ZAHB, defined by $Y_{cen} = 1 - Z - 0.06$.}
\label{fig:grid_mets}
\end{figure}

\begin{figure}
\includegraphics[width=8 cm]{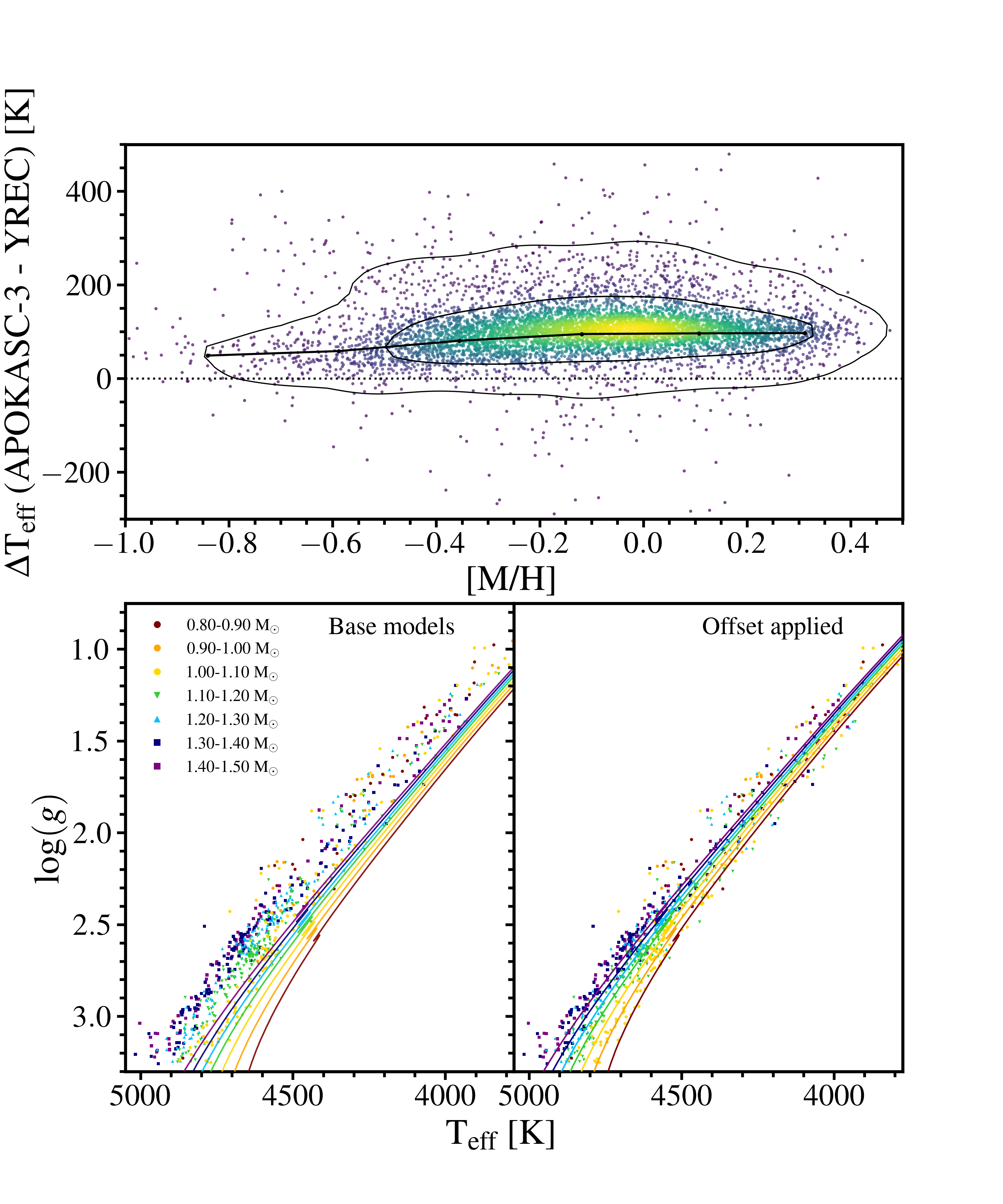}
\caption{\textit{Top:} Difference between APOKASC-3 and YREC effective temperatures for the RGB sample in the APOKASC-3 catalog. Contours represent 68\% and 95\% of stars in the sample. \textit{Bottom:} Kiel diagrams of APOKASC-3 RGB stars with metallicities in the range -0.05 $<$ [M/H] $<$ +0.05. The right panel shows the sample overlaid with the base grid solar metallicity tracks. The track colors correspond to the lower boundary of the mass bins. The left panel shows the same tracks after applying the temperature offset shown in the top panel.}
\label{fig:deltaT}
\end{figure}

Our final stop before the core He-burning phase is the red giant phase, where we now have rich and precise data sets due to asteroseismology. Stellar models make strong predictions about the metallicity dependence of stellar properties, as illustrated in Figure \ref{fig:grid_mets}. We are now in a position to test these predictions. 
Modern time domain surveys have revealed a rich landscape of stellar variability. One of the highest impact results was the discovery of solar-like oscillations in  cool and evolved giant stars. The global oscillation properties can be combined with spectra to infer mass, radius and age for large numbers of stars. In Figure \ref{fig:deltaT}, we compare our models with data from the APOKASC-3 catalog \citep{pinsonneault2025a}. There is a clear offset, which reflects the limitations of the solar calibration for stars very different from the Sun. See \citet{Tayar2017a} for a discussion.

\subsubsection{Core He-burning models}

The core He-burning phase is complex, and there can be significant differences between evolutionary codes. These complications arise in earnest even at helium ignition. The onset of He burning is a smooth process in high mass stars, though it does not proceed smoothly for low-mass stars. For stars below $\approx 2M_{\odot}$, He-burning ignites in a highly degenerate medium. Evolving a model through this helium flash is numerically challenging because the usual stabilizing effects of the ideal gas equation of state are short-circuited. The assumptions of the code that work well for most conditions - in particular, instantaneous mixing of convective regions and exact hydrostatic equilibrium - become poor ones during the helium flash. Great care with numerics is also required; the standard mesh allocation procedure in YREC works well for smooth and stable evolution, but less so for domains with rapid structural change. It is a testament to the strengths of the MESA code that it can routinely be evolved through the He flash \citep{Bildsten2012}. It is worth noting that YREC \textit{can} solve these issues, and in fact it was one of the first codes to carry evolution through the helium flash \citep{cole1985}. However, this capability greatly complicated the code, and was removed from modern versions over many years of revisions. 

Instead, YREC uses a two-step procedure for low-mass horizontal branch stars that experience the helium flash. The equations of stellar structure can be solved in a time-independent fashion for stars with nuclear energy sources. Given a composition profile, it is therefore possible for these stars to solve for a core He-burning model, similar to the approach used for the core H-burning main sequence. The most substantive difference is that the helium core mass, in addition to the total mass and metallicity, needs to be specified. For a self-consistent run, one evolves to the RGB tip, infers the core mass as a function of total mass and metallicity, and then rescales a seed core He burning model to the desired specifications. See Section 4.1 for a discussion of the procedure that we use. The current suite of seed models for core He-burning were built with the \cite{grevesse_noels1993} mixture, and some namelists in the test suites also use the corresponding atomic opacity tables. It is not entirely trivial to adjust the non-uniform ZAHB abundances to a different mix; we intend to expand the suite to different mixtures in later releases. The default, however, is to use atomic opacity tables calculated with the GS98 mixture. Neutrino cooling rates also differ from the default, taken from \cite{1989neas.book.....B} in order to be consistent with the seed models.  The helium flash models described in this section ($< 2M_{\odot}$) use this two-step procedure. 

The precise physics appropriate for modeling the regions above the convective cores of the HB are not settled, which we discuss briefly here. The fundamental complication arises from carbon- and oxygen-rich material in the convective core `overshooting' the nominal convective boundary due to inertia from convective motions at the boundary. Since this material is more opaque than the material it mixes into, this region becomes unstable to convection according to the Schwarzschild criterion. Nevertheless, the region is stable against convection when accounting for the composition gradient in the Ledoux stability criterion.

It was therefore first proposed by \cite{1947ApJ...105..305L} that such a `semi-convection' region would experience strong enough convection to mix abundances, but not strong enough to significantly transport energy; i.e., the region would transport energy via radiation. Discussed in detail in the context of very massive stars \citep{1958ApJ...128..348S} and later in core helium core burning stars \citep[e.g.,][]{1970AcA....20..195P}, the typical assumption is that the semi-convection region settles into a composition at every point such that $\nabla_{\mathrm{rad}} = \nabla_{\mathrm{ad}}$.

An alternate formulation to the mixing dilemma above a core helium-burning model is overshooting, in which the convective boundary is only extended to a certain extent that depends on the motion of the convective elements as they pass the nominal convective boundary. Both prescriptions predict that the lifetime of the horizontal branch will be extended as the core grows, since it has access to more helium than it otherwise would. This results in a different ratio in the number of horizontal branch stars to asymptotic giant branch stars in clusters compared to a case when new helium is not introduced into the core \citep[e.g.][]{1989ApJ...340..241C,constantino+2016}. In detail, the evolution of the helium abundance proceeds differently under these two scenarios, which produces different predictions for asteroseismic gravity mode frequencies. Unfortunately, neither standard overshooting nor semi-convection reproduce satisfactorily, though non-standard models can \citep{constantino+2015,2015MNRAS.453.2290B,2017MNRAS.469.4718B}.

YREC adopts an overshooting formalism to model the physics of mixing around the core helium-burning core. We compare our horizontal branch models against PARSEC v2.0, since it, too, adopts an overshoot formalism \citep{bressan_chiosi_bertelli1981} that is different in detail, but similar in extent. 

\begin{figure}
\includegraphics[width=8 cm]{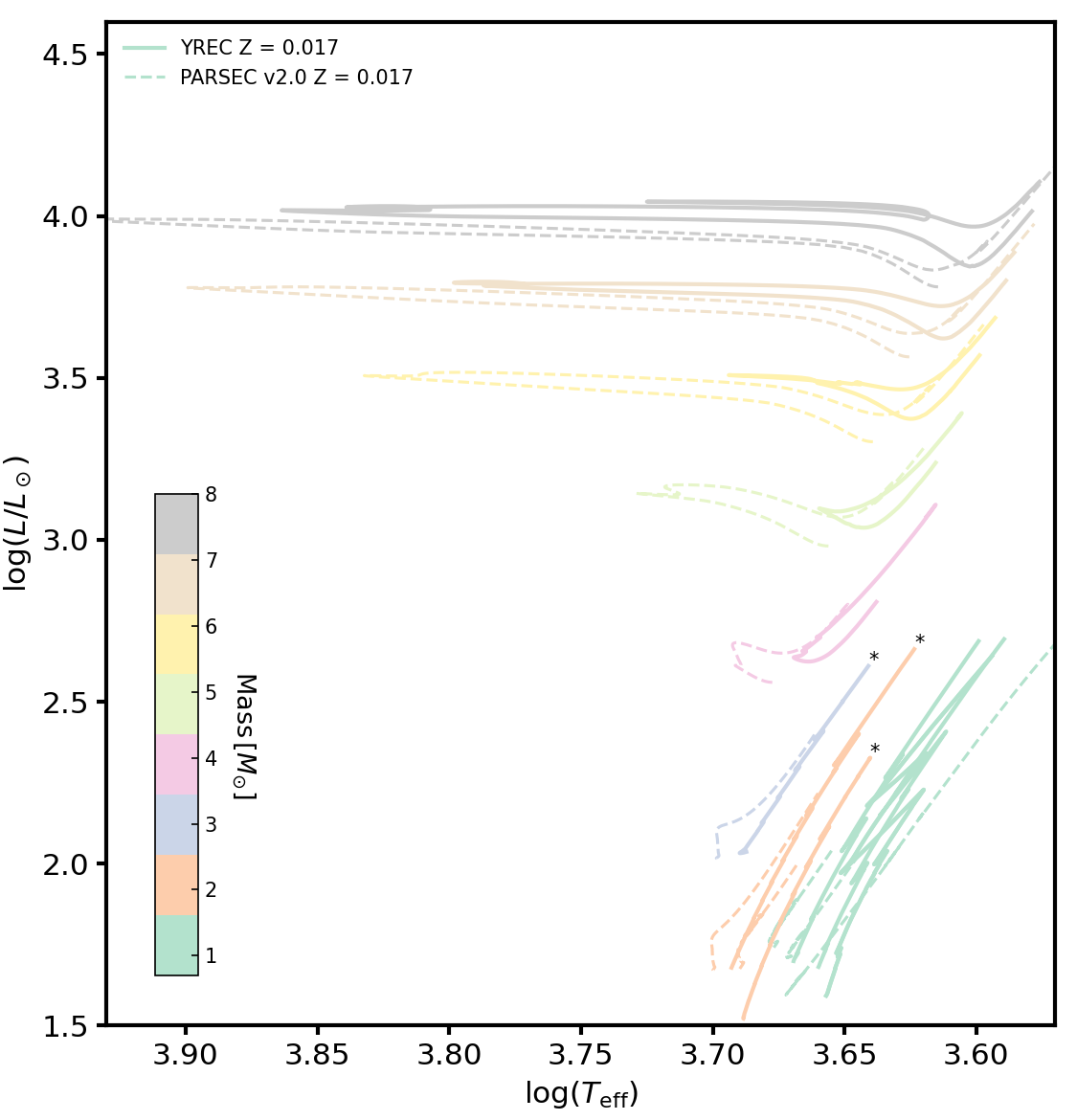}
\caption{Comparison of the low-mass core helium burning phase from YREC (solid) and PARSEC v2.0 (dashed) from beginning of central core helium burning to $Y_{\mathrm{cen}} = 10^{-3}$ except where noted with $^*$. The tracks include masses between and including $0.8M_{\odot}$ and $2.4M_{\odot}$ in steps of $0.4M_{\odot}$ and $3M_{\odot}$ and $8M_{\odot}$ in steps of $1M_{\odot}$. The low-mass PARSEC v2.0 models, which have suppressed helium ingestion episodes, generally reach core helium exhaustion at lower luminosities than YREC models, which experience helium ingestion episodes that allow for continued growth of the core and extension to larger luminosities.}
\label{fig:clumphr}
\end{figure}

We illustrate the evolution of our solar metallicity horizontal branch models from the ZAHB (defined to be when $Y_{\mathrm{cen}} < 1-Z-0.06$) in Figure~\ref{fig:clumphr}, with tracks also shown from rotating PARSEC v2.0 \citep{costa+2019,costa+2019b,nguyen+2022,nguyen+2025} models, which cover the low-mass ZAHB via a similar core mass rescaling as YREC uses. The stepwise evolution in the core of low-mass stars is evident in YREC models, which reflects episodes where fresh fuel is ingested into the He-burning core. These are less evident in the PARSEC v2.0 models at the scale of the figure, athough the qualitative behavior is present, as we show below when looking at the central helium evolution. Modest differences in luminosity ($\approx 10\%$) and in temperature ($\approx 3\%$) between low-mass YREC and PARSEC v2.0 ZAHB locations in luminosity--effective temperature space are consistent with previous findings of agreement among publicly-available solar metallicity ZAHB models \citep{an+2019}. 

YREC models proceed to much higher luminosities and lower temperatures than PARSEC. This can be traced to a higher ingestion rate of helium into the core during overshoot in YREC, which lengthens the lifetime of the horizontal branch. Importantly, PARSEC does not instantaneously mix abundances in convective regions, instead following a diffusion approach \citep{nguyen+2022}. This latter difference would reduce the amount of helium that enters into the core at any given time, and correspondingly reduce the severity of mixing zone mergers compared to those in YREC, which assumes instantaneous mixing. (We discuss these mixing episodes below, which are also responsible for the sharp decreases in luminosity in Figure~\ref{fig:clumphr}.)

\begin{figure}
\includegraphics[width=8 cm]{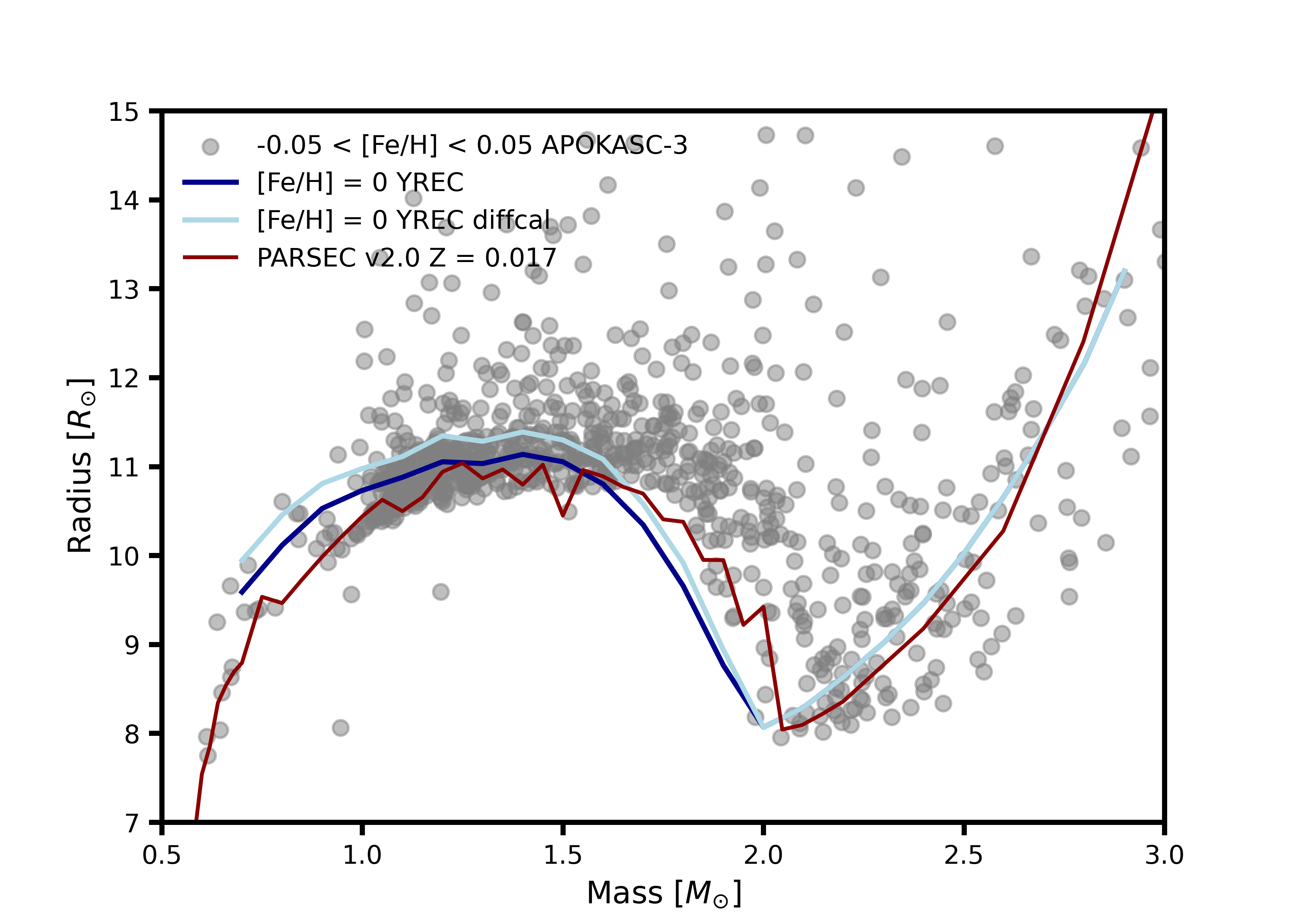}
\caption{Comparison of the core helium burning phase from YREC and PARSEC v2.0 against APOKASC-3 masses and radii for solar-metallicity stars. Cases with and without solar diffusion-calibrated $Y=.273$ and $Z=0.0188$ are shown. YREC models with $M \geq 2M_{\odot}$ use a two-step procedure. See text for details. Although there is qualitative agreement, asteroseismic constraints suggest opportunities for understanding the physics of the horizontal branch better in both YREC and PARSEC v2.0 models.}
\label{fig:clumpmr}
\end{figure}


Asteroseismic data is a powerful test of stellar models, and we compare ZAHB models with data in Figure~\ref{fig:clumpmr}. We note that we define the ZAHB as when $Y_{\mathrm{cen}} = 1 - Z - 0.06$ for YREC, which corrects for the energy required to lift degeneracy in the core. A sharp boundary in radius as a function of mass is observed in the data, which corresponds to the ZAHB. The handful of stars with radii below the edge are the products of binary star interactions \citep{Li2022}. The predicted mass range for the onset of degenerate helium ignition agrees well with the local minimum in radius at $M \approx 2.0 M_{\odot}$. In detail, none of the models trace the lower boundary precisely.
The core mass in the YREC case is constant to within 1\%  ($M_{\mathrm{core}} \approx 0.459 M_{\odot}$) for $0.7 M_{\odot} \leq M \le 1.2 M_{\odot}$. PARSEC v2.0 models have slightly higher core masses, sitting within 1\% of $0.475M_{\odot}$ until $M \approx 1.35 M_{\odot}$. 
Above $M \approx 1.4 M_{\odot}$ ($M \approx 1.35 M_{\odot}$), YREC (PARSEC v2.0) the core mass begins decreasing as neutrino cooling becomes less efficient, which is reflected in the turnover of the mass-radius relation. Neither YREC nor PARSEC v2.0 completely agrees with the ZAHB location in mass-radius space. YREC models at higher masses may be brought into better agreement by adopting an initial helium mass fraction consistent with the PARSEC choice (smaller than YREC's by 0.015), while adopting a higher helium abundance consistent with solar-calibrated diffusion models 
worsens the agreement (light blue in Fig.~\ref{fig:clumpmr}).
Further work is clearly needed to reconcile models with the powerful asteroseismic constraints.

\begin{figure}
\includegraphics[width=8 cm]{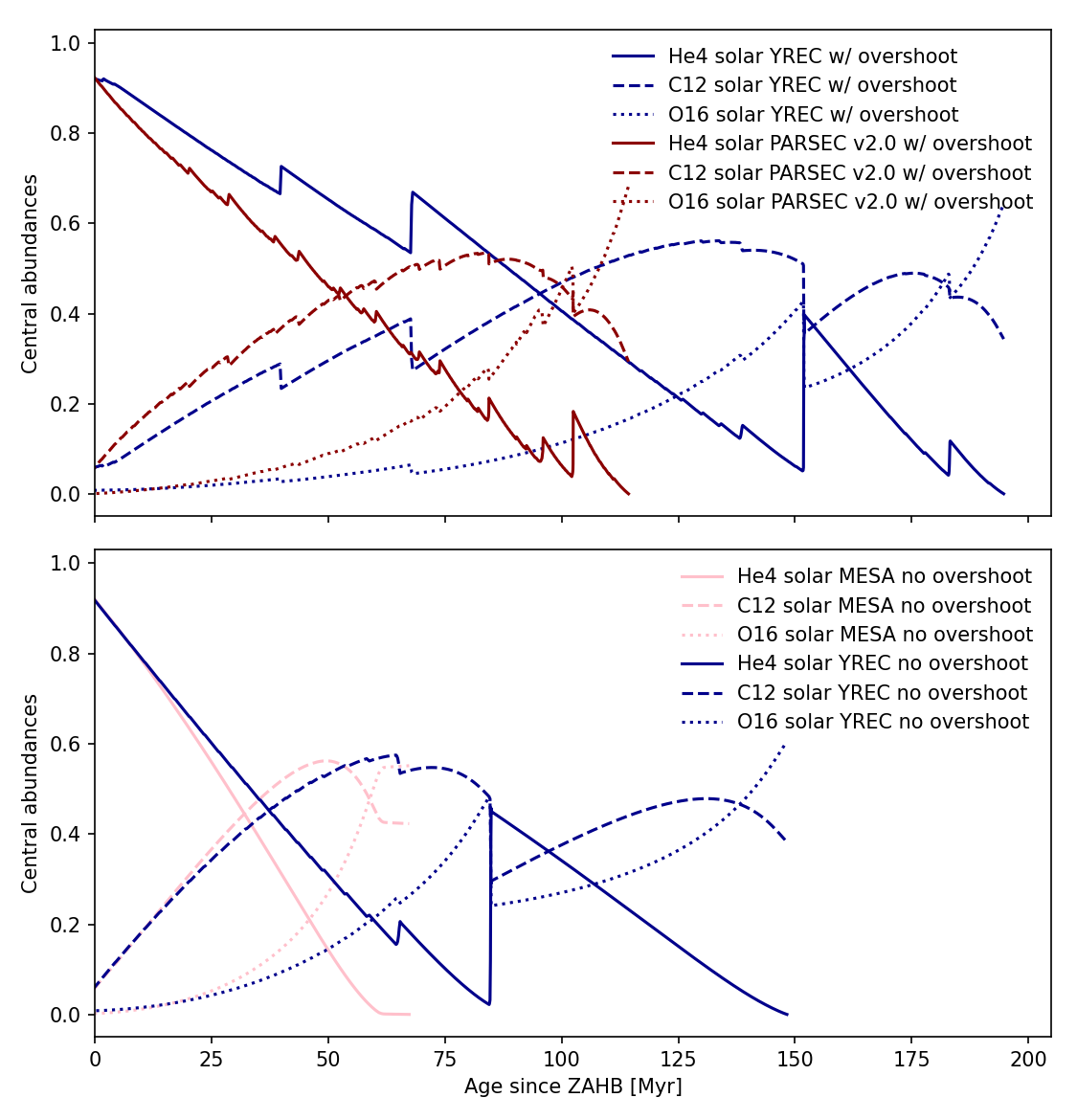}
\caption{Comparison of the core helium burning phase for a solar model from YREC (dark blue) with and without overshooting, matched to the core masses of PARSEC v2.0 (red; top) and MESA (pink; bottom). Note that MESA and PARSEC predict slightly different core masses: 0.443$M_{\odot}$ and 0.469$M_{\odot}$, respectively, though most of the lifetime differences in the top and bottom panels is due to the difference in helium ingestion due to overshooting. A significant core breathing pulse is observed at $\approx 80$\,Myr in the no overshooting case. The evolution of the region outside the convective core in horizontal branch models even with similar treatments can differ.}
\label{fig:clump}
\end{figure}

Turning to a closer inspection of the growth of the core with time, we show in Figure~\ref{fig:clump} the PARSEC v2.0 and YREC central abundance evolution of a solar horizontal branch model in the presence of overshooting. Initially helium is converted into carbon; at higher carbon abundance, $C^{12}+\alpha$ becomes more efficient than triple $\alpha$, leading to higher oxygen production.

The small stepwise increases in helium visible in Figure~\ref{fig:clump} and in the corresponding sharp decreases in luminosity in low-mass tracks shown in Figure~\ref{fig:clumphr} are due to the subtleties of the physics exterior to the convective core mentioned above. The emergence of a semi-convection region will split the convective core into two if using the Schwarzschild stability criterion and adopting the radiative temperature gradient in this region, as PARSEC v2.0 and YREC both do. The outer convection zone shrinks in size even as the convective core grows with the opacity increase from nuclear burning. Overshooting helps bridge the gap at which point the core reaches the convective shell discontinuity, ingesting helium, shrinking due to the temporary decrease in opacity this causes, and splitting again. This split does not occur if the temperature gradient is assumed to be adiabatic and not radiative as is commonly assumed in semi-convection prescriptions, and these episodes do not occur. The lifetime of the horizontal branch is still increased in this case, though, as the core can mix with the semi-convection region. Classical breathing pulses as described in \citealt{1973ASSL...36..221S} and occurring at $Y_\mathrm{cen} < 0.1$ are also present in YREC models for both the overshooting and no overshooting cases. Such pulses are related to the smaller mixing zone merger episodes, though in this case the increased temperature of the core at this phase in the evolution can produce luminosities in the core large enough to overcome decreases in opacity due to helium ingestion; the result is a large increase in core size with helium ingestion followed by quiescent increase in core size due to nuclear burning \citep[e.g.,][]{Castellani1971}.

We compare against MESA r24.03.1 to further demonstrate differences in models, even in the absence of the details of overshooting. For this purpose, we modify the 1MS\_pre\_ms\_to\_wd inlists from the MESA test suite, turning off mass loss as well as semi-convection and overshooting, and adopting GN93 abundances with Z=0.02, Y=0.28. We also enforce the same mixing length formalism as YREC \citep{BohmVitense1958} and do not use Type 2 opacity tables in order to be consistent with YREC interior opacity tables. For this comparison, our YREC models use GN93 OPAL atomic opacities to be as consistent as possible with the YREC ZAHB seed models, which were initially generated using these abundances.

As seen in Fig.~\ref{fig:clump}, whereas for YREC we see small increases in central helium content as the convective core grows in size and ingests new helium, the same does not occur in MESA. The large jump at $\approx 80$\,Myr would correspond to the classical breathing pulses that occur at low helium content (Y$ < 0.1$). There is also a small difference in the luminosity and the corresponding duration of the core He-burning phase from central $\Delta$Y$=0.06$ until reaching $Y_{\mathrm{cen}} = 10^{-3}$. 

\subsection{Sample Cases}

It can be useful for readers to see examples of use cases for YREC, and we present some examples here. For solar models, and ones on the lower and upper MS, we discuss the relevant models from the test suite. We also include models with rotation and starspots, which include grids not given in the testsuite. The namelists for these cases can be found in the sample cases directory and are listed in Table \ref{table:samplecases}.

\begin{deluxetable}{p{6cm}p{3cm}p{6cm}} 
\tabletypesize{\scriptsize}
\tablecaption{Families of sample cases discussed in the paper. The case prefix is presented in the first column. The location of the namelists (test suite or sample cases) is in the second, and a description is in the third. ``Both'' for the location is for situations where a subset of the cases are in the test suite.}
\label{table:samplecases}
\tablehead{
    \colhead{Case Name} & \colhead{Location}& \colhead{Descriptions}}
\startdata
Test solar & Test Suite & Solar models calibrated with a range of input physics and birth mixtures   \\
Sample base ZAMS & Sample Cases/LowerMS & Models of $0.09-1 M_{\odot}$ evolved from the DBl to the ZAMS.   \\
Sample allard ZAMS& Sample Cases/LowerMS & Models of $0.03-1 M_{\odot}$, evolved from the Dbl to 15 Gyr($0.03-0.08 M_{\odot}$) or the ZAMS ($0.09-1 M_{\odot}$) with Allard atmosphere SBCs   \\
Sample kurucz ZAMS& Sample Cases/LowerMS & Models of $0.15-1 M_{\odot}$, evolved from the Dbl to the ZAMS with Kurucz atmosphere SBCs   \\
Sample yaleeos ZAMS & Sample Cases/LowerMS & Models of $0.3-1 M_{\odot}$, evolved from the Dbl to the ZAMS with the Yale EoS   \\
Sample m9p0feh0+0GS98 Cases & Both & $9 M_{\odot}$ models with different EoS, degress of overshoot, with and without semi-convection  \\
Sample Gyrochronology & Sample Cases & Models of $1.4 M_{\odot}$ and below, including rotation, evolved from the Dbl to an age of 4 Gyr with different assumptions about angular momentum loss and internal angular momentum transport.  \\
Sample Starspots & Sample Cases & Models of $1.4 M_{\odot}$ and below, including star spots, evolved from the Dbl to the ZAMS with different star spot filling factors.  \\
Sample ZAHB & Sample Cases/TAHB & Models of $1.8 M_{\odot}$ and below, evolved from the ZAHB with core sizes consistent with the base case model grid of \S\ref{sec:base_case}, with overshooting.  \\
ZAHB diffcal & Sample Cases/TAHB & Same as ZAHB, but with diffusion-calibrated $Y$ and $Z$.  \\
Solar PARSEC & Sample Cases/TAHB & Solar model horizontal branch evolution matched to PARSEC v2.0, with overshooting. \\
Solar GN93 MESA & Sample Cases/TAHB & Solar model horizontal branch evolution matched to MESA, with GN93 opacity tables. \\
\enddata

\end{deluxetable}

\subsubsection {Solar Models}

YREC has a long, and high-impact, history of usage for solar models starting in the 1960s \citep{Demarque1964}. The first generation of solar models including rotation \citep{Pinsonneault1989} was also achieved with YREC. The ''standard'' solar model for neutrinos in the 1990s was produced with YREC \citep{Bahcall1992,Bahcall1995}, and these papers added gravitational settling as a key ingredient for solar models. YREC models also established the tension between a low solar metallicity and helioseismic data \citet{Bahcall2005,Delahaye2006,Villante2014} YREC solar model work continues to the present day \citet{Basinger2024}, including work on the faint young Sun problem. Our test suite includes a number of different solar models, calibrated under commonly used physical scenarios. Our sample cases account for differences in the mixture of heavy elements, the equation of state, nuclear reaction rates, surface boundary conditions, microscopic diffusion, the treatment of rotation, the initial angular momentum, and the associated mixing. 

These are precise models, which require tight numerical controls that are not required for general purpose models. For example, the convergence criteria for these models can be difficult to achieve in the general case. Small timesteps in the pre-MS are required for precise measurements of lithium depletion, but not for a precise pre-MS lifetime or HR diagram evolution. For our solar models, this is achieved by lowering the maximum change in the structure variables across a timestep relative to the general case. The inner fitting point is also moved inwards, for more precise solar neutrino calculations. This change has a negligible impact on broader classes of stellar models.

\begin{figure}[htbp]
    
    \centering
    \includegraphics[width=8 cm]{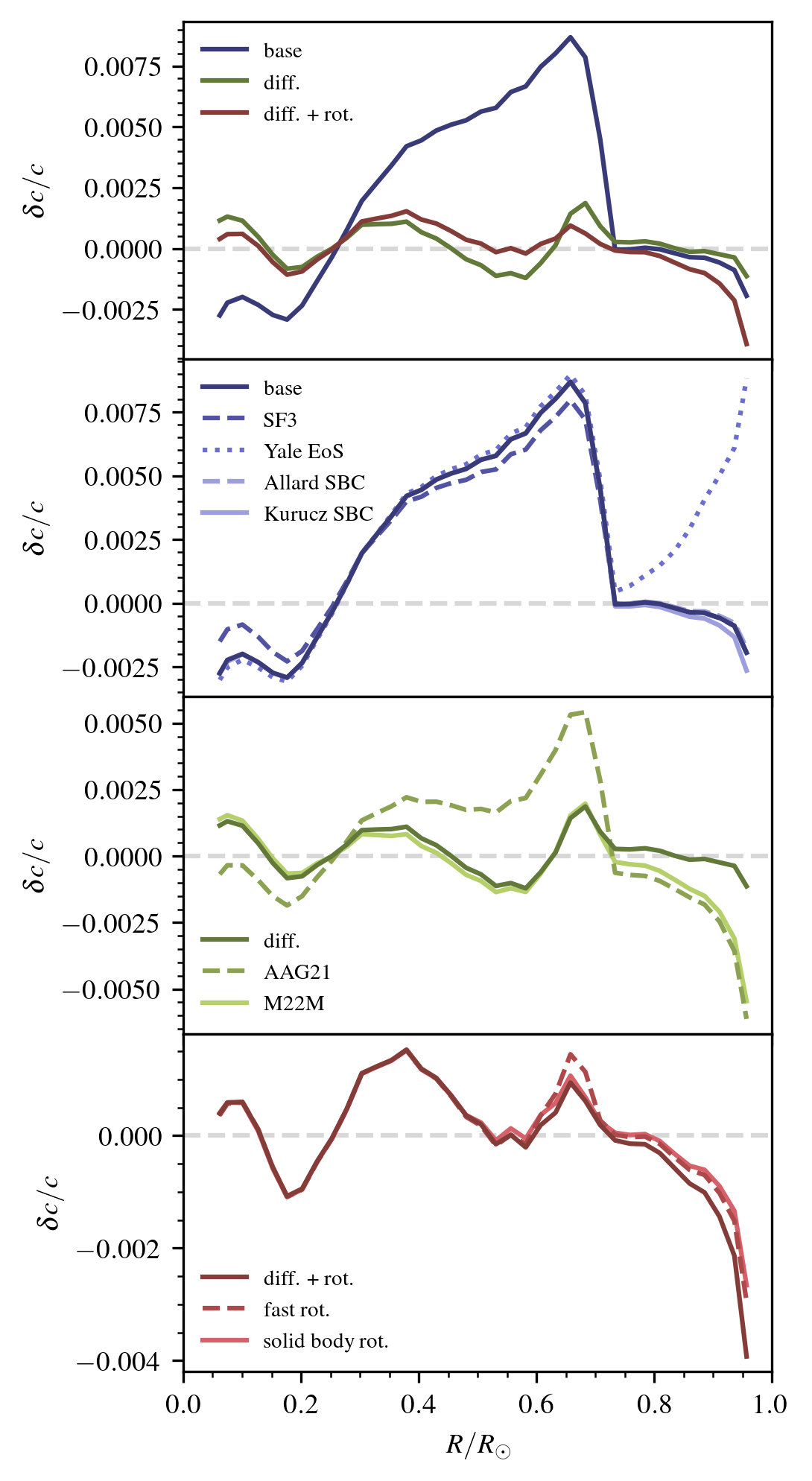}
    \caption{Fractional differences in sound speed between our test suite solar models and solar data from \citet{Basu2009}. The top panel shows differences between models with the same mixture that include different physical processes. Differences in nuclear reaction rates and surface boundary conditions are compared in the second panel. The third panel shows the impact of changes in the heavy element mixture; all of these models included diffusion but not rotation. The impact of different initial rotation rates and the transport model are compared in the bottom panel.}

\label{fig:csprofile}

\end{figure}

We show the sound speed profiles of our calibrated solar models in Figure \ref{fig:csprofile}. These results, as well as model solar neutrino fluxes, are in excellent agreement with literature values.  See \citet{Basinger2024} for a detailed discussion. The solar models without microscopic diffusion show large deviations from the true solar profile, a well-known literature result \citep{Bahcall1992,CD1993}. The choice of nuclear reaction rates and SBCs has a small impact, but the older Yale EoS is a worse fit to the data in the envelope. The birth mixture has a significant impact on the sound speed profile, although it is less than that for diffusion. The most metal poor mixture, AAG21, has the largest sound speed deviations. Rotationally induced mixing has a measurable effect, but broadly follows the same patterns as solar models with diffusion alone. Different initial angular momenta and models of internal angular momentum transport yield similar results.

\subsubsection{The Lower MS}
\label{sec:lowerMS}

\begin{figure}[htbp]
    
    \centering
    \includegraphics[width=8 cm]{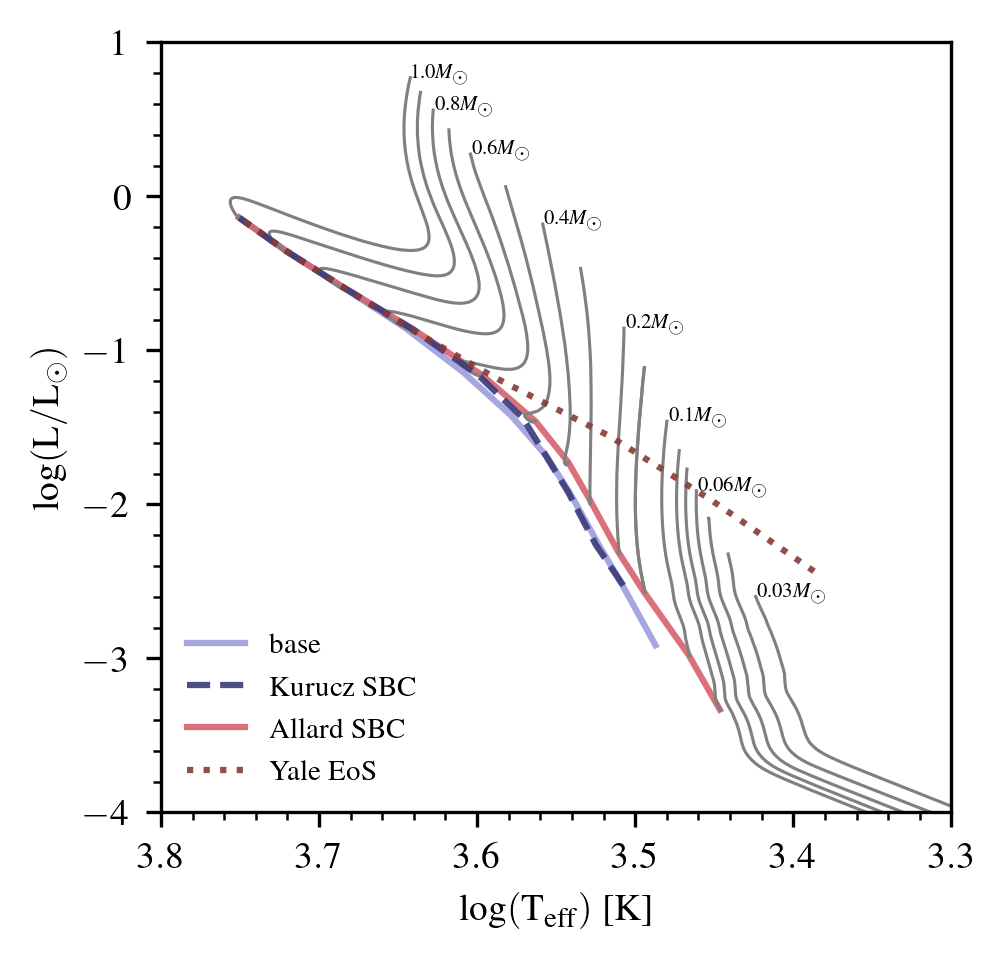}
    \caption{The impact of the choice of surface boundary conditions and equation of state on lower MS solar metallicity models. ZAMS curves are shown for the base models, Allard model atmospheres, Kurucz model atmospheres, and the Yale EoS. Model tracks to 15 Gyr (solid black lines) are shown for the Allard case. }

\label{fig:lowerZAMS}

\end{figure}

The choices for the surface boundary condition and the inclusion of star spots are quite important for lower MS stars, and we show the impact on the ZAMS for representative choices in Figure \ref{fig:lowerZAMS}. The Kurucz and Allard table edges define the cool end of the runs. The general trend is that the more simplified models (Yale EoS and gray atmospheres) depart significantly from those with more sophisticated SBCs (Kurucz and Allard) and equations of state (OPAL and SCV) below about 4000 K. We only show the Yale EoS down to 0.3 solar masses, and the Kurucz atmospheres down to 0.15. For 0.5 $M_{\odot}$ and above, the Yale EoS and gray atmosphere case performs quite well on the main sequence. Because this combination is much faster to compute than the full case, this is a good choice for educational purposes. We also show brown dwarf tracks here, along with the onset of the brown dwarf cooling curves.

\subsubsection{The Upper MS}

\begin{figure}[htbp]
    \centering
    \includegraphics[width=8 cm]{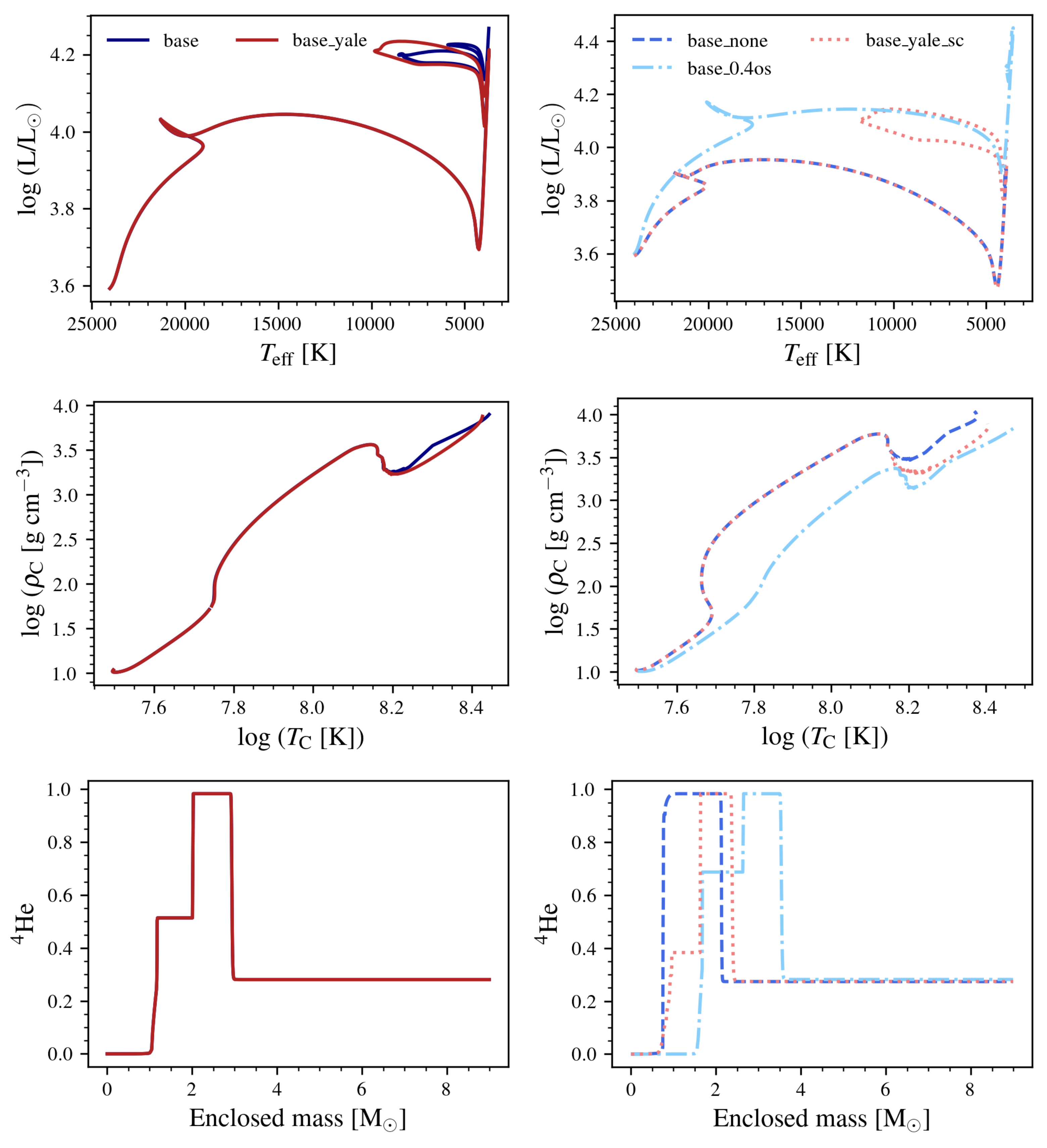}
    \caption{The evolution and final helium profile of 9 $M_\odot$ models run with different assumptions about convection and two different EoSs. The left-hand panels compare the two test suite models, which largely overlap, and the right-hand panels compare the sample case models. Blue models are for the default OPAL EoS and red models are for the default Yale EoS.}
    \label{fig:9m}
\end{figure}

Example high-mass models are spread across the test suite and sample cases. Test suite models include two 9 $M_\odot$ stars evolved to the TAHB, one with the base set of input physics (\texttt{base}) and the other with the base set of input physics but the Yale EoS (\texttt{base\_yale}). Sample case models, selected to showcase different types of extra convective mixing, include three 9 $M_\odot$ stars evolved with base case physics and either 1) no overshooting and no semi-convection (\texttt{base\_none}), 2) overshooting with 0.4 $H_p$ (\texttt{base\_0.4os}), and 3) semi-convection and the Yale EoS (\texttt{base\_yale\_sc}). All models are run with all four levels of iterations. High-mass models evolving beyond the main sequence are very sensitive to adopted physics and numerical choices. Presently, the central fitting point must be moved in during the initial phase of evolution for the later stages of evolution to run properly. Models with semi-convection are particularly sensitive. Presently, a 9 $M_\odot$ model with semi-convection and the OPAL EoS does not reach the TAHB. 

Figure \ref{fig:9m} shows the evolution in an HR diagram, the evolution of central density and temperature, and the final helium profile of the models. The test suite models, where only the EoS is changed between them, shown on the left-hand side of the figure, have nearly identical evolution, though the \texttt{base\_yale} model lives $\sim$1 Myr longer. Central density and temperature diverge during core helium burning, corresponding to differences in the blue loops the models exhibit. Despite differences in core helium burning, the models end up with the same helium profile at the TAHB. 

Sample case models largely behave as expected. The \texttt{base\_none} model evolves the fastest, does not exhibit a blue loop, and has the lowest luminosity and helium-core mass of the set. The \texttt{base\_yale\_sc} model evolves nearly identically to the \texttt{base\_none} the model, deviating during core helium-burning, when semi-convection becomes important in the region surrounding the core. The \texttt{base\_0.4os} model has the highest luminosity and core mass, the longest lifetime, and the hottest and least dense core.

Note the two-tiered helium abundance in the cores of most models. This behavior, though peculiar, is linked to extra convective mixing, as it is seen in all models with semi-convection or overshooting and not in the model without them. There ends up being a partially helium-depleted layer around the core where helium-rich material from the base of the helium shell was mixed into the core and burned. This partial He-depletion in models with overshooting or semi-convection only shows up at intermediate masses (4-5 $M_\odot$, depending on adopted physics) and above.

\subsubsection{Main Sequence Gyrochronology} \label{sec:gyro}

\begin{figure}
    
    \centering
    \includegraphics[]{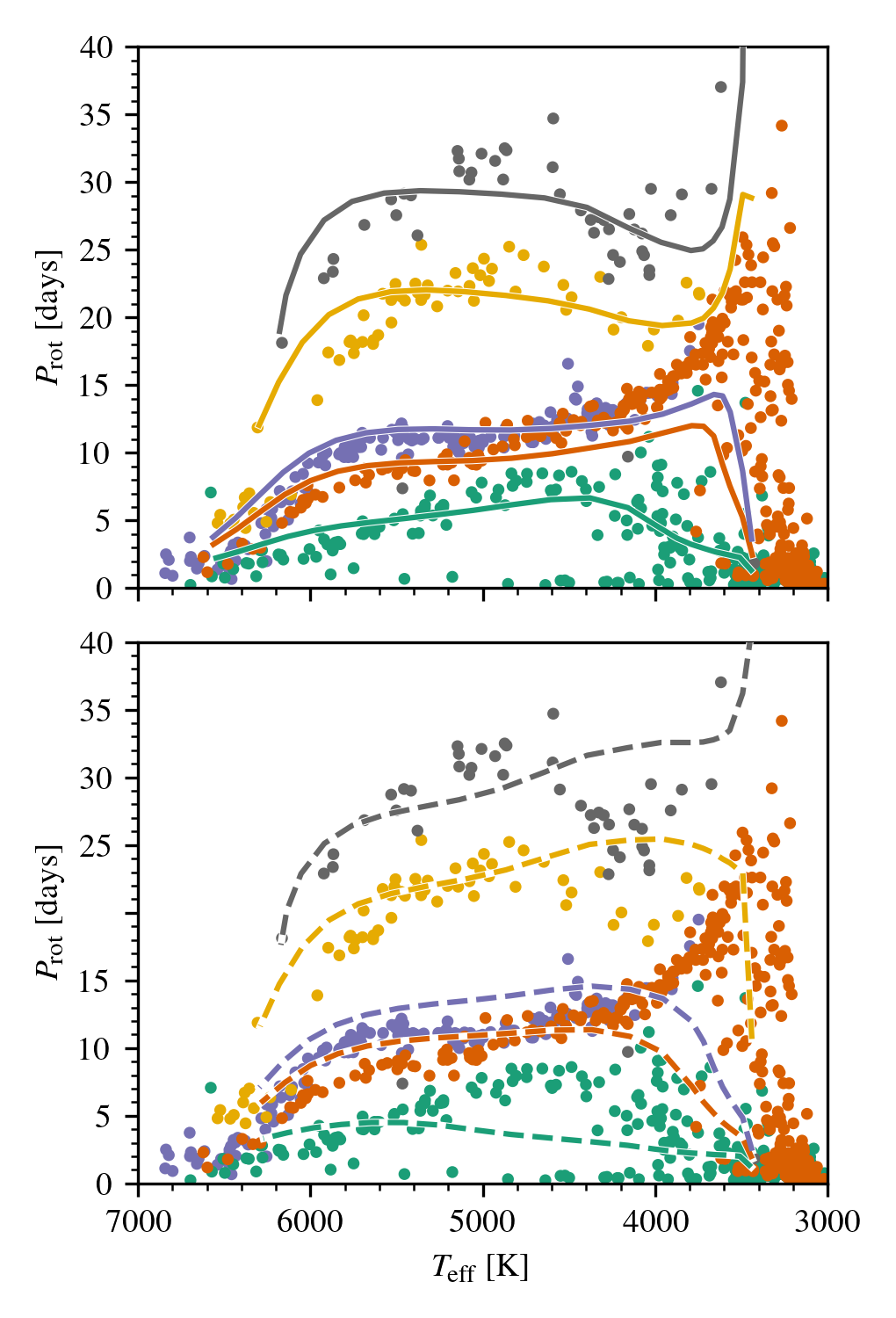}
    \caption{Rotating models compared against a select sample of open cluster rotation sequences from \citet{Curtis2020} and \citet{Gruner2023}. Solid curves (top panel) denote models that include the full treatment of microscopic diffusion, internal AM transport, and wind braking. Dashed curves (bottom panel) represent strongly magnetically coupled models, where solid body rotation is enforced at all times.}

\label{fig:gyro}

\end{figure}

The use of open cluster rotation sequences to calibrate both theoretical and empirical angular momentum loss models has a long history. Empirical fits to cluster sequences generally provide physics-agnostic interpolating functions for period as a function of age and mass \citep[e.g.][]{mamajek2008,barnes2010, lanzafame2015, bouma2023}. Theoretical approaches leverage stellar structure models coupled to a physically motivated magnetic braking law, but remain fundamentally anchored to cluster (and other) calibrators \citep[e.g.][]{ Krishnamurthi1997,bouvier1997, Sills2000, denissenkov2010, gallet2015, vansaders2013, Somers2016, Amard2020, Roquette2021}. We present an example of this calibration exercise with YREC rotating models in Figure \ref{fig:gyro}. Two sets of physical assumptions are shown: one in which models include microscopic diffusion, internal angular momentum transport, and magnetized braking from the surface. A second suite shows the behavior of strongly magnetically coupled models, approximated by enforcing solid body rotation at all times. Both sets of models are independently calibrated by varying a normalization parameter $f_{k}$ to match the rotational behavior at 0.9 solar masses in $<3$ Gyr old clusters \citep{Curtis2020}. 

Models are run with mass-dependent initial rotation periods following \citet{Somers2017} and subject to saturated spin-down below Rossby numbers of $0.1 \rm Ro_{\odot}$. For those models allowed to differentially rotate, AM transport is performed using both hydrodynamical instabilities and an extra AM diffusion term following \citet{Somers2016} that better reproduces both the chemical and rotational histories of cluster stars. This diffusion term is constant in time, but a strongly varying function of stellar mass; we adopt the mass-dependence determined in \citet{Somers2016} for display here.  The braking is performed with the \citet{vansaders2013} braking law.

Departures from the cluster sequences shown in Figure \ref{fig:gyro} are generically areas of active research; the gyrochrones we show here represent a reasonable but not exhaustive exploration of the possible parameter choices and their optimization. 

\subsubsection{ Young Stars and Star Spots}

Starspots are frequently observed on the surfaces of cool, active stars. The presence of spots has long been associated with anomalies in color \citep{Campbell1984, Stauffer2003}, abundances \citep{Soderblom1993}, and spectra \citep{GullySantiago2017}. As a magnetic activity proxy, it shows a clear relationship with the stellar Rossby number, similar to other activity proxies thought to have an origin in the stellar dynamo \citep{Cao2022b}. Spots have also associated with surface temperature systematics \citep{Cao2022b, perezpaolino2025}, lithium depletion anomalies in young stars \citep[e.g.][]{Jeffries2021, Jeffries2023, Cao2026a}, age systematics \citep{David2019, Cao2022a}, and radii excesses due to magnetic radius inflation \citep{Jackson2018, cao2025}. In YREC, starspots were initially implemented in \citet{Somers2015} using the formalism from \citet{Spruit1986}, and presented as a set of starspot isochrones in \citet{Somers2020}. Here, we explore the present capabilities of YREC to study cool stars with starspots.

\begin{figure}
    
    \centering
    \includegraphics[width=\linewidth]{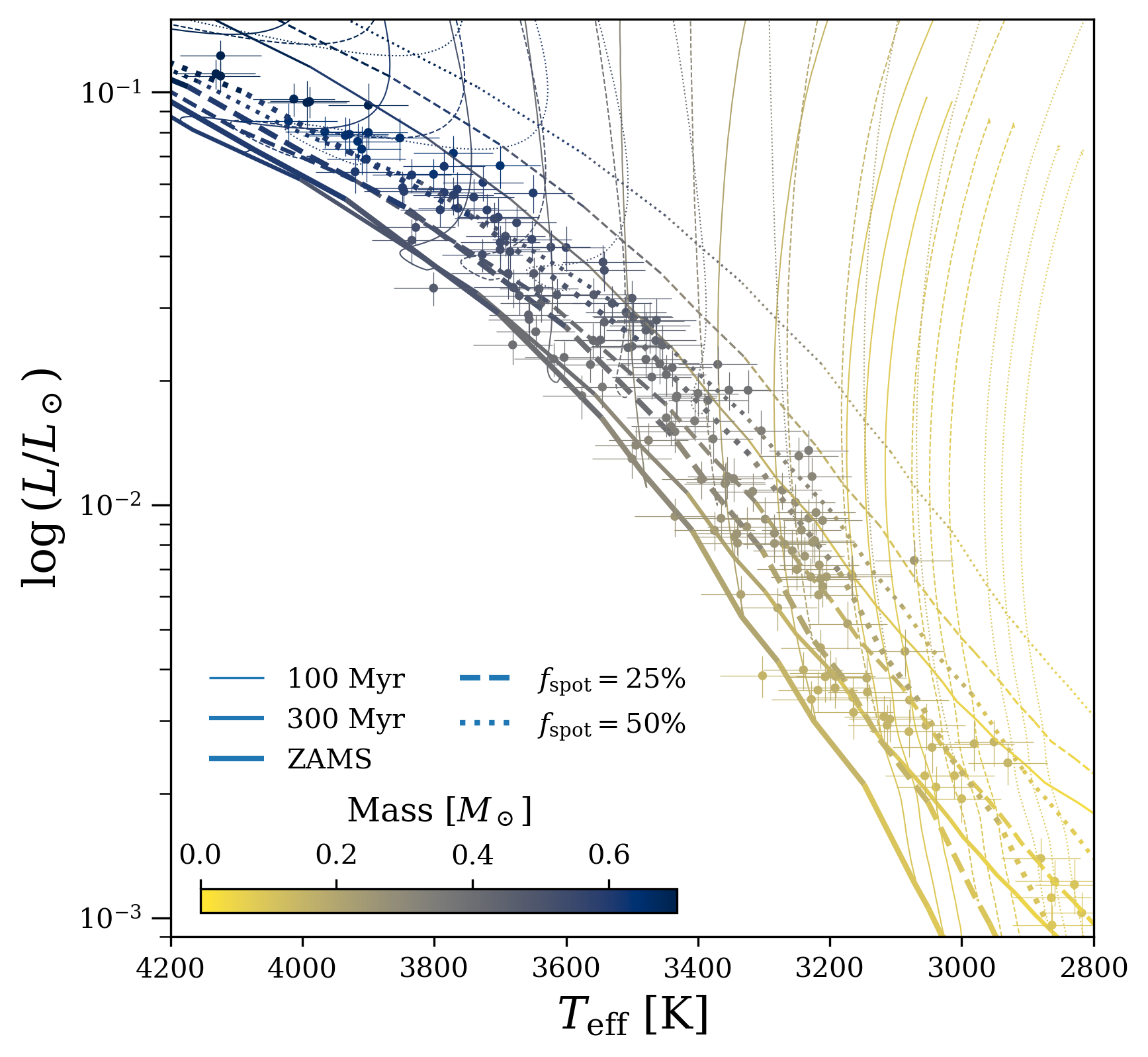}
    \caption{Starspot isochrones at solar composition evaluated at 100 Myr, 300 Myr, and the ZAMS (defined where $X_\mathrm{cen}/X_{\mathrm{cen, 0}} = 0.993$); and 0 (solid), 25\% (dashed) and 50\% (dotted) spot filling factor. Mass tracks are shown for constituent tracks evolved to the ZAMS. The \citet{Mann2015} sample is overplotted with reported errors. Colors are assigned by the stellar mass, as shown in the colorbar.}

\label{fig:starspots_HRD}

\end{figure}

In Figure \ref{fig:starspots_HRD}, we show the \citet{Mann2015} sample of nearby field stars with empirically-calibrated masses and radii, with a set of solar-metallicity starspot isochrones overplotted at several representative ages (100 Myr, 300 Myr, and at ZAMS) and spot filling factors (0, 25\%, and 50\%). These models have gray atmospheres, are non-rotating, and have diffusion turned off, similar to the test suite case. The starspot isochrones trace the \citet{Mann2015} sample very well; though there is some metallicity variation which is not captured in this plot, many stars in the \citet{Mann2015} sample appear to depart from the solid lines, and their locations on this diagram are qualitatively better fit with some degree of magnetic inflation from starspots.

\begin{figure}
    
    \centering
    \includegraphics[width=\linewidth]{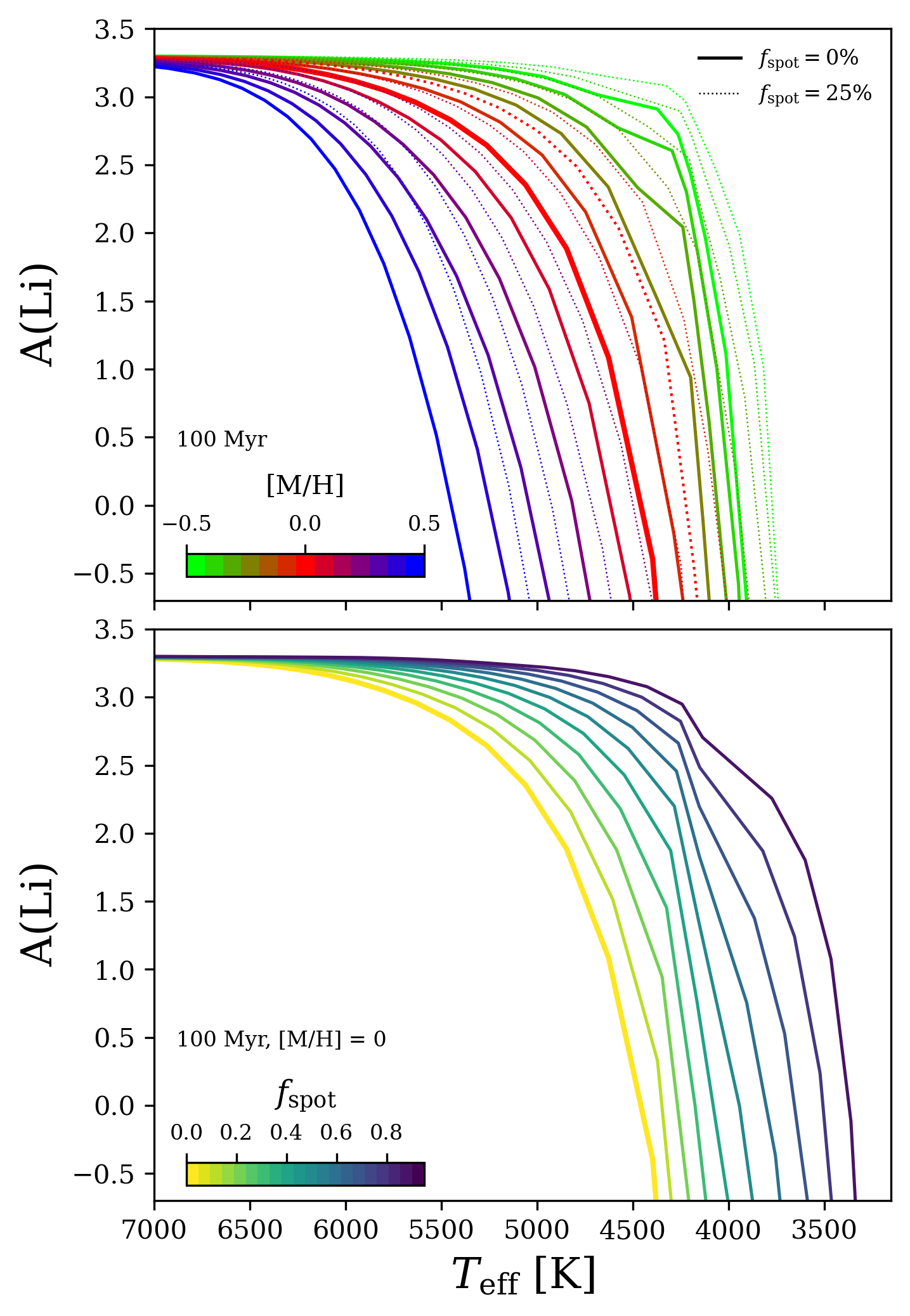}
    \caption{Starspot lithium isochrones evaluated at 100 Myr for various metallicities and starspot filling factors. Top plot shows the variation of the lithium abundance pattern with metallicity (between -0.5 and +0.5, stepping by 0.1 dex), for unspotted (solid) and 25\% spotted (dotted) cases. Bottom plot shows the variation of a 100 Myr solar metallicity lithium abundance pattern with starspot filling factor from $f_{\mathrm{spot}}$ = 0.0 to 0.9, stepping by 0.1.}

\label{fig:starspots_Li}

\end{figure}

We subsequently run a series of runs observing the same microphysics conventions from the \citet{Somers2020} paper to produce Figure \ref{fig:starspots_Li}: a set of lithium depletion patterns at 100 Myr, using the convention that $A(\mathrm{Li}) = \log_{\mathrm{10}}($\texttt{Li7\_sur}$)+12$. In this plot, we show non-rotating non-diffusive YREC models with a linear ramp from the Allard models at 0.4 $M_\odot$ to the Castelli--Kurucz models at 0.6 $M_\odot$. In the top panel, we show the variation of the lithium depletion curve with respect to metallicity, with an unspotted and a spot-saturated model ($f_{\mathrm{spot}}$ = 25\% plotted). We see that the effect of spots is similar in some respects to a decrease in metallicity, though the difference in the shapes of the distributions is more complicated in detail. The bottom plot shows lithium depletion patterns for a solar metallicity case with different starspot filling factors, showing that the effect of spots is also roughly to mimic a depletion pattern at cooler temperature. This is expected, because spotted model stars have cooler convective zone bases as a consequence of magnetic inflation \citep{Cao2026a}.

\subsubsection{ Rotation Profiles in Evolved Stars}
\begin{figure}
    
    \centering
    \includegraphics[]{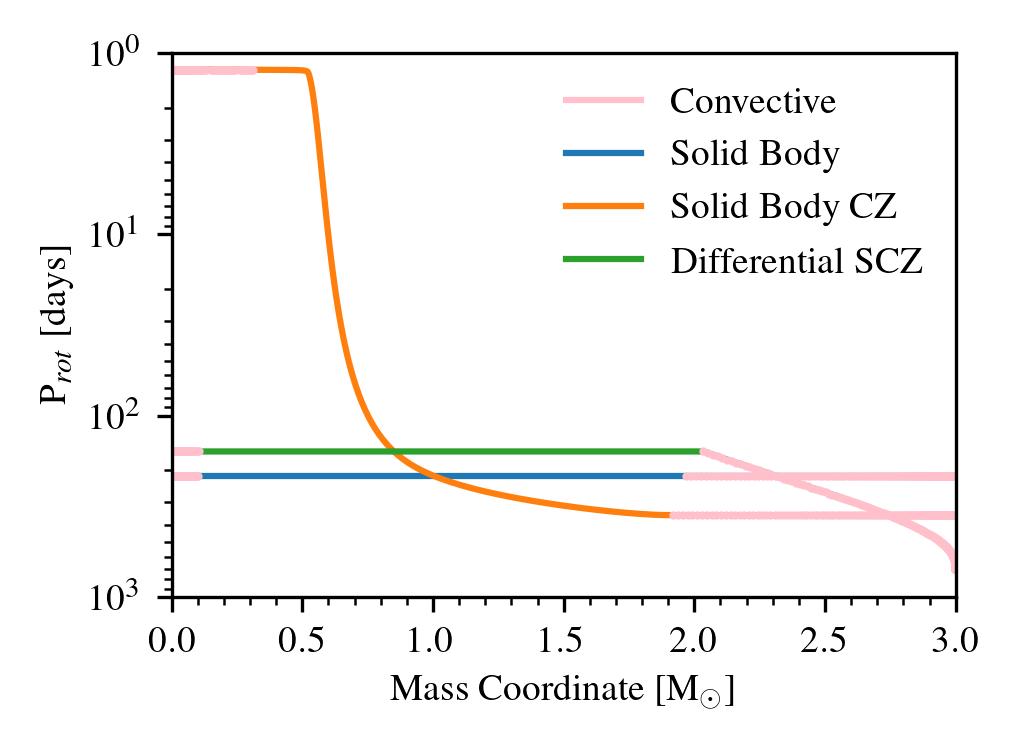}
    \caption{Radial rotation profiles for a 3 M$_\odot$ star halfway through helium burning. Convective regions are pink in all cases. We enforce various rotation profiles including solid body for the whole star (blue), a solid body interior and a differentially rotation surface convection zone (green), and solid body convection zones but a differentially rotating radiative region (orange). }

\label{fig:diffrot}

\end{figure}

While many main sequence stars appear to be rotating as close to solid bodies by the end of the main sequence, it is not clear that red giant branch stars with large envelopes and shorter lifetimes should do the same. In fact, the available asteroseismic results suggest that the cores of these stars are rotating significantly faster than their envelopes \citep{Tayar2019b, LiG2024}. It is therefore useful to be able to parameterize the radial rotation profiles in these stars in order to explore various limiting cases. YREC offers the option to force the entire star to rotate as a solid body or conversely, to conserve angular momentum within each zone \citep{Tayar2013}. There are also options to enforce an angular momentum transport timescale, or allow only hydrodynamic mechanisms to act \citep{Somers2016}. Finally, YREC also has the option to allow differential rotation in convection zones, where the differential rotation profile has the rotation frequency $\omega$ scaling with the radial coordinate $r$ to some power between -2.0 (constant specific angular momentum) and 0.0 (solid body rotation). When setting up the differentially rotating convection zone, the code also has the option to enforce solid body rotation below the surface convection zone, as well as the option to force the entire interior to rotate at the rate of the base of the convection zone. We show examples of these various cases for a 3 M$_\odot$ star halfway through core-helium burning in Figure \ref{fig:diffrot}. 

\section{Scripts}

\label{section:Scripts}

The YREC code was designed to run single initial value problems to a specified endpoint. However, there are a number of cases where scripts are essential or convenient. We provide tools and scripts in the YREC Wrappers folder\footnote{\url{https://urldefense.com/v3/__https://github.com/yreclab/yrec/tree/main/wrappers/modelgrid_tools}} of the YREC GitHub repository. Within this repository are the \texttt{modelgrid\_tools/main\_tools} repository, where the base tools for new users are found. In addition, we feature the \texttt{modelgrid\_tools/alternate\_tools} repository, which houses our more advanced tools for experienced users and large workflows. We recommend starting with the \texttt{main\_tools} repository and using the \texttt{alternate\_tools} for large workflows, parallel processing, and backups in case the file readers in the \texttt{main\_tools} fail. All scripts and wrappers are written in Python unless otherwise noted. We provide links to and brief descriptions of these tools in Appendix A. The specific tools released with this paper include: 

\begin{itemize}

\item A means of building the executable code for different compilers and common compile options.

\item A namelist converter, to customize the output and input directories for the code.

\item A YREC output file reader.

\item Scripts for running grids of models with different initial mass, composition, or rotation rate.

\item A script to calibrate solar models, using the results of prior runs to estimate best parameters for future ones.

\item Checks on the internal consistency of the input physics tables, used prior to running the code.

\item Checks on the numerical precision of the calculations.

\item Scripts for running the code in parallel, on multiple processors.

\end{itemize}

There are three other families of tools that we discuss in more detail here. The first is our procedure for restarting models after the He flash, which is more involved than typical YREC use cases. The second is the linkage of YREC with the \texttt{kiauhoku} tool, in the context of building isochrones from user-supplied grids of stellar models. We also discuss the \texttt{rotevol} code, which is a fast and efficient tracer code for generating angular momentum evolution scenarios from grids of standard stellar models. 

Our final sub-section contains some sample exercises, illustrating the power of YREC as a teaching tool.

\subsection{RGB Tip to ZAHB}
Users may generate custom zero-age horizontal branch models by re-scaling the envelope mass and the core mass of the YREC seed ZAHB models to match those of a user-generated tip-of-the-RGB model. In this way, a user may run a physically consistent calculation from the pre-MS to the ZAHB, bypassing the helium flash. An example of the envelope mass rescaling procedure is provided in the test suite (see Table~\ref{Table:testsuite}). An example is also provided for the core mass rescaling procedure in the context of the solar mass evolution run listed in Table~\ref{Table:testsuite}. This mass rescaling approach has been used in the latest PARSEC v2.0 models \citep{2025A&A...701A.258N}, and may be important, depending on the use case, to preserve consistency of the ZAHB models with those from previous stages of evolution (e.g., for isochrone generation purposes).

\subsection{Tracer Angular Momentum Evolution Grids}
\begin{figure}
    \centering
    \includegraphics[width=1\linewidth]{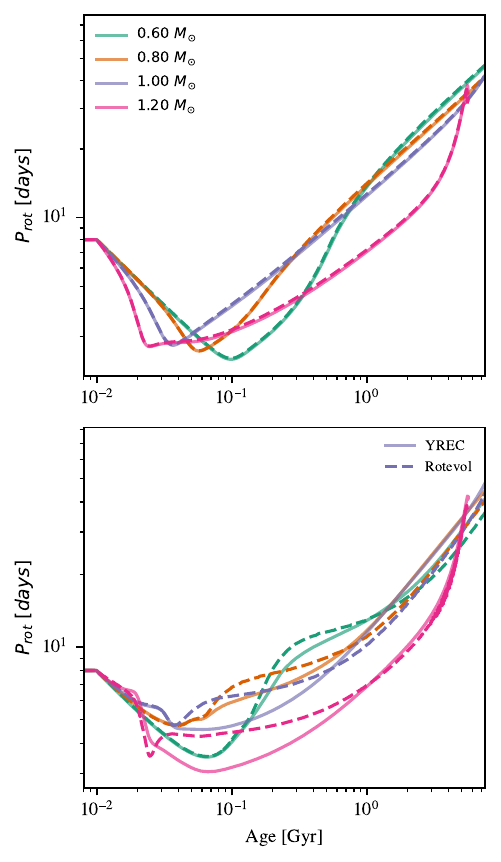}
    \caption{A comparison between full YREC rotation models and \texttt{rotevol} forward models for tracks with masses $0.6$, $0.8$, $1.0$, and $1.2$ $M_{\odot}$. The upper panel shows the comparison between the predicted rotation periods under the solid body approximation. The lower panel shows the difference in predicted rotation periods for non-solid body rotation with core-envelope coupling timescales given by \citet{2010ApJ...716.1269D}.}
    \label{fig:rotevol}
\end{figure}
For low mass stars there are a number of degrees of freedom in the model for angular momentum loss and in the initial conditions for rotation. The parameters in the model are usually constrained empirically, by matching the observed time evolution in star clusters against suites of theoretical models with multiple degrees of freedom. 

For example, angular momentum loss is a steeply increasing function of rotation rate with a saturation threshold. Slow rotators are insensitive to this saturation threshold for the angular momentum loss law, but rapid rotator histories are strongly influenced by it. Core-envelope decoupling, by contrast, is a mass-dependent process whose primary observational signature is the spin down of slower rotators. At late stages, the asymptotic slope depends on the assumed dynamo scalings, while the zero point of the loss model sets the absolute value of the asymptotic trend. It can therefore be useful to quickly generate large suites of models, with the intention of obtaining a best fit solution in a higher dimensional space. For slowly rotating stars the structural effects of rotation can be usefully treated as modest perturbations to the structure. We have therefore developed a program, \texttt{rotevol}, which reads in model tracks generated without rotation, and generates what-if scenarios for surface rotation as a function of time. \texttt{rotevol}, in turn, can be called by scripts that perturb the assumptions in the physical model and search for a best-fit solution relative to a user-defined figure of merit. \texttt{\texttt{rotevol}} has already seen extensive use in the literature \citep{vansaders2013, vansaders2016, vansaders2019, Somers2017, metcalfe2020, saunders2024, chiti2024, li2025b, Cao2023, cao2025}; this represents the first public release of these scripts. 

In addition to the test suites, we have included scripts which allow the user to run \texttt{rotevol} on pre-computed model grids. These scripts include functionalities for generating rotation tracks for different physical parameters and running a grid of models through \texttt{rotevol}. 
As \texttt{rotevol} forward models the rotational evolution of a pre-computed stellar track, one of its main advantages is speed. Compared to a full YREC run with rotation, \texttt{rotevol} reduces the computation time by 2 orders of magnitude. \texttt{rotevol} takes on order a tenth of a second to compute the rotation evolution of a mass track. 

Using \texttt{rotevol}, we have computed comparison cases between rotating YREC models and \texttt{rotevol} forward models for stellar masses between $0.2 M_{\odot}$ and $1.2 M_{\odot}$. Figure \ref{fig:rotevol} compares the predicted rotation periods from YREC and \texttt{rotevol} for the two rotation scenarios discussed in Section \ref{sec:gyro}: solid body rotation and core-envelope decoupling. The figure shows \texttt{rotevol} runs with identical braking law parameters on tracks computed with identical physics to those in Section \ref{sec:gyro}, but with rotation off. The solid body case produces excellent agreement (within$\approx 5\%$) in the predicted rotation periods compared to fully rotating YREC models. The \texttt{rotevol} models with core-envelope decoupling show disagreements up to $\approx 20 \%$ for stellar masses $\ge 1.1 \ M_{\odot}$---discrepancies that are not unexpected, given that the underlying physical treatment of the internal angular momentum transport is different \citep[see also][]{denissenkov2010, Somers2016, spada2020}.  

Users can find the \texttt{rotevol} code in the examples folder of the YREC release code. This folder includes scripts for running \texttt{rotevol} in non-rotating YREC tracks found in the sample output folder. 

\subsection{ISTARY - The YREC Stellar Isochrone Set}
\begin{figure}[h]
    \centering
    \includegraphics[width=1\linewidth]{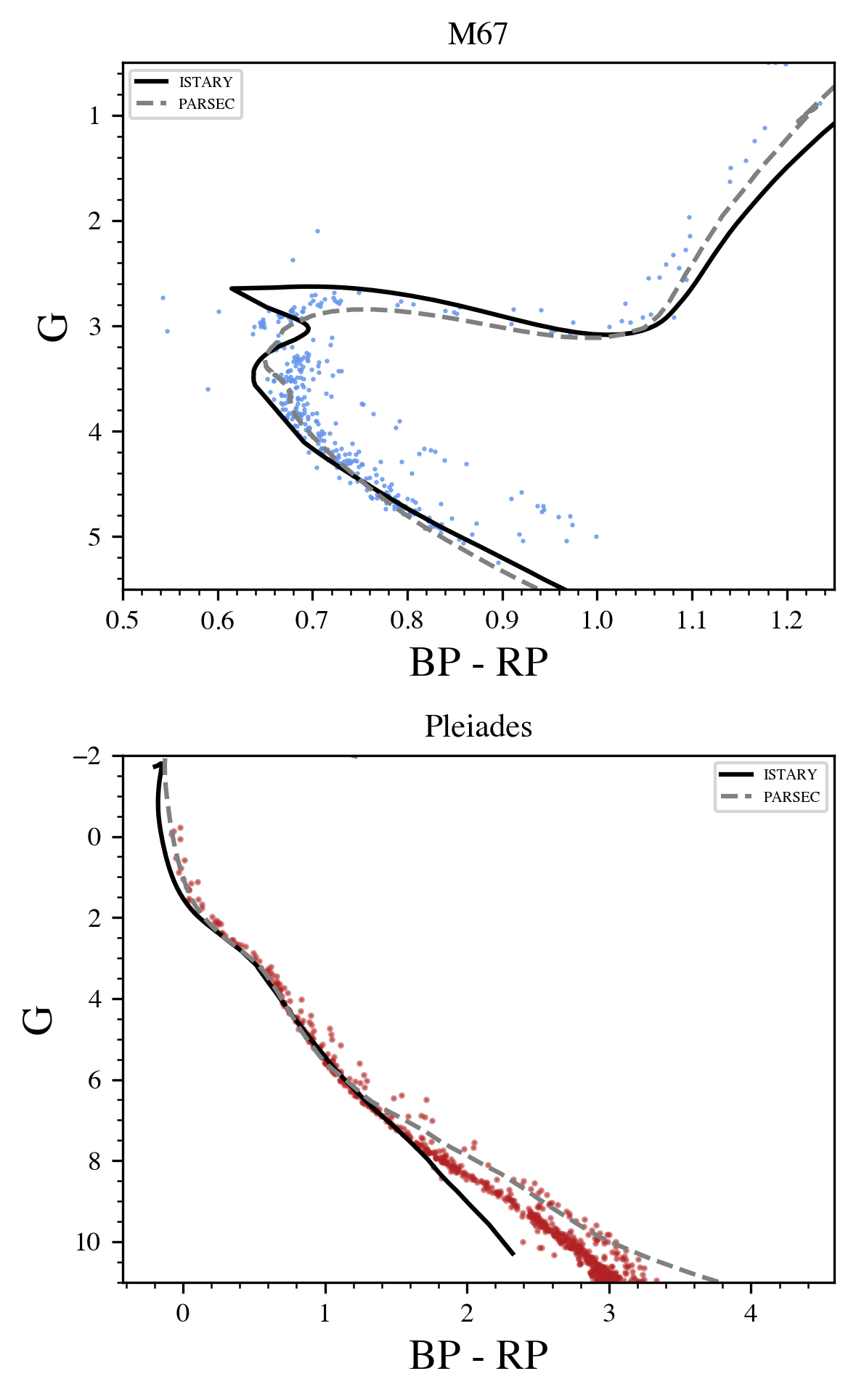}
    \caption{A CMD example of the isochrone set generated from the YREC sample grid. We plot the absolute Gaia DR3 G mag and Gaia DR3 BP-RP. Isochrones (black) were computed using the \texttt{kiauhoku}-generated EEPs derived from the YREC main grid. We adopted solar metallicity for M67 and an age of 3.95 Gyr \citep{claudiam67}, and Fe/H = 0.03 with an age of 0.125 Myr \citep{pleiadesage} for the Pleiades. Data for M67 and the Pleiades were taken from \citet{claudiam67,m67pleiadespaper}, respectively. Isochrones are compared with Pleiades data (bottom panel) and M67 data (top panel).}
    \label{fig:isochrones}
\end{figure}

Isochrones are elegant tools for studying stellar populations. Here we present a sample set derived from our grid, a guide for isochrone creation in the form of a Jupyter .ipynb file, and a visualization of the results. The \textbf{I}sochrones and \textbf{S}tellar \textbf{T}racks for \textbf{A}strophysics \textbf{R}esearch by the \textbf{Y}REC Collaboration (ISTARY)\footnote{\url{https://github.com/avincesmedile/yrec_isochrones}} isochrone set was produced using our custom isochrone functions in conjunction with the \texttt{kiauhoku} stellar interpolator code from \citet{kiauhokupaper}. \texttt{kiauhoku} takes a grid of YREC .track files and converts it into a multi-index pandas dataframe by common initial mass, mixing length, model number, and metallicity. Then, it converts this dataframe into equivalent evolutionary phases (EEPs) using the method by \citet{dottermist}. We then create isochrones using a custom wrapper, \texttt{make\_iso()}. ISTARY and our isochrone creation guide can be found in the YREC Github repository.

Our sample isochrones are for $\rm{[Fe/H]} = 0.0$ and cover $0.250\le \rm{Gyr} \le 13$, although others can be generated with the same machinery. We compare with data in the Pleiades and M67 open clusters, which we show in Figure \ref{fig:isochrones}. The isochrones here were converted into bolometric magnitudes using the YBC Code \citet{ybcpaper}. For the Pleiades, we employ individual distances for members derived from Gaia DR3 parallaxes, and utilize an extinction value of $A_{v} = 0.1$. 
We find good agreement between  ISTARY, PARSEC, and the data up to the lower MS, where known discrepancies between the Pleiades and isochrones occur \citet{ybcpaper}. For M67, we used the distances (an average distance modulus of 9.614) and extinctions from \citet{claudiam67}. We find good agreement between ISTARY, PARSEC, and the M67 data.

\subsection{YREC as a Tool for Stellar Structure and Evolution Classes}

\begin{figure}
    
    \centering
    \includegraphics[width=\linewidth]{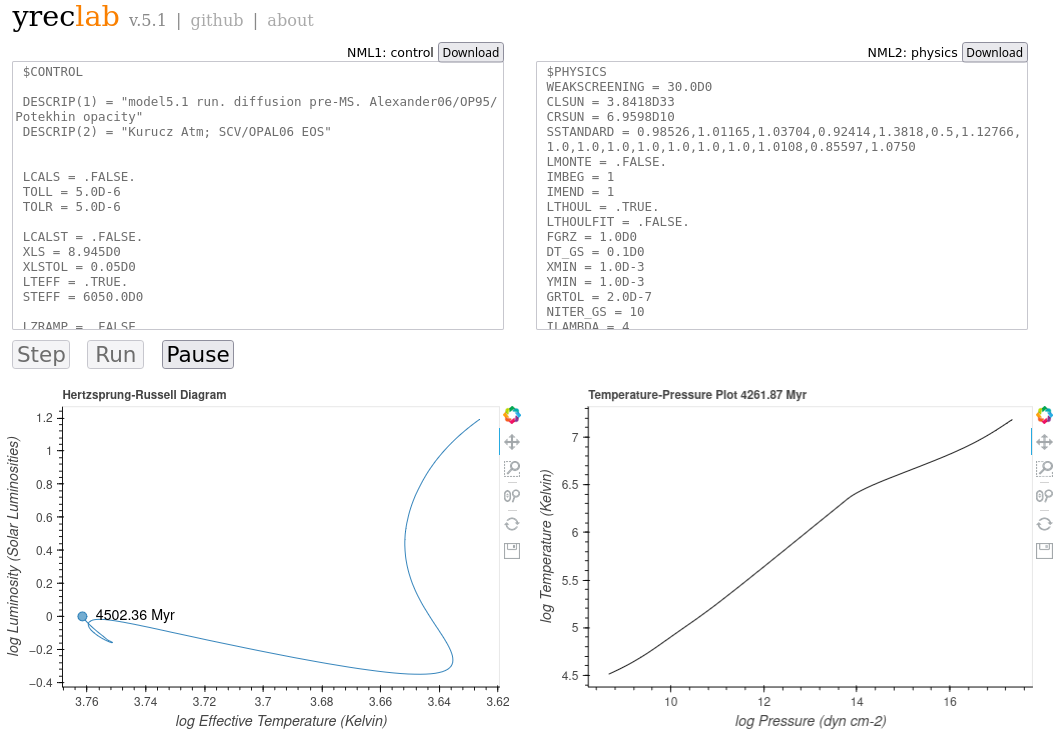}
    \caption{The \texttt{yreclab} web interface, showing a run in progress, with toggles to ``step'' (evaluate one timestep), ``run'' (evaluate until paused or execution ends), and ``pause'' the execution.}

\label{fig:yreclab_demo}

\end{figure}

Stellar structure and evolution is a core topic for undergraduate and graduate education in astronomy. The two topics---structure and evolution---demand different approaches. Stellar structure is deeply tied to the physics of stars, and most instructors adopt a semi-analytic approach close in nature to the tactics used in core physics courses. Stellar evolution, however, is more typically treated as a distillation of results. YREC is an excellent tool for both structure and evolution. Students can compare the idealized results from analytic structure calculations to those from full models, and they can generate their own evolutionary histories. We have provided some exercises in the examples folder. The first exercise compares pressure as a function of density, temperature and composition to that predicted by analytic expressions used in classes, such as those for radiation pressure or the ideal gas law. The second exercise illustrates the effects of significant variations in entire families of rates (CNO and pp) as opposed to selective changes in single rates (pp and $N^{14}+p$.). The third has students compare analytic estimates for the equilibrium abundances of elements that participate in CNO burning to their model abundances. 

Another unique teaching application of YREC is the web laboratory, also known as \texttt{yreclab}. Due to the efficient design of the code and its lightweight nature, it has been possible to compile YREC to WebAssembly with the code \texttt{emscripten} \citep{emscripten}, which allows it to be run interactively client-side as a modern single-page web application hosted on Github Pages\footnote{\url{https://yreclab.github.io/}} \citep{yreclab}. An example run is shown in Figure \ref{fig:yreclab_demo}. \texttt{yreclab} fully supports the current YREC implementation and its namelists can be modified just like the desktop version of YREC. Modifications to the code\footnote{\url{https://github.com/yreclab/yreclab}} allow it to be run in an interpreted state, and the stellar evolution code can be directly controlled via the Javascript scripting language---allowing it to be paused and restarted to examine the models of a run in progress. During any point of the evolutionary run, the virtual file system allows all files created during the run to be downloaded.

These architectural choices make \texttt{yreclab} uniquely suited for teaching \& experimentation in stellar astrophysics \citep{yreclab}. Existing efforts such as \texttt{MESA-Web} \citep{Fields2023} have been successful at providing a platform for teaching stellar physics, but require calculations to be submitted for later download by email, and have a limited subset of toggles available on their calculation submission page. With \texttt{yreclab}, the scriptable WebAssembly platform allows calculations to be supervised and interacted with at real-time, at near native desktop speeds (within a factor of two), with full support for the same namelists available in the desktop version of YREC.

\section{Summary}

Stellar astrophysics is a rich topic, and it has deep connections across astronomy writ large. Stellar evolution codes are natural tools for understanding the life cycles of stars, which is why a number of were developed over the years.  However, the list of public codes is short, and different codes are optimized for different kinds of stars.
In this paper we present a public release of the YREC stellar evolution code. YREC has been primarily used for the evolution of low and intermediate mass stars, and in that context it is a state-of-the-art code with sophisticated options. It can be run for a wide range of conditions, but caution should be taken when using any tools outside of the domains where development has been focused. We begin by summarizing the key domains for use, followed by the road map for improvements and updates. For the discussion below, we distinguish between brown dwarfs ($M < \sim85 M_J$); low mass stars with deeper convective envelopes ($\sim1.3 M_{\odot} > M > \sim85 M_J$), intermediate mass stars ($\sim8 M_{\odot} > M > \sim1.3 M_{\odot}$), and high mass stars ($M > \sim8 M_{\odot}$), with these ranges defined at solar metallicity.

YREC has been extensively used for low mass stars, starting from the pre-MS through core He burning. It can be used for hotter brown dwarfs, but more realistic models are limited by the surface boundary condition and equation of state tables to either younger ages or higher mass. In its current form we recommend caution below $30 M_J.$ 

The code requires a starting model, but fortunately rescaling allows users to generate starting models with mass and composition different from those in the library. There is also no rule against exporting models from other codes to YREC. Our initial conditions, on the upper Hayashi track, begin in a fictional place. Fortunately, it's also a useful one, as these models can be rescaled and evolved to a more physically reasonable deuterium birthline. Even in this case, real models would accrete up the birthline, so the real very early evolution could be quite different than that in the models. This has limited impact on classical model properties, but it's important for evolving models with rotation. Artificially large starting models will have extravagant angular momenta, so care needs to be taken to choose a proper point for initializing such models.

The code allows a range of angular momentum evolution models, including the structural effects of star spots, which will be especially useful for gyrochronology and rotationally induced mixing. Strong magnetic coupling, such as that inferred from the Taylor-Spruit instability, can be effectively modeled as solid body rotation for these purposes. Wave-driven angular momentum transport is not currently supported in the code. YREC makes state of the art solar models. 

The code also treats microscopic diffusion well in the low mass star domain. The primary limitation is that metals are diffused at the same rate. We note, however, that this is of order a 1\% level effect for solar models. The code crashes at the He flash, but our scripting tools can be used to restart evolution in the core He-burning phase.  The current code does not include support for the thermally pulsing AGB phase, nor does it include thermohaline mixing. The code does include mass loss, but it is being re-worked and the current version is not supported.

For intermediate mass stars, numerical issues arise with microscopic diffusion in stars with thin surface convection zones. These can be avoided by setting the outer fitting point deep enough, at the cost of shutting the microscopic diffusion calculations off. The detailed behavior of the code in the outer layers will depend on the location of the outer fitting point, which should be chosen with care. Note that this has very little impact on ages, or standard isoschrones, as radiative envelope stars are insensitive to the treatment of the outer boundary condition. YREC has limited options for the treatment of convective boundaries, which can become important in this mass domain. Users interested in rotation should ensure that the initial angular momenta of the models are reasonable for the stars under consideration.

The upper mass range of YREC is set by the onset of radiation-pressure driven winds. These induce negative pressure in the atmosphere solutions above 16 solar masses - a physical, but regrettable, feature in a system with logarithmic co-ordinates. YREC at present does not include radiation pressure driven winds, nor does it include nuclear burning stages beyond helium burning. It does capture the broad properties of models in the domain where it runs; above 12 solar masses, the code should be used with caution.

YREC will run with a wide range of birth mixtures, although the input physics tables in general do not support changes in the heavy element mixture. The maximum Z for the tables is 0.1, and users should be prepared to face challenges in the limit where the metal content is taken to zero. The isotopes and species that the code can track are hard-coded at present, although work is underway to generalize this.

The code continues to be actively developed. We intend to overhaul and bring the code up to date for the treatment of convective boundaries and mass loss. More generally, we are looking at adding in well-posed treatments of important physical processes. We are also in the process of evaluating and updating the basic input physics. This is an effort where we welcome external effort. Over a longer term, we intend to make the code easier to modify, and to generalize some of the hard-coded components that could make such changes more challenging. It is our hope that this code proves to be useful, and we welcome feedback.

\begin{acknowledgments}
We thank an anonymous referee and Earl Bellinger for helpful comments on the manuscript. We thank Yaguang Li for example MESA inlists.

The YREC stellar evolution code has been the product of decades of effort and improvements from a variety of contributors. Many, but not all, of these contributions are preserved in comments in the code.  Major contributors in addition to the authors on this paper include Pierre Demarque, Richard Larson, Michael Prather, Don van den Berg, David Guenther, Grant Newsham, Brian Chaboyer, and John Bahcall. Lawrence Pinsonneault also made major contributions to the organization and documentation of the code.  

MHP acknowledges support from NASA grants 80NSSC24K0622 and 80NSSC24K0637. JT and LM acknowledge support from 80NSSC22K0812. JvS acknowledges support from NSF grants AST-2205888 and AST-2507789. ZRC acknowledges support from NASA ROSES 80NSSC24K0081. LC acknowledges support from the Vanderbilt Initiative in Data-intensive Astrophysics (VIDA) and NASA Grant 80NSSC24K0622. 

\end{acknowledgments}

\begin{contribution}

MP is responsible for a large fraction of the YREC stellar evolution code. He also took primary responsibility for writing the text and developing the test suite, as well as documenting the code. JvS contributed to code testing and development, gyrochronology example cases, and documentation. LC contributed to code development, documentation, starspots example cases, test suite validation, and the \texttt{yreclab} web interface. JT contributed to the development of the radial differential rotation modules, the available atmosphere boundary condition options, assisted with grid tool development, and participated in writing and documentation. FD contributed opacity tables, some writing of the code, and he participated in the writing of the manuscript. LM contributed to the grid tool development, generation of the base stellar model grid, and writing of the relevant sections of the manuscript. RAP tested the structural and evolutionary behavior of high-mass models with various modifications to convection and contributed to associated portions of the manuscript. VAS wrote the scripts for installing the base model grid into the \texttt{kiauhoku} stellar model interpolator, the guide to creating isochrones with either a custom or the base YREC grid, and the ISTARY code/library. In addition, VAS organized the user tools repository, contributed to grid tool development, and wrote relevant sections of the manuscript. SB contributed to the grid tool development. AA contributed to the scripts for the angular momentum tracer code and the relevant manuscript sections. KC participated in the maintenance and modernization of the YREC code and related discussions. MR developed the build system, test suite executor, hunted down bugs in the YREC code, and contributed to relevant sections of the manuscript. ZRC authored the \texttt{kiauhoku} package, advised on its usage, and provided feedback on the manuscript.  JCZ contributed to horizontal branch model testing and associated manuscript writing.

\end{contribution}

\software{astropy \citep{2013A&A...558A..33A}
\texttt{kiauhoku} \citep{kiauhokupaper, kiauhoku_ascl},
          }

\appendix

\section{Scripts and Tools}

\subsection{Building YREC}
Users can download YREC from \url{https://github.com/yreclab/yrec}. YREC is implemented entirely in Fortran 77 and is straightforward to build. The code is Fortran 90 compliant. A simple build system is provided to create binaries of YREC for several different purposes. Two major compiler families are currently supported, the GNU Fortran compiler (gfortran) and the Intel Fortran compiler (ifort, ifx). Building the software is supported on Linux, Mac OS, and Windows. The default build configuration is set to generate a program with the best performance for large scale model runs. Development and debug build configurations are also provided which may be useful to those who wish to modify or further develop YREC.

\subsection{Namelist Converters}
 The YREC control namelists (nml) contain file paths to input files containing physical parameters and input/output controls. Users can run examples using the relative paths provided, or edit the locations of input and output files manually or by utilizing the \texttt{change\_nml()} or \texttt{update\_nml} codes. We provide a simpler namelist tool for new users, \texttt{update\_nml}\footnote{\url{https://github.com/avincesmedile/YREC-Wrappers/blob/main/modelgrid_tools/main_tools/update_nml.py}}, which allows users to modify namelists one variable at a time using command-line prompts and inputs. In addition, we provide the more advanced \texttt{change\_nml}\footnote{\url{https://github.com/avincesmedile/YREC-Wrappers/blob/main/modelgrid_tools/alternate_tools/change_nml.py}} tool, which can intelligently modify multiple namelist files at once to specify input and output data locations for YREC. After modification, users can run their YREC namelists either from the command line or via the Python wrappers introduced here for coordinated parallel runs, with the outputs deposited in an organized fashion. 

\subsubsection{YREC Output Reader}

YREC returns a set of output files, such as the .track files, which detail the evolution of the star from the designated starting point. The ability to read batches of such files can be useful.  We provide functionality to locate and read large batches of files into Python programs for manipulation with pandas or astropy. We note that some minor modifications of the source code, discussed in the package, are required to use this option.

The wrappers themselves---\texttt{load\_yrec\_tracks()}\footnote{\url{https://github.com/avincesmedile/YREC-Wrappers/blob/main/modelgrid_tools/main_tools/load_yrec_tracks.py}} and \texttt{tracker()}\footnote{\url{https://github.com/avincesmedile/YREC-Wrappers/blob/main/modelgrid_tools/main_tools/Tracker.py}}---are provided to users to read in .track files. \texttt{The tracker()} function can be used together with \texttt{load\_yrec\_tracks()} or on its own. Users provide a filepath to a .track file and it is read into Python as a Pandas DataFrame. The load\_yrec\_tracks() function takes a filepath to a folder containing all the outputted files. The code either uses the immediate directory or searches all subdirectories for .track files (to be used if the nml were modified using change\_nml() function). Optional functionality is given to manipulate the stored data to create EEPs and Isochrones and return them alongside the main dataset. 

\subsection{Model Grid Generators}

We provide wrappers that can be used to load the input physics and namelists and execute YREC to generate various grids of models. We provide the means to re-create our sample model grid (\texttt{make\_modelgrid.py})\footnote{\url{https://github.com/avincesmedile/YREC-Wrappers/blob/main/modelgrid_tools/main_tools/make_modelgrid.py}}, as well as generate a grid of models with the same input physics but a range of masses and metallicities via Python (\texttt{batchrunner.py})\footnote{\url{https://github.com/avincesmedile/YREC-Wrappers/blob/main/modelgrid_tools/main_tools/batchrunner.py}} and Slurm (\texttt{run\_yrec\_grid.slurm})\footnote{\url{https://github.com/avincesmedile/YREC-Wrappers/tree/main/modelgrid_tools/slurm_tools}} scripts to run a grid. In addition to this, we also provide the \texttt{yrec\_parallel.py}\footnote{\url{https://github.com/avincesmedile/YREC-Wrappers/blob/main/modelgrid_tools/alternate_tools/yrec_parallel.py}} script for running large batches of models on different CPUs of the users' native machine. This tool supports systematic variation of input values to explore a given physics parameter space across many independent YREC runs. For all batch scripts, users provide the path information for their YREC installation and namelists, then specify the masses and Fe/H they wish the grid to cover. The codes run the YREC executable with the corresponding namelists. Forthcoming additions include an option to generate ranges of birth rotation rates, which requires custom nml2 files as well as custom nml1 files. 

\subsection{Solar Model Calibration}

Our solar model calibrations illustrate the power of scripting YREC execution. We provide the \texttt{solar\_rot\_calibrated.py} script\footnote{\url{https://github.com/avincesmedile/YREC-Wrappers/blob/main/modelgrid_tools/main_tools/solar_rot_calibrated.py}} to this end. The code performs a template solar model run, including rotation, with user-specified initial X, Z, mixing length $\alpha$, birth rotation rate, angular momentum loss scale factor $f_k$, and mixing efficiency parameter $f_c$. The user also specifies the solar M, Z/X, L, R, age, surface rotation period, and the target solar lithium depletion.  In the classical or diffusion model case, the script evaluates L and R, and iterates X, Z, and $\alpha$ until they match the solar values within a user-specified tolerance. This is done with partial derivatives of L and R with respect to X and $\alpha$ taken from prior solar model runs; Z is adjusted to the user-specified Z/X. Users can also specify their own partial derivatives if desired. In the case where diffusion is included, the script adds a check on the surface Z/X to ensure that it matches the user-specified current value. In a classical model, the birth value is the same as the current value, so no such check is performed. Finally, the model with rotation adds the requirement that the model surface rotation rate and logarithmic lithium depletion match the solar values. This is done by adjusting the zero point of the angular momentum loss rate $f_k$ and the efficiency of mixing with respect to angular momentum transport $f_c$, respectively. 

\subsection{Tolerance Calibration}

\begin{figure}[h]
    \centering
    \includegraphics[width=8 cm]{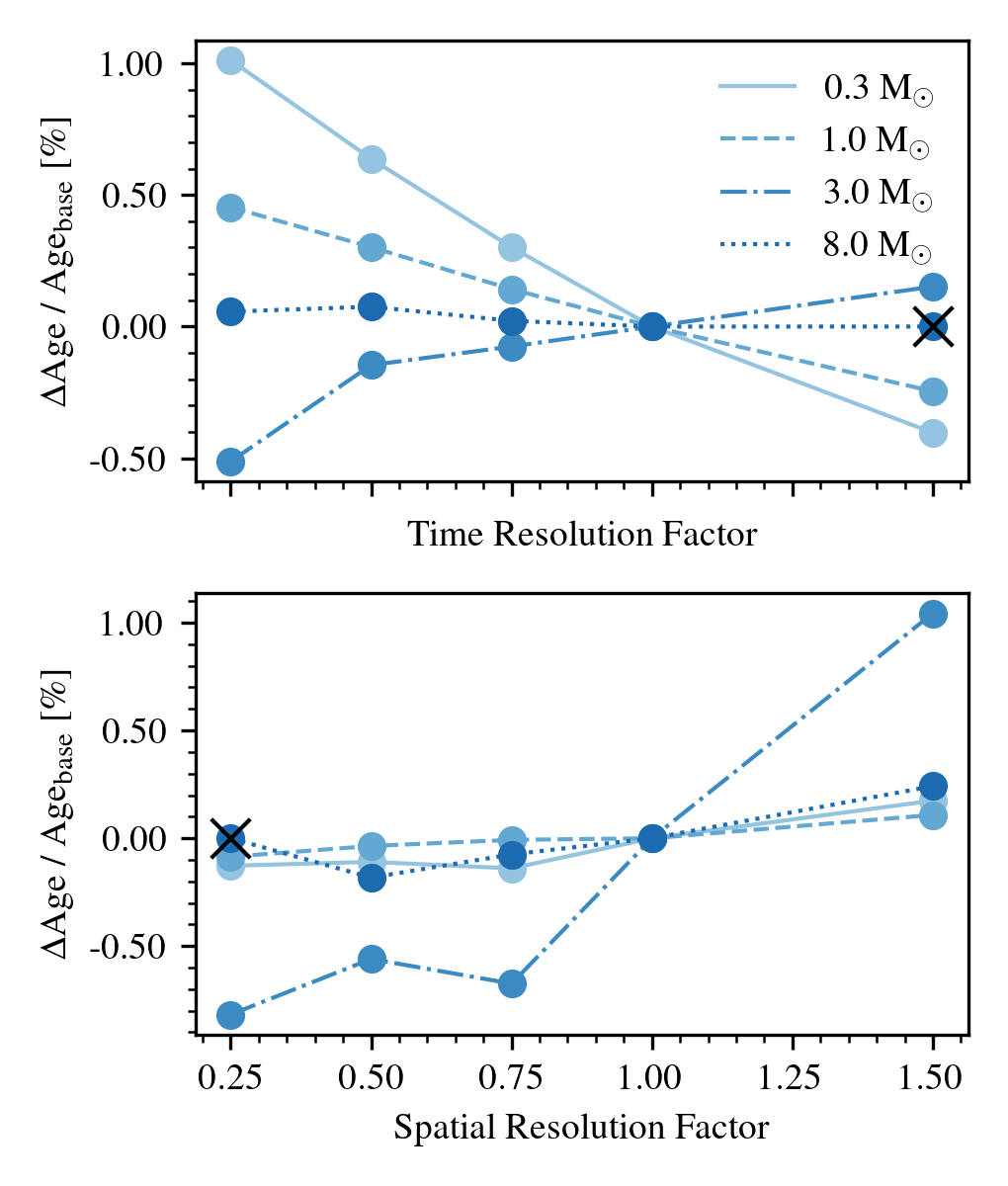}
    \caption{Spatial and temporal resolution tests for solar metallicity models at 0.3, 1.0, 3.0, and 8.0 M$_\odot$. The tolerances from the grid namelists were multiplied by the indicated factors, and the resulting percent change in the TAMS age is show relative to the base-resolution model. X's denote failure to converge.}
    \label{fig:toltest}
\end{figure}

These scripts perturb the spatial and temporal resolution of a specified run and infer the change in a figure of merit. This can be used to define the numerical controls required to meet a goal. In Figure \ref{fig:toltest} we show an example for the models used to construct the base grid; in this case, our figure of merit was the age at TAMS. We recommend using the \texttt{update\_nml()} or \texttt{change\_nml()} functions to make a model run and change the spatial and temporal tolerances as we do in Figure \ref{fig:toltest}, and choose a metric of merit (such as Age in our case) to compare the end-of-run values to. 

\section{Troubleshooting YREC}

The YREC code will fail to execute in some cases, and it can be helpful to understand the origin of some of the failure modes. The .short output file contains very useful diagnostic data in case of a failure, and it should be consulted.  Here are some common problems and solutions:

\subsection{Failure to converge}
The code will apply corrections to the structure variables iteratively. If it has not converged within the tolerances in a user-specified number of attempts, it will stop. Sometimes this can be solved by simply increasing the maximum iterations; this may be fruitful if the corrections are decreasing slowly in amplitude. More often, they reach a flat amplitude and change sign. The most common cause is that the spatial or temporal tolerances are too loose, such that the solution cannot reach the desired precision. In some cases, changing the size of the envelope triangle can also help, as this influences the accuracy of the surface boundary conditon interpolator. Users may also relax the convergence criteria; be cautioned, however, that tolerances looser than the $10^{-4}$ level can produce erratic behavior in global properties.

Sometimes the code will cascade down to an extremely small timestep, which always results in a convergence failure. In this case, the solution is counter-intuitive: the timestep criteria need to be relaxed. In the limit of very small timesteps, the corrections become very sensitive to numerical noise.

The third and fourth levels of iteration can produce convergence failures in evolved red giant branch models; we recommend that users set these to 0 unless they are explicitly needed to study rotational mixing.

\subsection{Hard failure modes.}

Other failures are tied to the assumptions in the code, and they are not numerical in nature. The current code fails if both overshoot and semi-convection are used at the same time. Above $\sim16$ solar masses on the main sequence, gray atmospheres models will have a net negative gas pressure due to the onset of radiation pressure driven winds, which YREC does not currently account for. The code will also fail if rotation is so rapid that material at the equator is unbound; users should check that the internal rotation profile does not fail this criterion. Frequently this condition is triggered by a poor choice of starting conditions (see below). Finally, the code cannot evolve through a degenerate helium flash. This failure mode will manifest as runaway corrections, and it cannot be solved without changes to the code itself.

\subsection{Challenging domains}

Stellar physics is the most uncertain at very high and very low masses. We are actively working on the code, and anticipate progress on the phase space where it can be used. Below we document the current domains where YREC is either difficult to run or not well suited given its current assumptions.

For high mass stars, a combination of missing physics and numerical challenges makes it difficult to run models above 9 solar masses through core He exhaustion. 

Models with rotation can be run across the full mass range. However, rotationally induced mixing is too efficient for stars above the Kraft break, and we do not recommend running models with mixing in this domain.

Models with microscopic diffusion encounter numerical difficulties when the hydrogen or helium mass fractions approach zero in the envelope. We therefore also recommend that the code not be used with diffusion enabled above $\sim 1.4$ solar masses at solar metallicity.

Brown dwarfs and stars close to the hydrogen-burning limit are numerically challenging, both due to the uncertain EoS and the importance of the surface boundary conditions for the solution. We do not recommend using the Yale EoS below $0.5 M_{\odot}$, or using a gray atmosphere below $0.3 M_{\odot}$. The \citet{Allard1995} grid is the currently recommended choice for low mass stars.

We also note that models close to $0.1 M_{\odot}$ can encounter challenges because their physical conditions skirt close to the edge of published equation of state tables, making their structure sensitive to the ramping between different EoS sources.

Core helium burning stars may not achieve $Y_{\mathrm{cen}} = 10^{-4}$, reaching temperatures that may be outside the bounds of interior opacity tables before reaching low central helium abundance. The equilibrium solution for the seed ZAHB models used for the low-mass two-step rescaling procedure is also sensitive to the choice of neutrino cooling physics, and users are cautioned that their model may not run if attempting to use \citet{Itoh1996} rates, which are otherwise the standard for YREC. All low-mass core helium burning model namelists in the test suite set \texttt{LNULOS1} to false, which results in defaulting to rates from \cite{1989neas.book.....B}.

\subsection{Choices that can lead to sorrow.}

Our digital constructs lack imagination and foresight, merely doing what they are asked to do. It is therefore in the nature of software to fail. Some of these failures reflect poor design. There are times, however, where the fault lies not in the stars, but in ourselves (Shakespeare 1599). Here we list, from personal experience, lessons that users of the code should take to heart.

There may be a time when computers can read our minds, doing the things that we wish to be done. We do not yet live in such a time. If you do not tell YREC where to read from, or where to write to, it will fail. Check your pointers. 

The code features a lovely variety of options, but human imagination is far greater in scope. If you include variables that the code does not recognize in the namelists, it will quit in disgust. Ensure that your comments are clearly labelled as such, and that the version of the code that you are using supports the choices that you are making.

It is tempting to use a single template to clone a series of runs, changing the output pointers to produce large grids of models. It is important to check that you are also changing the inputs, such that the files that you produce are what you think they are. In a better world, you would always get what you desire. In this one, you get what you deserve.

The code does, in fact, provide diagnostic outputs.  When it fails, or misbehaves, reading the messages on the screen, or those stored in the short file, can lead to enlightenment. Or not, of course.

If you assign the wrong input to the wrong name, the code will become confused at this offense to the natural order of things and quit. However, if you assign a file with the right format, but the wrong physics, to the right place---say, for example, an atmosphere table with a metal content different from that of the star---it will assume that you have wisdom, and it will allow you to proceed without comment. Ask yourself whether this trust is warranted.

Conservation laws are the foundation of modern physics, and YREC respects these iron rules. If you start with a very large star, and give it a modest spin, it will quickly become unruly as it shrinks and spins up. At some point it may become unbound at the equator---a situation that YREC finds intolerable, as it cannot conceive of imaginary numbers. Check your starting point for models with rotation.  It is easy to give stars far more angular momentum than they can find a use for.

While we are on the subject of imaginary numbers, recall that the code uses logarithmic variables. Zero is a special number. Before you set something in the code to zero, some introspection is required. Do you really need that number to be zero, or will a very small number suffice instead? Do you feel lucky? Be prepared to experiment and learn. In some cases, the code will anticipate folly and quit. In other, more devious cases, it will not, and it may break in unusual ways.

Finally, a successful run of the code is the beginning, not the end, of the journey. You should plot your results and see if they appear sensible. You should change your numerics and see whether they change your answer to any interesting degree. If they do, ponder your life choices, and adjust. Ensure that the tables that you use are the ones you actually want to use. Ensure that the physics that you think that you are using are the same as the physics that you are, in fact, using. 
 
\bibliography{Bibliography.bib}{}

@ARTICLE{2020Planck,
       author = {{Planck Collaboration} and {Aghanim}, N. and {Akrami}, Y. and {Ashdown}, M. and {Aumont}, J. and {Baccigalupi}, C. and {Ballardini}, M. and {Banday}, A.~J. and {Barreiro}, R.~B. and {Bartolo}, N. and {Basak}, S. and {Battye}, R. and {Benabed}, K. and {Bernard}, J.-P. and {Bersanelli}, M. and {Bielewicz}, P. and {Bock}, J.~J. and {Bond}, J.~R. and {Borrill}, J. and {Bouchet}, F.~R. and {Boulanger}, F. and {Bucher}, M. and {Burigana}, C. and {Butler}, R.~C. and {Calabrese}, E. and {Cardoso}, J.-F. and {Carron}, J. and {Challinor}, A. and {Chiang}, H.~C. and {Chluba}, J. and {Colombo}, L.~P.~L. and {Combet}, C. and {Contreras}, D. and {Crill}, B.~P. and {Cuttaia}, F. and {de Bernardis}, P. and {de Zotti}, G. and {Delabrouille}, J. and {Delouis}, J.-M. and {Di Valentino}, E. and {Diego}, J.~M. and {Dor{\'e}}, O. and {Douspis}, M. and {Ducout}, A. and {Dupac}, X. and {Dusini}, S. and {Efstathiou}, G. and {Elsner}, F. and {En{\ss}lin}, T.~A. and {Eriksen}, H.~K. and {Fantaye}, Y. and {Farhang}, M. and {Fergusson}, J. and {Fernandez-Cobos}, R. and {Finelli}, F. and {Forastieri}, F. and {Frailis}, M. and {Fraisse}, A.~A. and {Franceschi}, E. and {Frolov}, A. and {Galeotta}, S. and {Galli}, S. and {Ganga}, K. and {G{\'e}nova-Santos}, R.~T. and {Gerbino}, M. and {Ghosh}, T. and {Gonz{\'a}lez-Nuevo}, J. and {G{\'o}rski}, K.~M. and {Gratton}, S. and {Gruppuso}, A. and {Gudmundsson}, J.~E. and {Hamann}, J. and {Handley}, W. and {Hansen}, F.~K. and {Herranz}, D. and {Hildebrandt}, S.~R. and {Hivon}, E. and {Huang}, Z. and {Jaffe}, A.~H. and {Jones}, W.~C. and {Karakci}, A. and {Keih{\"a}nen}, E. and {Keskitalo}, R. and {Kiiveri}, K. and {Kim}, J. and {Kisner}, T.~S. and {Knox}, L. and {Krachmalnicoff}, N. and {Kunz}, M. and {Kurki-Suonio}, H. and {Lagache}, G. and {Lamarre}, J.-M. and {Lasenby}, A. and {Lattanzi}, M. and {Lawrence}, C.~R. and {Le Jeune}, M. and {Lemos}, P. and {Lesgourgues}, J. and {Levrier}, F. and {Lewis}, A. and {Liguori}, M. and {Lilje}, P.~B. and {Lilley}, M. and {Lindholm}, V. and {L{\'o}pez-Caniego}, M. and {Lubin}, P.~M. and {Ma}, Y.-Z. and {Mac{\'\i}as-P{\'e}rez}, J.~F. and {Maggio}, G. and {Maino}, D. and {Mandolesi}, N. and {Mangilli}, A. and {Marcos-Caballero}, A. and {Maris}, M. and {Martin}, P.~G. and {Martinelli}, M. and {Mart{\'\i}nez-Gonz{\'a}lez}, E. and {Matarrese}, S. and {Mauri}, N. and {McEwen}, J.~D. and {Meinhold}, P.~R. and {Melchiorri}, A. and {Mennella}, A. and {Migliaccio}, M. and {Millea}, M. and {Mitra}, S. and {Miville-Desch{\^e}nes}, M.-A. and {Molinari}, D. and {Montier}, L. and {Morgante}, G. and {Moss}, A. and {Natoli}, P. and {N{\o}rgaard-Nielsen}, H.~U. and {Pagano}, L. and {Paoletti}, D. and {Partridge}, B. and {Patanchon}, G. and {Peiris}, H.~V. and {Perrotta}, F. and {Pettorino}, V. and {Piacentini}, F. and {Polastri}, L. and {Polenta}, G. and {Puget}, J.-L. and {Rachen}, J.~P. and {Reinecke}, M. and {Remazeilles}, M. and {Renzi}, A. and {Rocha}, G. and {Rosset}, C. and {Roudier}, G. and {Rubi{\~n}o-Mart{\'\i}n}, J.~A. and {Ruiz-Granados}, B. and {Salvati}, L. and {Sandri}, M. and {Savelainen}, M. and {Scott}, D. and {Shellard}, E.~P.~S. and {Sirignano}, C. and {Sirri}, G. and {Spencer}, L.~D. and {Sunyaev}, R. and {Suur-Uski}, A.-S. and {Tauber}, J.~A. and {Tavagnacco}, D. and {Tenti}, M. and {Toffolatti}, L. and {Tomasi}, M. and {Trombetti}, T. and {Valenziano}, L. and {Valiviita}, J. and {Van Tent}, B. and {Vibert}, L. and {Vielva}, P. and {Villa}, F. and {Vittorio}, N. and {Wandelt}, B.~D. and {Wehus}, I.~K. and {White}, M. and {White}, S.~D.~M. and {Zacchei}, A. and {Zonca}, A.},
        title = "{Planck 2018 results. VI. Cosmological parameters}",
      journal = {\aap},
         year = 2020,
        month = sep,
       volume = {641},
          eid = {A6},
        pages = {A6},
          doi = {10.1051/0004-6361/201833910},
archivePrefix = {arXiv},
       eprint = {1807.06209},
 primaryClass = {astro-ph.CO},
       adsurl = {https://ui.adsabs.harvard.edu/abs/2020A&A...641A...6P}
}

@BOOK{1989neas.book.....B,
       author = {{Bahcall}, John N.},
        title = "{Neutrino Astrophysics}",
         year = 1989,
         publisher = {Cambridge University Press},
       adsurl = {https://ui.adsabs.harvard.edu/abs/1989neas.book.....B}
}

@ARTICLE{2015MNRAS.453.2290B,
       author = {{Bossini}, Diego and {Miglio}, Andrea and {Salaris}, Maurizio and {Pietrinferni}, Adriano and {Montalb{\'a}n}, Josefina and {Bressan}, Alessandro and {Noels}, Arlette and {Cassisi}, Santi and {Girardi}, L{\'e}o and {Marigo}, Paola},
        title = "{Uncertainties on near-core mixing in red-clump stars: effects on the period spacing and on the luminosity of the AGB bump}",
      journal = {\mnras},
         year = 2015,
        month = nov,
       volume = {453},
       number = {3},
        pages = {2290-2301},
          doi = {10.1093/mnras/stv1738},
archivePrefix = {arXiv},
       eprint = {1507.07797},
 primaryClass = {astro-ph.SR},
       adsurl = {https://ui.adsabs.harvard.edu/abs/2015MNRAS.453.2290B}
}

@ARTICLE{2017MNRAS.469.4718B,
       author = {{Bossini}, D. and {Miglio}, A. and {Salaris}, M. and {Vrard}, M. and {Cassisi}, S. and {Mosser}, B. and {Montalb{\'a}n}, J. and {Girardi}, L. and {Noels}, A. and {Bressan}, A. and {Pietrinferni}, A. and {Tayar}, J.},
        title = "{Kepler red-clump stars in the field and in open clusters: constraints on core mixing}",
      journal = {\mnras},
         year = 2017,
        month = aug,
       volume = {469},
       number = {4},
        pages = {4718-4725},
          doi = {10.1093/mnras/stx1135},
archivePrefix = {arXiv},
       eprint = {1705.03077},
 primaryClass = {astro-ph.GA},
       adsurl = {https://ui.adsabs.harvard.edu/abs/2017MNRAS.469.4718B}
}

@ARTICLE{1989ApJ...340..241C,
       author = {{Caputo}, F. and {Castellani}, V. and {Chieffi}, A. and {Pulone}, L. and {Tornambe}, Jr., A.},
        title = "{The ``Red Giant Clock'' as an Indicator for the Efficiency of Central Mixing in Horizontal-Branch Stars}",
      journal = {\apj},
         year = 1989,
        month = may,
       volume = {340},
        pages = {241},
          doi = {10.1086/167387},
       adsurl = {https://ui.adsabs.harvard.edu/abs/1989ApJ...340..241C}
}

@article{constantino+2016,
	author = {{Constantino}, Thomas and {Campbell}, Simon W. and {Lattanzio}, John C. and {van Duijneveldt}, Adam},
	journal = {\mnras},
	month = mar,
	number = {4},
	pages = {3866-3885},
	title = {{The treatment of mixing in core helium burning models - II. Constraints from cluster star counts}},
	volume = {456},
	year = 2016}

@ARTICLE{1958ApJ...128..348S,
       author = {{Schwarzschild}, M. and {H{\"a}rm}, R.},
        title = "{Evolution of Very Massive Stars.}",
      journal = {\apj},
         year = 1958,
        month = sep,
       volume = {128},
        pages = {348},
          doi = {10.1086/146548},
       adsurl = {https://ui.adsabs.harvard.edu/abs/1958ApJ...128..348S}
}

@ARTICLE{1947ApJ...105..305L,
       author = {{Ledoux}, P.},
        title = "{Stellar Models with Convection and with Discontinuity of the Mean Molecular Weight}",
      journal = {\apj},
         year = 1947,
        month = mar,
       volume = {105},
        pages = {305},
          doi = {10.1086/144905},
       adsurl = {https://ui.adsabs.harvard.edu/abs/1947ApJ...105..305L}
}

@ARTICLE{1970AcA....20..195P,
       author = {{Paczy{\'n}ski}, B.},
        title = "{Evolution of Single Stars. II. Core Helium Burning in Population I Stars}",
      journal = {\actaa},
         year = 1970,
        month = jan,
       volume = {20},
        pages = {195},
       adsurl = {https://ui.adsabs.harvard.edu/abs/1970AcA....20..195P}
}

@article{constantino+2015,
	author = {{Constantino}, Thomas and {Campbell}, Simon W. and {Christensen-Dalsgaard}, J{\o}rgen and {Lattanzio}, John C. and {Stello}, Dennis},
	journal = {\mnras},
	month = sep,
	number = {1},
	pages = {123-145},
	title = {{The treatment of mixing in core helium burning models - I. Implications for asteroseismology}},
	volume = {452},
	year = 2015}

@article{bressan_chiosi_bertelli1981,
	author = {{Bressan}, A.~G. and {Chiosi}, C. and {Bertelli}, G.},
	journal = {\aap},
	month = sep,
	number = {1},
	pages = {25-30},
	title = {{Mass loss and overshooting in massive stars}},
	volume = {102},
	year = 1981}

@article{nguyen+2022,
	author = {{Nguyen}, C.~T. and {Costa}, G. and {Girardi}, L. and {Volpato}, G. and {Bressan}, A. and {Chen}, Y. and {Marigo}, P. and {Fu}, X. and {Goudfrooij}, P.},
	journal = {\aap},
	month = sep,
	pages = {A126},
	title = {{PARSEC V2.0: Stellar tracks and isochrones of low- and intermediate-mass stars with rotation}},
	volume = {665},
	year = 2022}

@article{costa+2019,
	author = {{Costa}, Guglielmo and {Girardi}, L{\'e}o and {Bressan}, Alessandro and {Marigo}, Paola and {Rodrigues}, Tha{\'\i}se S. and {Chen}, Yang and {Lanza}, Antonio and {Goudfrooij}, Paul},
	journal = {\mnras},
	month = jun,
	number = {4},
	pages = {4641-4657},
	title = {{Mixing by overshooting and rotation in intermediate-mass stars}},
	volume = {485},
	year = 2019}

@article{costa+2019b,
	author = {{Costa}, Guglielmo and {Girardi}, L{\'e}o and {Bressan}, Alessandro and {Chen}, Yang and {Goudfrooij}, Paul and {Marigo}, Paola and {Rodrigues}, Tha{\'\i}se S. and {Lanza}, Antonio},
	journal = {\aap},
	month = nov,
	pages = {A128},
	title = {{Multiple stellar populations in NGC 1866. New clues from Cepheids and colour-magnitude diagram}},
	volume = {631},
	year = 2019}

@article{nguyen+2025,
	author = {{Nguyen}, C.~T. and {Costa}, G. and {Bressan}, A. and {Girardi}, L. and {Cescutti}, G. and {Korn}, A.~J. and {Volpato}, G. and {Chen}, Y. and {Pastorelli}, G. and {Trabucchi}, M. and {Shepherd}, K.~G. and {Ettorre}, G. and {Zaggia}, S.},
	journal = {\aap},
	month = sep,
	pages = {A258},
	title = {{PARSEC V2.0: Rotating tracks and isochrones for seven additional metallicities in the range Z = 0.0001{\textendash}0.03}},
	volume = {701},
	year = 2025}

@INPROCEEDINGS{1973ASSL...36..221S,
       author = {{Sweigart}, A.~V. and {Demarque}, P.},
        title = "{Semiconvection and the RR Lyrae Variables}",
    booktitle = {IAU Colloquium 21: Variable Stars in Globular Clusters and in Related Systems},
         year = 1973,
       editor = {{Fernie}, J.~D.},
       series = {Astrophysics and Space Science Library},
       volume = {36},
        month = jan,
        pages = {221},
          doi = {10.1007/978-94-010-2590-4_32},
       adsurl = {https://ui.adsabs.harvard.edu/abs/1973ASSL...36..221S}
}

@inproceedings{grevesse_noels1993,
	author = {{Grevesse}, N. and {Noels}, A.},
	booktitle = {Perfectionnement de l'Association Vaudoise des Chercheurs en Physique},
	editor = {{Hauck}, B. and {Paltani}, S. and {Raboud}, D.},
	month = jan,
	pages = {205-257},
	title = {{La composition chimique du Soleil.}},
	year = 1993}

@ARTICLE{2025A&A...701A.258N,
       author = {{Nguyen}, C.~T. and {Costa}, G. and {Bressan}, A. and {Girardi}, L. and {Cescutti}, G. and {Korn}, A.~J. and {Volpato}, G. and {Chen}, Y. and {Pastorelli}, G. and {Trabucchi}, M. and {Shepherd}, K.~G. and {Ettorre}, G. and {Zaggia}, S.},
        title = "{PARSEC V2.0: Rotating tracks and isochrones for seven additional metallicities in the range Z = 0.0001{\textendash}0.03}",
      journal = {\aap},
         year = 2025,
        month = sep,
       volume = {701},
          eid = {A258},
        pages = {A258},
          doi = {10.1051/0004-6361/202556005},
archivePrefix = {arXiv},
       eprint = {2508.02393},
 primaryClass = {astro-ph.SR},
       adsurl = {https://ui.adsabs.harvard.edu/abs/2025A&A...701A.258N}
}

@ARTICLE{2010Torres,
       author = {{Torres}, G. and {Andersen}, J. and {Gim{\'e}nez}, A.},
        title = "{Accurate masses and radii of normal stars: modern results and applications}",
      journal = {\aapr},
         year = 2010,
        month = feb,
       volume = {18},
       number = {1-2},
        pages = {67-126},
          doi = {10.1007/s00159-009-0025-1},
archivePrefix = {arXiv},
       eprint = {0908.2624},
 primaryClass = {astro-ph.SR},
       adsurl = {https://ui.adsabs.harvard.edu/abs/2010A&ARv..18...67T}
}

@ARTICLE{1993Kurucz,
       author = {{Kurucz}, Robert},
        title = "{SYNTHE Spectrum Synthesis Programs and Line Data.}",
      journal = {Robert Kurucz CD-ROM},
         year = 1993,
        month = jan,
       volume = {18},
       adsurl = {https://ui.adsabs.harvard.edu/abs/1993KurCD..18.....K}
}

@article{an+2019,
	author = {{An}, Deokkeun and {Pinsonneault}, Marc H. and {Terndrup}, Donald M. and {Chung}, Chul},
	journal = {\apj},
	month = {Jul},
	number = {2},
	pages = {81},
	title = {{Comparison of the Asteroseismic Mass Scale of Red Clump Giants with Photometric Mass Estimates}},
	volume = {879},
	year = {2019}}

@ARTICLE{Demarque2008,
       author = {{Demarque}, P. and {Guenther}, D.~B. and {Li}, L.~H. and {Mazumdar}, A. and {Straka}, C.~W.},
        title = "{YREC: the Yale rotating stellar evolution code. Non-rotating version, seismology applications}",
      journal = {\apss},
         year = 2008,
        month = aug,
       volume = {316},
       number = {1-4},
        pages = {31-41},
          doi = {10.1007/s10509-007-9698-y},
archivePrefix = {arXiv},
       eprint = {0710.4003},
 primaryClass = {astro-ph},
       adsurl = {https://ui.adsabs.harvard.edu/abs/2008Ap&SS.316...31D}
}

@ARTICLE{Rogers1996,
       author = {{Rogers}, Forrest J. and {Swenson}, Fritz J. and {Iglesias}, Carlos A.},
        title = "{OPAL Equation-of-State Tables for Astrophysical Applications}",
      journal = {\apj},
         year = 1996,
        month = jan,
       volume = {456},
        pages = {902},
          doi = {10.1086/176705},
       adsurl = {https://ui.adsabs.harvard.edu/abs/1996ApJ...456..902R}
}

@article{iglesias_rogers1996,
	author = {{Iglesias}, Carlos A. and {Rogers}, Forrest J.},
	journal = {\apj},
	month = jun,
	pages = {943},
	title = {{Updated Opal Opacities}},
	volume = {464},
	year = 1996}

@article{grevesse_sauval1998,
	author = {{Grevesse}, N. and {Sauval}, A.~J.},
	journal = {\ssr},
	month = may,
	pages = {161-174},
	title = {{Standard Solar Composition}},
	volume = 85,
	year = 1998}

@ARTICLE{LiG2024,
       author = {{Li}, Gang and {Deheuvels}, S{\'e}bastien and {Ballot}, J{\'e}r{\^o}me},
        title = "{Asteroseismic measurement of core and envelope rotation rates for 2006 red giant branch stars}",
      journal = {\aap},
         year = 2024,
        month = aug,
       volume = {688},
          eid = {A184},
        pages = {A184},
          doi = {10.1051/0004-6361/202449882},
archivePrefix = {arXiv},
       eprint = {2405.12116},
 primaryClass = {astro-ph.SR},
       adsurl = {https://ui.adsabs.harvard.edu/abs/2024A&A...688A.184L}
}

@article{paxton+2011,
	author = {{Paxton}, B. and {Bildsten}, L. and {Dotter}, A. and {Herwig}, F. and {Lesaffre}, P. and {Timmes}, F.},
	journal = {\apjs},
	month = jan,
	pages = {3},
	title = {{Modules for Experiments in Stellar Astrophysics (MESA)}},
	volume = 192,
	year = 2011}

@article{paxton+2013,
	author = {{Paxton}, B. and {Cantiello}, M. and {Arras}, P. and {Bildsten}, L. and {Brown}, E.~F. and {Dotter}, A. and {Mankovich}, C. and {Montgomery}, M.~H. and {Stello}, D. and {Timmes}, F.~X. and {Townsend}, R.},
	journal = {\apjs},
	month = sep,
	pages = {4},
	title = {{Modules for Experiments in Stellar Astrophysics (MESA): Planets, Oscillations, Rotation, and Massive Stars}},
	volume = 208,
	year = 2013}

@article{paxton+2015,
	author = {{Paxton}, B. and {Marchant}, P. and {Schwab}, J. and {Bauer}, E.~B. and {Bildsten}, L. and {Cantiello}, M. and {Dessart}, L. and {Farmer}, R. and {Hu}, H. and {Langer}, N. and {Townsend}, R.~H.~D. and {Townsley}, D.~M. and {Timmes}, F.~X.},
	journal = {\apjs},
	month = sep,
	pages = {15},
	title = {{Modules for Experiments in Stellar Astrophysics (MESA): Binaries, Pulsations, and Explosions}},
	volume = 220,
	year = 2015}

@ARTICLE{Gehan2018,
       author = {{Gehan}, C. and {Mosser}, B. and {Michel}, E. and {Samadi}, R. and {Kallinger}, T.},
        title = "{Core rotation braking on the red giant branch for various mass ranges}",
      journal = {\aap},
         year = 2018,
        month = aug,
       volume = {616},
          eid = {A24},
        pages = {A24},
          doi = {10.1051/0004-6361/201832822},
archivePrefix = {arXiv},
       eprint = {1802.04558},
 primaryClass = {astro-ph.SR},
       adsurl = {https://ui.adsabs.harvard.edu/abs/2018A&A...616A..24G}
}

@ARTICLE{Tayar2017a,
       author = {{Tayar}, Jamie and {Somers}, Garrett and {Pinsonneault}, Marc H. and {Stello}, Dennis and {Mints}, Alexey and {Johnson}, Jennifer A. and {Zamora}, O. and {Garc{\'\i}a-Hern{\'a}ndez}, D.~A. and {Maraston}, Claudia and {Serenelli}, Aldo and {Allende Prieto}, Carlos and {Bastien}, Fabienne A. and {Basu}, Sarbani and {Bird}, J.~C. and {Cohen}, R.~E. and {Cunha}, Katia and {Elsworth}, Yvonne and {Garc{\'\i}a}, Rafael A. and {Girardi}, Leo and {Hekker}, Saskia and {Holtzman}, Jon and {Huber}, Daniel and {Mathur}, Savita and {M{\'e}sz{\'a}ros}, Szabolcs and {Mosser}, B. and {Shetrone}, Matthew and {Silva Aguirre}, Victor and {Stassun}, Keivan and {Stringfellow}, Guy S. and {Zasowski}, Gail and {Roman-Lopes}, A.},
        title = "{The Correlation between Mixing Length and Metallicity on the Giant Branch: Implications for Ages in the Gaia Era}",
      journal = {\apj},
         year = 2017,
        month = may,
       volume = {840},
       number = {1},
          eid = {17},
        pages = {17},
          doi = {10.3847/1538-4357/aa6a1e},
archivePrefix = {arXiv},
       eprint = {1704.01164},
 primaryClass = {astro-ph.SR},
       adsurl = {https://ui.adsabs.harvard.edu/abs/2017ApJ...840...17T}
}

@ARTICLE{Li2022,
       author = {{Li}, Yaguang and {Bedding}, Timothy R. and {Murphy}, Simon J. and {Stello}, Dennis and {Chen}, Yifan and {Huber}, Daniel and {Joyce}, Meridith and {Marks}, Dion and {Zhang}, Xianfei and {Bi}, Shaolan and {Colman}, Isabel L. and {Hayden}, Michael R. and {Hey}, Daniel R. and {Li}, Gang and {Montet}, Benjamin T. and {Sharma}, Sanjib and {Wu}, Yaqian},
        title = "{Discovery of post-mass-transfer helium-burning red giants using asteroseismology}",
      journal = {Nature Astronomy},
         year = 2022,
        month = apr,
       volume = {6},
        pages = {673-680},
          doi = {10.1038/s41550-022-01648-5},
archivePrefix = {arXiv},
       eprint = {2204.06203},
 primaryClass = {astro-ph.SR},
       adsurl = {https://ui.adsabs.harvard.edu/abs/2022NatAs...6..673L}
}

@ARTICLE{Larson1964,
       author = {{Larson}, R.~B. and {Demarque}, P.~R.},
        title = "{An Application of Henyey's Approach to the Integration of the Equations of Stellar Structure.}",
      journal = {\apj},
         year = 1964,
        month = aug,
       volume = {140},
        pages = {524},
          doi = {10.1086/147946},
       adsurl = {https://ui.adsabs.harvard.edu/abs/1964ApJ...140..524L}
}

@PHDTHESIS{Prather1976,
       author = {{Prather}, M.~J.},
        title = "{The Effect of a Brans-Dicke Cosmology upon Stellar Evolution and the Evolution of Galaxies.}",
       school = {Yale University, Connecticut},
         year = 1976,
        month = dec,
       adsurl = {https://ui.adsabs.harvard.edu/abs/1976PhDT.........8P}
}

@ARTICLE{vansaders2012,
       author = {{van Saders}, Jennifer L. and {Pinsonneault}, Marc H.},
        title = "{The Sensitivity of Convection Zone Depth to Stellar Abundances: An Absolute Stellar Abundance Scale from Asteroseismology}",
      journal = {\apj},
         year = 2012,
        month = feb,
       volume = {746},
       number = {1},
          eid = {16},
        pages = {16},
          doi = {10.1088/0004-637X/746/1/16},
archivePrefix = {arXiv},
       eprint = {1108.2273},
 primaryClass = {astro-ph.SR},
       adsurl = {https://ui.adsabs.harvard.edu/abs/2012ApJ...746...16V}
}

@ARTICLE{Bahcall1992,
       author = {{Bahcall}, J.~N. and {Pinsonneault}, M.~H.},
        title = "{Standard solar models, with and without helium diffusion, and the solar neutrino problem}",
      journal = {Reviews of Modern Physics},
         year = 1992,
        month = oct,
       volume = {64},
       number = {4},
        pages = {885-926},
          doi = {10.1103/RevModPhys.64.885},
       adsurl = {https://ui.adsabs.harvard.edu/abs/1992RvMP...64..885B}
}

@ARTICLE{Bahcall1995,
       author = {{Bahcall}, John N. and {Pinsonneault}, M.~H. and {Wasserburg}, G.~J.},
        title = "{Solar models with helium and heavy-element diffusion}",
      journal = {Reviews of Modern Physics},
         year = 1995,
        month = oct,
       volume = {67},
       number = {4},
        pages = {781-808},
          doi = {10.1103/RevModPhys.67.781},
archivePrefix = {arXiv},
       eprint = {hep-ph/9505425},
 primaryClass = {hep-ph},
       adsurl = {https://ui.adsabs.harvard.edu/abs/1995RvMP...67..781B}
}

@ARTICLE{Somers2020,
       author = {{Somers}, Garrett and {Cao}, Lyra and {Pinsonneault}, Marc H.},
        title = "{The SPOTS Models: A Grid of Theoretical Stellar Evolution Tracks and Isochrones for Testing the Effects of Starspots on Structure and Colors}",
      journal = {\apj},
         year = 2020,
        month = mar,
       volume = {891},
       number = {1},
          eid = {29},
        pages = {29},
          doi = {10.3847/1538-4357/ab722e},
archivePrefix = {arXiv},
       eprint = {2002.10644},
 primaryClass = {astro-ph.SR},
       adsurl = {https://ui.adsabs.harvard.edu/abs/2020ApJ...891...29S}
}

@ARTICLE{Somers2015,
       author = {{Somers}, Garrett and {Pinsonneault}, Marc H.},
        title = "{Older and Colder: The Impact of Starspots on Pre-main-sequence Stellar Evolution}",
      journal = {\apj},
         year = 2015,
        month = jul,
       volume = {807},
       number = {2},
          eid = {174},
        pages = {174},
          doi = {10.1088/0004-637X/807/2/174},
archivePrefix = {arXiv},
       eprint = {1506.01393},
 primaryClass = {astro-ph.SR},
       adsurl = {https://ui.adsabs.harvard.edu/abs/2015ApJ...807..174S}
}

@ARTICLE{Somers2016,
       author = {{Somers}, Garrett and {Pinsonneault}, Marc H.},
        title = "{Lithium Depletion is a Strong Test of Core-envelope Recoupling}",
      journal = {\apj},
         year = 2016,
        month = sep,
       volume = {829},
       number = {1},
          eid = {32},
        pages = {32},
          doi = {10.3847/0004-637X/829/1/32},
archivePrefix = {arXiv},
       eprint = {1606.00004},
 primaryClass = {astro-ph.SR},
       adsurl = {https://ui.adsabs.harvard.edu/abs/2016ApJ...829...32S}
}

@ARTICLE{Dotter2008,
       author = {{Dotter}, Aaron and {Chaboyer}, Brian and {Jevremovi{\'c}}, Darko and {Kostov}, Veselin and {Baron}, E. and {Ferguson}, Jason W.},
        title = "{The Dartmouth Stellar Evolution Database}",
      journal = {\apjs},
         year = 2008,
        month = sep,
       volume = {178},
       number = {1},
        pages = {89-101},
          doi = {10.1086/589654},
archivePrefix = {arXiv},
       eprint = {0804.4473},
 primaryClass = {astro-ph},
       adsurl = {https://ui.adsabs.harvard.edu/abs/2008ApJS..178...89D}
}

@ARTICLE{Pinsonneault1989,
       author = {{Pinsonneault}, M.~H. and {Kawaler}, Steven D. and {Sofia}, S. and {Demarque}, P.},
        title = "{Evolutionary Models of the Rotating Sun}",
      journal = {\apj},
         year = 1989,
        month = mar,
       volume = {338},
        pages = {424},
          doi = {10.1086/167210},
       adsurl = {https://ui.adsabs.harvard.edu/abs/1989ApJ...338..424P}
}

@ARTICLE{Krishnamurthi1997,
       author = {{Krishnamurthi}, Anita and {Pinsonneault}, M.~H. and {Barnes}, S. and {Sofia}, S.},
        title = "{Theoretical Models of the Angular Momentum Evolution of Solar-Type Stars}",
      journal = {\apj},
         year = 1997,
        month = may,
       volume = {480},
       number = {1},
        pages = {303-323},
          doi = {10.1086/303958},
       adsurl = {https://ui.adsabs.harvard.edu/abs/1997ApJ...480..303K}
}

@ARTICLE{Pinsonneault2002,
       author = {{Pinsonneault}, M.~H. and {Steigman}, G. and {Walker}, T.~P. and {Narayanan}, V.~K.},
        title = "{Stellar Mixing and the Primordial Lithium Abundance}",
      journal = {\apj},
         year = 2002,
        month = jul,
       volume = {574},
       number = {1},
        pages = {398-411},
          doi = {10.1086/340119},
archivePrefix = {arXiv},
       eprint = {astro-ph/0105439},
 primaryClass = {astro-ph},
       adsurl = {https://ui.adsabs.harvard.edu/abs/2002ApJ...574..398P}
}

@ARTICLE{vansaders2013,
       author = {{van Saders}, Jennifer L. and {Pinsonneault}, Marc H.},
        title = "{Fast Star, Slow Star; Old Star, Young Star: Subgiant Rotation as a Population and Stellar Physics Diagnostic}",
      journal = {\apj},
         year = 2013,
        month = oct,
       volume = {776},
       number = {2},
          eid = {67},
        pages = {67},
          doi = {10.1088/0004-637X/776/2/67},
archivePrefix = {arXiv},
       eprint = {1306.3701},
 primaryClass = {astro-ph.SR},
       adsurl = {https://ui.adsabs.harvard.edu/abs/2013ApJ...776...67V}
}

@ARTICLE{Angulo1999,
       author = {{Angulo}, C. and {Arnould}, M. and {Rayet}, M. and {Descouvemont}, P. and {Baye}, D. and {Leclercq-Willain}, C. and {Coc}, A. and {Barhoumi}, S. and {Aguer}, P. and {Rolfs}, C. and {Kunz}, R. and {Hammer}, J.~W. and {Mayer}, A. and {Paradellis}, T. and {Kossionides}, S. and {Chronidou}, C. and {Spyrou}, K. and {degl'Innocenti}, S. and {Fiorentini}, G. and {Ricci}, B. and {Zavatarelli}, S. and {Providencia}, C. and {Wolters}, H. and {Soares}, J. and {Grama}, C. and {Rahighi}, J. and {Shotter}, A. and {Lamehi Rachti}, M.},
        title = "{A compilation of charged-particle induced thermonuclear reaction rates}",
      journal = {\nphysa},
         year = 1999,
        month = aug,
       volume = {656},
       number = {1},
        pages = {3-183},
          doi = {10.1016/S0375-9474(99)00030-5},
       adsurl = {https://ui.adsabs.harvard.edu/abs/1999NuPhA.656....3A}
}

@ARTICLE{Caughlin1988,
       author = {{Caughlan}, Georgeann R. and {Fowler}, William A.},
        title = "{Thermonuclear Reaction Rates V}",
      journal = {Atomic Data and Nuclear Data Tables},
         year = 1988,
        month = jan,
       volume = {40},
        pages = {283},
          doi = {10.1016/0092-640X(88)90009-5},
       adsurl = {https://ui.adsabs.harvard.edu/abs/1988ADNDT..40..283C}
}

@ARTICLE{Adelberger2011,
       author = {{Adelberger}, E.~G. and {Garc{\'\i}a}, A. and {Robertson}, R.~G. Hamish and {Snover}, K.~A. and {Balantekin}, A.~B. and {Heeger}, K. and {Ramsey-Musolf}, M.~J. and {Bemmerer}, D. and {Junghans}, A. and {Bertulani}, C.~A. and {Chen}, J. -W. and {Costantini}, H. and {Prati}, P. and {Couder}, M. and {Uberseder}, E. and {Wiescher}, M. and {Cyburt}, R. and {Davids}, B. and {Freedman}, S.~J. and {Gai}, M. and {Gazit}, D. and {Gialanella}, L. and {Imbriani}, G. and {Greife}, U. and {Hass}, M. and {Haxton}, W.~C. and {Itahashi}, T. and {Kubodera}, K. and {Langanke}, K. and {Leitner}, D. and {Leitner}, M. and {Vetter}, P. and {Winslow}, L. and {Marcucci}, L.~E. and {Motobayashi}, T. and {Mukhamedzhanov}, A. and {Tribble}, R.~E. and {Nollett}, Kenneth M. and {Nunes}, F.~M. and {Park}, T. -S. and {Parker}, P.~D. and {Schiavilla}, R. and {Simpson}, E.~C. and {Spitaleri}, C. and {Strieder}, F. and {Trautvetter}, H. -P. and {Suemmerer}, K. and {Typel}, S.},
        title = "{Solar fusion cross sections. II. The pp chain and CNO cycles}",
      journal = {Reviews of Modern Physics},
         year = 2011,
        month = jan,
       volume = {83},
       number = {1},
        pages = {195-246},
          doi = {10.1103/RevModPhys.83.195},
archivePrefix = {arXiv},
       eprint = {1004.2318},
 primaryClass = {nucl-ex},
       adsurl = {https://ui.adsabs.harvard.edu/abs/2011RvMP...83..195A}
}

@ARTICLE{Acharya2025,
       author = {{Acharya}, B. and {Aliotta}, M. and {Balantekin}, A.~B. and {Bemmerer}, D. and {Bertulani}, C.~A. and {Best}, A. and {Brune}, C.~R. and {Buompane}, R. and {Gialanella}, L. and {Cavanna}, F. and {Chen}, J.~W. and {Colgan}, J. and {Czarnecki}, A. and {Davids}, B. and {deBoer}, R.~J. and {Delahaye}, F. and {Depalo}, R. and {Guglielmetti}, A. and {Garc{\'\i}a}, A. and {Robertson}, R.~G.~H. and {Gatu Johnson}, M. and {Gazit}, D. and {Greife}, U. and {Guffanti}, D. and {Hambleton}, K. and {Haxton}, W.~C. and {Herrera}, Y. and {Serenelli}, A. and {Huang}, M. and {Iliadis}, C. and {Kravvaris}, K. and {La Cognata}, M. and {Langanke}, K. and {Marcucci}, L.~E. and {Nagayama}, T. and {Nollett}, K.~M. and {Odell}, D. and {Orebi Gann}, G.~D. and {Piatti}, D. and {Pinsonneault}, M. and {Platter}, L. and {Rupak}, G. and {Sferrazza}, M. and {Sz{\"u}cs}, T. and {Tang}, X. and {Tumino}, A. and {Villante}, F.~L. and {Walker-Loud}, A. and {Zhang}, X. and {Zuber}, K.},
        title = "{Solar fusion III: New data and theory for hydrogen-burning stars}",
      journal = {Reviews of Modern Physics},
         year = 2025,
        month = jul,
       volume = {97},
       number = {3},
          eid = {035002},
        pages = {035002},
          doi = {10.1103/8lm7-gs18},
archivePrefix = {arXiv},
       eprint = {2405.06470},
 primaryClass = {astro-ph.SR},
       adsurl = {https://ui.adsabs.harvard.edu/abs/2025RvMP...97c5002A}
}

@ARTICLE{Xu2013,
       author = {{Xu}, Y. and {Takahashi}, K. and {Goriely}, S. and {Arnould}, M. and {Ohta}, M. and {Utsunomiya}, H.},
        title = "{NACRE II: an update of the NACRE compilation of charged-particle-induced thermonuclear reaction rates for nuclei with mass number A<16}",
      journal = {\nphysa},
         year = 2013,
        month = nov,
       volume = {918},
        pages = {61-169},
          doi = {10.1016/j.nuclphysa.2013.09.007},
archivePrefix = {arXiv},
       eprint = {1310.7099},
 primaryClass = {nucl-th},
       adsurl = {https://ui.adsabs.harvard.edu/abs/2013NuPhA.918...61X}
}

@ARTICLE{Cyburt2010,
       author = {{Cyburt}, Richard H. and {Amthor}, A. Matthew and {Ferguson}, Ryan and {Meisel}, Zach and {Smith}, Karl and {Warren}, Scott and {Heger}, Alexander and {Hoffman}, R.~D. and {Rauscher}, Thomas and {Sakharuk}, Alexander and {Schatz}, Hendrik and {Thielemann}, F.~K. and {Wiescher}, Michael},
        title = "{The JINA REACLIB Database: Its Recent Updates and Impact on Type-I X-ray Bursts}",
      journal = {\apjs},
         year = 2010,
        month = jul,
       volume = {189},
       number = {1},
        pages = {240-252},
          doi = {10.1088/0067-0049/189/1/240},
       adsurl = {https://ui.adsabs.harvard.edu/abs/2010ApJS..189..240C}
}

@ARTICLE{Itoh1996,
       author = {{Itoh}, Naoki and {Hayashi}, Hiroshi and {Nishikawa}, Akinori and {Kohyama}, Yasuharu},
        title = "{Neutrino Energy Loss in Stellar Interiors. VII. Pair, Photo-, Plasma, Bremsstrahlung, and Recombination Neutrino Processes}",
      journal = {\apjs},
         year = 1996,
        month = feb,
       volume = {102},
        pages = {411},
          doi = {10.1086/192264},
       adsurl = {https://ui.adsabs.harvard.edu/abs/1996ApJS..102..411I}
}

@BOOK{Burgers1969,
       author = {{Burgers}, J.~M.},
        title = "{Flow Equations for Composite Gases}",
         year = 1969,
         publisher = {Academic Press},
       adsurl = {https://ui.adsabs.harvard.edu/abs/1969fecg.book.....B}
}

@ARTICLE{Thoul1994,
       author = {{Thoul}, Anne A. and {Bahcall}, John N. and {Loeb}, Abraham},
        title = "{Element Diffusion in the Solar Interior}",
      journal = {\apj},
         year = 1994,
        month = feb,
       volume = {421},
        pages = {828},
          doi = {10.1086/173695},
archivePrefix = {arXiv},
       eprint = {astro-ph/9304005},
 primaryClass = {astro-ph},
       adsurl = {https://ui.adsabs.harvard.edu/abs/1994ApJ...421..828T}
}

@ARTICLE{Farag2024,
       author = {{Farag}, Ebraheem and {Fontes}, Christopher J. and {Timmes}, F.~X. and {Bellinger}, Earl P. and {Guzik}, Joyce A. and {Bauer}, Evan B. and {Wood}, Suzannah R. and {Mussack}, Katie and {Hakel}, Peter and {Colgan}, James and {Kilcrease}, David P. and {Sherrill}, Manolo E. and {Raecke}, Tryston C. and {Chidester}, Morgan T.},
        title = "{An Expanded Set of Los Alamos OPLIB Tables in MESA: Type-1 Rosseland-mean Opacities and Solar Models}",
      journal = {\apj},
         year = 2024,
        month = jun,
       volume = {968},
       number = {2},
          eid = {56},
        pages = {56},
          doi = {10.3847/1538-4357/ad4355},
archivePrefix = {arXiv},
       eprint = {2406.02845},
 primaryClass = {astro-ph.SR},
       adsurl = {https://ui.adsabs.harvard.edu/abs/2024ApJ...968...56F}
}

@ARTICLE{Ferguson2005,
       author = {{Ferguson}, Jason W. and {Alexander}, David R. and {Allard}, France and {Barman}, Travis and {Bodnarik}, Julia G. and {Hauschildt}, Peter H. and {Heffner-Wong}, Amanda and {Tamanai}, Akemi},
        title = "{Low-Temperature Opacities}",
      journal = {\apj},
         year = 2005,
        month = apr,
       volume = {623},
       number = {1},
        pages = {585-596},
          doi = {10.1086/428642},
archivePrefix = {arXiv},
       eprint = {astro-ph/0502045},
 primaryClass = {astro-ph},
       adsurl = {https://ui.adsabs.harvard.edu/abs/2005ApJ...623..585F}
}

@ARTICLE{Cassisi2007,
       author = {{Cassisi}, S. and {Potekhin}, A.~Y. and {Pietrinferni}, A. and {Catelan}, M. and {Salaris}, M.},
        title = "{Updated Electron-Conduction Opacities: The Impact on Low-Mass Stellar Models}",
      journal = {\apj},
         year = 2007,
        month = jun,
       volume = {661},
       number = {2},
        pages = {1094-1104},
          doi = {10.1086/516819},
archivePrefix = {arXiv},
       eprint = {astro-ph/0703011},
 primaryClass = {astro-ph},
       adsurl = {https://ui.adsabs.harvard.edu/abs/2007ApJ...661.1094C}
}

@ARTICLE{Rogers2002,
       author = {{Rogers}, F.~J. and {Nayfonov}, A.},
        title = "{Updated and Expanded OPAL Equation-of-State Tables: Implications for Helioseismology}",
      journal = {\apj},
         year = 2002,
        month = sep,
       volume = {576},
       number = {2},
        pages = {1064-1074},
          doi = {10.1086/341894},
       adsurl = {https://ui.adsabs.harvard.edu/abs/2002ApJ...576.1064R}
}

@ARTICLE{Saumon1995,
       author = {{Saumon}, D. and {Chabrier}, G. and {van Horn}, H.~M.},
        title = "{An Equation of State for Low-Mass Stars and Giant Planets}",
      journal = {\apjs},
         year = 1995,
        month = aug,
       volume = {99},
        pages = {713},
          doi = {10.1086/192204},
       adsurl = {https://ui.adsabs.harvard.edu/abs/1995ApJS...99..713S}
}

@ARTICLE{Badnell2005,
       author = {{Badnell}, N.~R. and {Bautista}, M.~A. and {Butler}, K. and {Delahaye}, F. and {Mendoza}, C. and {Palmeri}, P. and {Zeippen}, C.~J. and {Seaton}, M.~J.},
        title = "{Updated opacities from the Opacity Project}",
      journal = {\mnras},
         year = 2005,
        month = jun,
       volume = {360},
       number = {2},
        pages = {458-464},
          doi = {10.1111/j.1365-2966.2005.08991.x},
archivePrefix = {arXiv},
       eprint = {astro-ph/0410744},
 primaryClass = {astro-ph},
       adsurl = {https://ui.adsabs.harvard.edu/abs/2005MNRAS.360..458B}
}

@ARTICLE{BohmVitense1958,
       author = {{B{\"o}hm-Vitense}, E.},
        title = "{{\"U}ber die Wasserstoffkonvektionszone in Sternen verschiedener Effektivtemperaturen und Leuchtkr{\"a}fte. Mit 5 Textabbildungen}",
      journal = {\zap},
         year = 1958,
        month = jan,
       volume = {46},
        pages = {108},
       adsurl = {https://ui.adsabs.harvard.edu/abs/1958ZA.....46..108B}
}

@ARTICLE{Castellani1971,
       author = {{Castellani}, V. and {Giannone}, P. and {Renzini}, A.},
        title = "{Induced Semi-Convection in Helium-Burning Horizontal-Branch Stars II}",
      journal = {\apss},
         year = 1971,
        month = mar,
       volume = {10},
       number = {3},
        pages = {355-362},
          doi = {10.1007/BF00649680},
       adsurl = {https://ui.adsabs.harvard.edu/abs/1971Ap&SS..10..355C}
}

@INPROCEEDINGS{Kippenhahn1970,
       author = {{Kippenhahn}, R. and {Thomas}, H. -C.},
        title = "{A Simple Method for the Solution of the Stellar Structure Equations Including Rotation and Tidal Forces}",
    booktitle = {IAU Colloquium 4: Stellar Rotation},
         year = 1970,
       editor = {{Slettebak}, Arne},
        month = jan,
        pages = {20},
       adsurl = {https://ui.adsabs.harvard.edu/abs/1970stro.coll...20K}
}

@ARTICLE{Heger2000,
       author = {{Heger}, A. and {Langer}, N. and {Woosley}, S.~E.},
        title = "{Presupernova Evolution of Rotating Massive Stars. I. Numerical Method and Evolution of the Internal Stellar Structure}",
      journal = {\apj},
         year = 2000,
        month = jan,
       volume = {528},
       number = {1},
        pages = {368-396},
          doi = {10.1086/308158},
archivePrefix = {arXiv},
       eprint = {astro-ph/9904132},
 primaryClass = {astro-ph},
       adsurl = {https://ui.adsabs.harvard.edu/abs/2000ApJ...528..368H}
}

@ARTICLE{Endal1976,
       author = {{Endal}, A.~S. and {Sofia}, S.},
        title = "{The evolution of rotating stars. I. Method and exploratory calculations for a 7 M sun star.}",
      journal = {\apj},
         year = 1976,
        month = nov,
       volume = {210},
        pages = {184-198},
          doi = {10.1086/154817},
       adsurl = {https://ui.adsabs.harvard.edu/abs/1976ApJ...210..184E}
}

@ARTICLE{Woo2001,
       author = {{Woo}, Jong-Hak and {Demarque}, Pierre},
        title = "{Empirical Constraints on Convective Core Overshoot}",
      journal = {\aj},
         year = 2001,
        month = sep,
       volume = {122},
       number = {3},
        pages = {1602-1606},
          doi = {10.1086/322122},
archivePrefix = {arXiv},
       eprint = {astro-ph/0106061},
 primaryClass = {astro-ph},
       adsurl = {https://ui.adsabs.harvard.edu/abs/2001AJ....122.1602W}
}

@ARTICLE{Matt2012,
       author = {{Matt}, Sean P. and {MacGregor}, Keith B. and {Pinsonneault}, Marc H. and {Greene}, Thomas P.},
        title = "{Magnetic Braking Formulation for Sun-like Stars: Dependence on Dipole Field Strength and Rotation Rate}",
      journal = {\apjl},
         year = 2012,
        month = aug,
       volume = {754},
       number = {2},
          eid = {L26},
        pages = {L26},
          doi = {10.1088/2041-8205/754/2/L26},
archivePrefix = {arXiv},
       eprint = {1206.2354},
 primaryClass = {astro-ph.SR},
       adsurl = {https://ui.adsabs.harvard.edu/abs/2012ApJ...754L..26M}
}

@ARTICLE{Matt2008,
       author = {{Matt}, Sean and {Pudritz}, Ralph E.},
        title = "{Accretion-powered Stellar Winds. II. Numerical Solutions for Stellar Wind Torques}",
      journal = {\apj},
         year = 2008,
        month = may,
       volume = {678},
       number = {2},
        pages = {1109-1118},
          doi = {10.1086/533428},
archivePrefix = {arXiv},
       eprint = {0801.0436},
 primaryClass = {astro-ph},
       adsurl = {https://ui.adsabs.harvard.edu/abs/2008ApJ...678.1109M}
}

@ARTICLE{Kawaler1988,
       author = {{Kawaler}, Steven D.},
        title = "{Angular Momentum Loss in Low-Mass Stars}",
      journal = {\apj},
         year = 1988,
        month = oct,
       volume = {333},
        pages = {236},
          doi = {10.1086/166740},
       adsurl = {https://ui.adsabs.harvard.edu/abs/1988ApJ...333..236K}
}

@ARTICLE{Wood2005,
       author = {{Wood}, B.~E. and {M{\"u}ller}, H. -R. and {Zank}, G.~P. and {Linsky}, J.~L. and {Redfield}, S.},
        title = "{New Mass-Loss Measurements from Astrospheric Ly{\ensuremath{\alpha}} Absorption}",
      journal = {\apjl},
         year = 2005,
        month = aug,
       volume = {628},
       number = {2},
        pages = {L143-L146},
          doi = {10.1086/432716},
archivePrefix = {arXiv},
       eprint = {astro-ph/0506401},
 primaryClass = {astro-ph},
       adsurl = {https://ui.adsabs.harvard.edu/abs/2005ApJ...628L.143W}
}

@ARTICLE{Pizzolato2003,
       author = {{Pizzolato}, N. and {Maggio}, A. and {Micela}, G. and {Sciortino}, S. and {Ventura}, P.},
        title = "{The stellar activity-rotation relationship revisited: Dependence of saturated and non-saturated X-ray emission regimes on stellar mass for late-type dwarfs}",
      journal = {\aap},
         year = 2003,
        month = jan,
       volume = {397},
        pages = {147-157},
          doi = {10.1051/0004-6361:20021560},
       adsurl = {https://ui.adsabs.harvard.edu/abs/2003A&A...397..147P}
}

@ARTICLE{Zahn1992,
       author = {{Zahn}, J. -P.},
        title = "{Circulation and turbulence in rotating stars.}",
      journal = {\aap},
         year = 1992,
        month = nov,
       volume = {265},
        pages = {115-132},
       adsurl = {https://ui.adsabs.harvard.edu/abs/1992A&A...265..115Z}
}

@ARTICLE{Maeder1997,
       author = {{Maeder}, A.},
        title = "{Stellar evolution with rotation. II. A new approach for shear mixing.}",
      journal = {\aap},
         year = 1997,
        month = may,
       volume = {321},
        pages = {134-144},
       adsurl = {https://ui.adsabs.harvard.edu/abs/1997A&A...321..134M}
}

@ARTICLE{Maeder1998,
       author = {{Maeder}, Andre and {Zahn}, Jean-Paul},
        title = "{Stellar evolution with rotation. III. Meridional circulation with MU -gradients and non-stationarity}",
      journal = {\aap},
         year = 1998,
        month = jun,
       volume = {334},
        pages = {1000-1006},
       adsurl = {https://ui.adsabs.harvard.edu/abs/1998A&A...334.1000M}
}

@ARTICLE{Chaboyer1992,
       author = {{Chaboyer}, B. and {Zahn}, J. -P.},
        title = "{Effect of horizontal turbulent diffusion on transport by meridional circulation.}",
      journal = {\aap},
         year = 1992,
        month = jan,
       volume = {253},
        pages = {173-177},
       adsurl = {https://ui.adsabs.harvard.edu/abs/1992A&A...253..173C}
}

@ARTICLE{Jermyn2023,
       author = {{Jermyn}, Adam S. and {Bauer}, Evan B. and {Schwab}, Josiah and {Farmer}, R. and {Ball}, Warrick H. and {Bellinger}, Earl P. and {Dotter}, Aaron and {Joyce}, Meridith and {Marchant}, Pablo and {Mombarg}, Joey S.~G. and {Wolf}, William M. and {Sunny Wong}, Tin Long and {Cinquegrana}, Giulia C. and {Farrell}, Eoin and {Smolec}, R. and {Thoul}, Anne and {Cantiello}, Matteo and {Herwig}, Falk and {Toloza}, Odette and {Bildsten}, Lars and {Townsend}, Richard H.~D. and {Timmes}, F.~X.},
        title = "{Modules for Experiments in Stellar Astrophysics (MESA): Time-dependent Convection, Energy Conservation, Automatic Differentiation, and Infrastructure}",
      journal = {\apjs},
         year = 2023,
        month = mar,
       volume = {265},
       number = {1},
          eid = {15},
        pages = {15},
          doi = {10.3847/1538-4365/acae8d},
archivePrefix = {arXiv},
       eprint = {2208.03651},
 primaryClass = {astro-ph.SR},
       adsurl = {https://ui.adsabs.harvard.edu/abs/2023ApJS..265...15J}
}

@ARTICLE{Somers2017,
       author = {{Somers}, Garrett and {Stauffer}, John and {Rebull}, Luisa and {Cody}, Ann Marie and {Pinsonneault}, Marc},
        title = "{M Dwarf Rotation from the K2 Young Clusters to the Field. I. A Mass-Rotation Correlation at 10 Myr}",
      journal = {\apj},
         year = 2017,
        month = dec,
       volume = {850},
       number = {2},
          eid = {134},
        pages = {134},
          doi = {10.3847/1538-4357/aa93ed},
archivePrefix = {arXiv},
       eprint = {1710.07638},
 primaryClass = {astro-ph.SR},
       adsurl = {https://ui.adsabs.harvard.edu/abs/2017ApJ...850..134S}
}

@ARTICLE{Curtis2020,
       author = {{Curtis}, Jason Lee and {Ag{\"u}eros}, Marcel A. and {Matt}, Sean P. and {Covey}, Kevin R. and {Douglas}, Stephanie T. and {Angus}, Ruth and {Saar}, Steven H. and {Cody}, Ann Marie and {Vanderburg}, Andrew and {Law}, Nicholas M. and {Kraus}, Adam L. and {Latham}, David W. and {Baranec}, Christoph and {Riddle}, Reed and {Ziegler}, Carl and {Lund}, Mikkel N. and {Torres}, Guillermo and {Meibom}, S{\o}ren and {Aguirre}, Victor Silva and {Wright}, Jason T.},
        title = "{When Do Stalled Stars Resume Spinning Down? Advancing Gyrochronology with Ruprecht 147}",
      journal = {\apj},
         year = 2020,
        month = dec,
       volume = {904},
       number = {2},
          eid = {140},
        pages = {140},
          doi = {10.3847/1538-4357/abbf58},
archivePrefix = {arXiv},
       eprint = {2010.02272},
 primaryClass = {astro-ph.SR},
       adsurl = {https://ui.adsabs.harvard.edu/abs/2020ApJ...904..140C}
}

@ARTICLE{2012Bressan,
       author = {{Bressan}, Alessandro and {Marigo}, Paola and {Girardi}, L{\'e}o. and {Salasnich}, Bernardo and {Dal Cero}, Claudia and {Rubele}, Stefano and {Nanni}, Ambra},
        title = "{PARSEC: stellar tracks and isochrones with the PAdova and TRieste Stellar Evolution Code}",
      journal = {\mnras},
         year = 2012,
        month = nov,
       volume = {427},
       number = {1},
        pages = {127-145},
          doi = {10.1111/j.1365-2966.2012.21948.x},
archivePrefix = {arXiv},
       eprint = {1208.4498},
 primaryClass = {astro-ph.SR},
       adsurl = {https://ui.adsabs.harvard.edu/abs/2012MNRAS.427..127B}
}

@ARTICLE{2014Chen,
       author = {{Chen}, Yang and {Girardi}, L{\'e}o and {Bressan}, Alessandro and {Marigo}, Paola and {Barbieri}, Mauro and {Kong}, Xu},
        title = "{Improving PARSEC models for very low mass stars}",
      journal = {\mnras},
         year = 2014,
        month = nov,
       volume = {444},
       number = {3},
        pages = {2525-2543},
          doi = {10.1093/mnras/stu1605},
archivePrefix = {arXiv},
       eprint = {1409.0322},
 primaryClass = {astro-ph.SR},
       adsurl = {https://ui.adsabs.harvard.edu/abs/2014MNRAS.444.2525C}
}

@ARTICLE{2016Choi,
       author = {{Choi}, Jieun and {Dotter}, Aaron and {Conroy}, Charlie and {Cantiello}, Matteo and {Paxton}, Bill and {Johnson}, Benjamin D.},
        title = "{Mesa Isochrones and Stellar Tracks (MIST). I. Solar-scaled Models}",
      journal = {\apj},
         year = 2016,
        month = jun,
       volume = {823},
       number = {2},
          eid = {102},
        pages = {102},
          doi = {10.3847/0004-637X/823/2/102},
archivePrefix = {arXiv},
       eprint = {1604.08592},
 primaryClass = {astro-ph.SR},
       adsurl = {https://ui.adsabs.harvard.edu/abs/2016ApJ...823..102C}
}

@ARTICLE{2022Nguyen,
       author = {{Nguyen}, C.~T. and {Costa}, G. and {Girardi}, L. and {Volpato}, G. and {Bressan}, A. and {Chen}, Y. and {Marigo}, P. and {Fu}, X. and {Goudfrooij}, P.},
        title = "{PARSEC V2.0: Stellar tracks and isochrones of low- and intermediate-mass stars with rotation}",
      journal = {\aap},
         year = 2022,
        month = sep,
       volume = {665},
          eid = {A126},
        pages = {A126},
          doi = {10.1051/0004-6361/202244166},
archivePrefix = {arXiv},
       eprint = {2207.08642},
 primaryClass = {astro-ph.SR},
       adsurl = {https://ui.adsabs.harvard.edu/abs/2022A&A...665A.126N}
}

@ARTICLE{2025Costa,
       author = {{Costa}, G. and {Shepherd}, K.~G. and {Bressan}, A. and {Addari}, F. and {Chen}, Y. and {Fu}, X. and {Volpato}, G. and {Nguyen}, C.~T. and {Girardi}, L. and {Marigo}, P. and {Mazzi}, A. and {Pastorelli}, G. and {Trabucchi}, M. and {Bossini}, D. and {Zaggia}, S.},
        title = "{Evolutionary tracks, ejecta, and ionizing photons from intermediate-mass to very massive stars with PARSEC}",
      journal = {\aap},
         year = 2025,
        month = feb,
       volume = {694},
          eid = {A193},
        pages = {A193},
          doi = {10.1051/0004-6361/202452573},
archivePrefix = {arXiv},
       eprint = {2501.12917},
 primaryClass = {astro-ph.SR},
       adsurl = {https://ui.adsabs.harvard.edu/abs/2025A&A...694A.193C}
}

@ARTICLE{2015Chen,
       author = {{Chen}, Yang and {Bressan}, Alessandro and {Girardi}, L{\'e}o and {Marigo}, Paola and {Kong}, Xu and {Lanza}, Antonio},
        title = "{PARSEC evolutionary tracks of massive stars up to 350 M$_{{\ensuremath{\odot}}}$ at metallicities 0.0001 {\ensuremath{\leq}} Z {\ensuremath{\leq}} 0.04}",
      journal = {\mnras},
         year = 2015,
        month = sep,
       volume = {452},
       number = {1},
        pages = {1068-1080},
          doi = {10.1093/mnras/stv1281},
archivePrefix = {arXiv},
       eprint = {1506.01681},
 primaryClass = {astro-ph.SR},
       adsurl = {https://ui.adsabs.harvard.edu/abs/2015MNRAS.452.1068C}
}

@ARTICLE{2018Fu,
       author = {{Fu}, Xiaoting and {Bressan}, Alessandro and {Marigo}, Paola and {Girardi}, L{\'e}o and {Montalb{\'a}n}, Josefina and {Chen}, Yang and {Nanni}, Ambra},
        title = "{New PARSEC data base of {\ensuremath{\alpha}}-enhanced stellar evolutionary tracks and isochrones - I. Calibration with 47 Tuc (NGC 104) and the improvement on RGB bump}",
      journal = {\mnras},
         year = 2018,
        month = may,
       volume = {476},
       number = {1},
        pages = {496-511},
          doi = {10.1093/mnras/sty235},
archivePrefix = {arXiv},
       eprint = {1801.07137},
 primaryClass = {astro-ph.SR},
       adsurl = {https://ui.adsabs.harvard.edu/abs/2018MNRAS.476..496F}
}

@ARTICLE{Salpeter1954,
   author = {{Salpeter}, E.~E.},
    title = "{Electrons Screening and Thermonuclear Reactions}",
  journal = {Australian Journal of Physics},
     year = 1954,
    month = sep,
   volume = 7,
    pages = {373},
      doi = {10.1071/PH540373},
   adsurl = {http://adsabs.harvard.edu/abs/1954AuJPh...7..373S}
}

@ARTICLE{Gruner2023,
       author = {{Gruner}, D. and {Barnes}, S.~A. and {Weingrill}, J.},
        title = "{New insights into the rotational evolution of near-solar age stars from the open cluster M 67}",
      journal = {\aap},
         year = 2023,
        month = apr,
       volume = {672},
          eid = {A159},
        pages = {A159},
          doi = {10.1051/0004-6361/202345942},
archivePrefix = {arXiv},
       eprint = {2305.16997},
 primaryClass = {astro-ph.SR},
       adsurl = {https://ui.adsabs.harvard.edu/abs/2023A&A...672A.159G}
}

@ARTICLE{Manchon25,
       author = {{Manchon}, L. and {Deal}, M. and {Marques}, J.~P.~C. and {Lebreton}, Y.},
        title = "{Cesam2k20: A code for a new generation of stellar evolution models. I. Description of the code}",
      journal = {arXiv e-prints},
         year = 2025,
        month = nov,
          eid = {arXiv:2511.02801},
        pages = {arXiv:2511.02801},
          doi = {10.48550/arXiv.2511.02801},
archivePrefix = {arXiv},
       eprint = {2511.02801},
 primaryClass = {astro-ph.SR},
       adsurl = {https://ui.adsabs.harvard.edu/abs/2025arXiv251102801M}
}

@ARTICLE{Morel97,
       author = {{Morel}, P.},
        title = "{CESAM: A code for stellar evolution calculations}",
      journal = {\aaps},
         year = 1997,
        month = sep,
       volume = {124},
        pages = {597-614},
          doi = {10.1051/aas:1997209},
       adsurl = {https://ui.adsabs.harvard.edu/abs/1997A&AS..124..597M}
}

@ARTICLE{Morel08,
       author = {{Morel}, P. and {Lebreton}, Y.},
        title = "{CESAM: a free code for stellar evolution calculations}",
      journal = {\apss},
         year = 2008,
        month = aug,
       volume = {316},
       number = {1-4},
        pages = {61-73},
          doi = {10.1007/s10509-007-9663-9},
archivePrefix = {arXiv},
       eprint = {0801.2019},
 primaryClass = {astro-ph},
       adsurl = {https://ui.adsabs.harvard.edu/abs/2008Ap&SS.316...61M}
}

@ARTICLE{Paxton2018,
       author = {{Paxton}, Bill and {Schwab}, Josiah and {Bauer}, Evan B. and {Bildsten}, Lars and {Blinnikov}, Sergei and {Duffell}, Paul and {Farmer}, R. and {Goldberg}, Jared A. and {Marchant}, Pablo and {Sorokina}, Elena and {Thoul}, Anne and {Townsend}, Richard H.~D. and {Timmes}, F.~X.},
        title = "{Modules for Experiments in Stellar Astrophysics (MESA): Convective Boundaries, Element Diffusion, and Massive Star Explosions}",
      journal = {\apjs},
         year = 2018,
        month = feb,
       volume = {234},
       number = {2},
          eid = {34},
        pages = {34},
          doi = {10.3847/1538-4365/aaa5a8},
archivePrefix = {arXiv},
       eprint = {1710.08424},
 primaryClass = {astro-ph.SR},
       adsurl = {https://ui.adsabs.harvard.edu/abs/2018ApJS..234...34P}
}

@ARTICLE{Paxton2019,
       author = {{Paxton}, Bill and {Smolec}, R. and {Schwab}, Josiah and {Gautschy}, A. and {Bildsten}, Lars and {Cantiello}, Matteo and {Dotter}, Aaron and {Farmer}, R. and {Goldberg}, Jared A. and {Jermyn}, Adam S. and {Kanbur}, S.~M. and {Marchant}, Pablo and {Thoul}, Anne and {Townsend}, Richard H.~D. and {Wolf}, William M. and {Zhang}, Michael and {Timmes}, F.~X.},
        title = "{Modules for Experiments in Stellar Astrophysics (MESA): Pulsating Variable Stars, Rotation, Convective Boundaries, and Energy Conservation}",
      journal = {\apjs},
         year = 2019,
        month = jul,
       volume = {243},
       number = {1},
          eid = {10},
        pages = {10},
          doi = {10.3847/1538-4365/ab2241},
archivePrefix = {arXiv},
       eprint = {1903.01426},
 primaryClass = {astro-ph.SR},
       adsurl = {https://ui.adsabs.harvard.edu/abs/2019ApJS..243...10P}
}

@ARTICLE{White2025,
       author = {{White}, Russel and {Pratt}, Jane and {Rieutord}, Michel},
        title = "{An Educational Guide for 2D Stellar Structure Calculations of Rapidly Rotating Stars using the ESTER code}",
      journal = {arXiv e-prints},
         year = 2025,
        month = sep,
          eid = {arXiv:2509.20264},
        pages = {arXiv:2509.20264},
          doi = {10.48550/arXiv.2509.20264},
archivePrefix = {arXiv},
       eprint = {2509.20264},
 primaryClass = {astro-ph.SR},
       adsurl = {https://ui.adsabs.harvard.edu/abs/2025arXiv250920264W}
}

@ARTICLE{Basinger2024,
       author = {{Basinger}, Connor and {Pinsonneault}, Marc and {Bastelberger}, Sandra T. and {Gaudi}, B. Scott and {Domagal-Goldman}, Shawn D.},
        title = "{Constraints on the early luminosity history of the Sun: applications to the Faint Young Sun problem}",
      journal = {\mnras},
         year = 2024,
        month = nov,
       volume = {534},
       number = {3},
        pages = {2968-2985},
          doi = {10.1093/mnras/stae2280},
archivePrefix = {arXiv},
       eprint = {2409.03823},
 primaryClass = {astro-ph.SR},
       adsurl = {https://ui.adsabs.harvard.edu/abs/2024MNRAS.534.2968B}
}

@ARTICLE{Iglesias1993,
       author = {{Iglesias}, Carlos A. and {Rogers}, Forrest J.},
        title = "{Radiative Opacities for Carbon- and Oxygen-rich Mixtures}",
      journal = {\apj},
         year = 1993,
        month = aug,
       volume = {412},
        pages = {752},
          doi = {10.1086/172958},
       adsurl = {https://ui.adsabs.harvard.edu/abs/1993ApJ...412..752I}
}

@ARTICLE{Konigl1991,
       author = {{Koenigl}, Arieh},
        title = "{Disk Accretion onto Magnetic T Tauri Stars}",
      journal = {\apjl},
         year = 1991,
        month = mar,
       volume = {370},
        pages = {L39},
          doi = {10.1086/185972},
       adsurl = {https://ui.adsabs.harvard.edu/abs/1991ApJ...370L..39K}
}

@ARTICLE{Shu1994,
       author = {{Shu}, Frank and {Najita}, Joan and {Ostriker}, Eve and {Wilkin}, Frank and {Ruden}, Steven and {Lizano}, Susana},
        title = "{Magnetocentrifugally Driven Flows from Young Stars and Disks. I. A Generalized Model}",
      journal = {\apj},
         year = 1994,
        month = jul,
       volume = {429},
        pages = {781},
          doi = {10.1086/174363},
       adsurl = {https://ui.adsabs.harvard.edu/abs/1994ApJ...429..781S}
}

@ARTICLE{Palla1991,
       author = {{Palla}, Francesco and {Stahler}, Steven W.},
        title = "{The Evolution of Intermediate-Mass Protostars. I. Basic Results}",
      journal = {\apj},
         year = 1991,
        month = jul,
       volume = {375},
        pages = {288},
          doi = {10.1086/170188},
       adsurl = {https://ui.adsabs.harvard.edu/abs/1991ApJ...375..288P}
}

@ARTICLE{dottermist,
       author = {{Dotter}, Aaron},
        title = "{MESA Isochrones and Stellar Tracks (MIST) 0: Methods for the Construction of Stellar Isochrones}",
      journal = {\apjs},
         year = 2016,
        month = jan,
       volume = {222},
       number = {1},
          eid = {8},
        pages = {8},
          doi = {10.3847/0067-0049/222/1/8},
archivePrefix = {arXiv},
       eprint = {1601.05144},
 primaryClass = {astro-ph.SR},
       adsurl = {https://ui.adsabs.harvard.edu/abs/2016ApJS..222....8D}
}

@ARTICLE{kiauhokupaper,
       author = {{Claytor}, Zachary R. and {van Saders}, Jennifer L. and {Santos}, {\^A}ngela R.~G. and {Garc{\'\i}a}, Rafael A. and {Mathur}, Savita and {Tayar}, Jamie and {Pinsonneault}, Marc H. and {Shetrone}, Matthew},
        title = "{Chemical Evolution in the Milky Way: Rotation-based Ages for APOGEE-Kepler Cool Dwarf Stars}",
      journal = {\apj},
         year = 2020,
        month = jan,
       volume = {888},
       number = {1},
          eid = {43},
        pages = {43},
          doi = {10.3847/1538-4357/ab5c24},
archivePrefix = {arXiv},
       eprint = {1911.04518},
 primaryClass = {astro-ph.SR},
       adsurl = {https://ui.adsabs.harvard.edu/abs/2020ApJ...888...43C}
}

@software{kiauhoku_ascl,
       author = {{Claytor}, Zachary R. and {van Saders}, Jennifer L. and {Santos}, {\^A}ngela R.~G. and {Garc{\'\i}a}, Rafael A. and {Mathur}, Savita and {Tayar}, Jamie and {Pinsonneault}, Marc H. and {Shetrone}, Matthew},
        title = "{kiauhoku: Stellar model grid interpolation}",
 howpublished = {Astrophysics Source Code Library, record ascl:2011.027},
         year = 2020,
        month = nov,
          eid = {ascl:2011.027},
archivePrefix = {ascl},
       eprint = {2011.027},
       adsurl = {https://ui.adsabs.harvard.edu/abs/2020ascl.soft11027C}
}

@ARTICLE{pleiadesage,
       author = {{Gossage}, Seth and {Conroy}, Charlie and {Dotter}, Aaron and {Choi}, Jieun and {Rosenfield}, Philip and {Cargile}, Philip and {Dolphin}, Andrew},
        title = "{Age Determinations of the Hyades, Praesepe, and Pleiades via MESA Models with Rotation}",
      journal = {\apj},
         year = 2018,
        month = aug,
       volume = {863},
       number = {1},
          eid = {67},
        pages = {67},
          doi = {10.3847/1538-4357/aad0a0},
archivePrefix = {arXiv},
       eprint = {1804.06441},
 primaryClass = {astro-ph.SR},
       adsurl = {https://ui.adsabs.harvard.edu/abs/2018ApJ...863...67G}
}

@ARTICLE{Pinsonneault1991,
       author = {{Pinsonneault}, Marc H. and {Deliyannis}, Constantine P. and {Demarque}, Pierre},
        title = "{Evolutionary Models of Halo Stars with Rotation. I. Evidence for Differential Rotation with Depth in Stars}",
      journal = {\apj},
         year = 1991,
        month = jan,
       volume = {367},
        pages = {239},
          doi = {10.1086/169623},
       adsurl = {https://ui.adsabs.harvard.edu/abs/1991ApJ...367..239P}
}

@ARTICLE{Sills2000,
       author = {{Sills}, Alison and {Pinsonneault}, M.~H.},
        title = "{Rotation of Horizontal-Branch Stars in Globular Clusters}",
      journal = {\apj},
         year = 2000,
        month = sep,
       volume = {540},
       number = {1},
        pages = {489-503},
          doi = {10.1086/309306},
archivePrefix = {arXiv},
       eprint = {astro-ph/9911024},
 primaryClass = {astro-ph},
       adsurl = {https://ui.adsabs.harvard.edu/abs/2000ApJ...540..489S}
}

@ARTICLE{Asplund2021,
       author = {{Asplund}, M. and {Amarsi}, A.~M. and {Grevesse}, N.},
        title = "{The chemical make-up of the Sun: A 2020 vision}",
      journal = {\aap},
         year = 2021,
        month = sep,
       volume = {653},
          eid = {A141},
        pages = {A141},
          doi = {10.1051/0004-6361/202140445},
archivePrefix = {arXiv},
       eprint = {2105.01661},
 primaryClass = {astro-ph.SR},
       adsurl = {https://ui.adsabs.harvard.edu/abs/2021A&A...653A.141A}
}

@ARTICLE{Magg2022,
       author = {{Magg}, Ekaterina and {Bergemann}, Maria and {Serenelli}, Aldo and {Bautista}, Manuel and {Plez}, Bertrand and {Heiter}, Ulrike and {Gerber}, Jeffrey M. and {Ludwig}, Hans-G{\"u}nter and {Basu}, Sarbani and {Ferguson}, Jason W. and {Gallego}, Helena Carvajal and {Gamrath}, S{\'e}bastien and {Palmeri}, Patrick and {Quinet}, Pascal},
        title = "{Observational constraints on the origin of the elements. IV. Standard composition of the Sun}",
      journal = {\aap},
         year = 2022,
        month = may,
       volume = {661},
          eid = {A140},
        pages = {A140},
          doi = {10.1051/0004-6361/202142971},
archivePrefix = {arXiv},
       eprint = {2203.02255},
 primaryClass = {astro-ph.SR},
       adsurl = {https://ui.adsabs.harvard.edu/abs/2022A&A...661A.140M}
}

@ARTICLE{Lodders2025,
       author = {{Lodders}, K. and {Bergemann}, M. and {Palme}, H.},
        title = "{Solar System Elemental Abundances from the Solar Photosphere and CI-Chondrites}",
      journal = {\ssr},
         year = 2025,
        month = mar,
       volume = {221},
       number = {2},
          eid = {23},
        pages = {23},
          doi = {10.1007/s11214-025-01146-w},
archivePrefix = {arXiv},
       eprint = {2502.10575},
 primaryClass = {astro-ph.SR},
       adsurl = {https://ui.adsabs.harvard.edu/abs/2025SSRv..221...23L}
}

@ARTICLE{Lodders2021,
       author = {{Lodders}, Katharina},
        title = "{Relative Atomic Solar System Abundances, Mass Fractions, and Atomic Masses of the Elements and Their Isotopes, Composition of the Solar Photosphere, and Compositions of the Major Chondritic Meteorite Groups}",
      journal = {\ssr},
         year = 2021,
        month = apr,
       volume = {217},
       number = {3},
          eid = {44},
        pages = {44},
          doi = {10.1007/s11214-021-00825-8},
       adsurl = {https://ui.adsabs.harvard.edu/abs/2021SSRv..217...44L}
}

@ARTICLE{Allard1995,
       author = {{Allard}, France and {Hauschildt}, Peter H.},
        title = "{Model Atmospheres for M (Sub)Dwarf Stars. I. The Base Model Grid}",
      journal = {\apj},
         year = 1995,
        month = may,
       volume = {445},
        pages = {433},
          doi = {10.1086/175708},
archivePrefix = {arXiv},
       eprint = {astro-ph/9601150},
 primaryClass = {astro-ph},
       adsurl = {https://ui.adsabs.harvard.edu/abs/1995ApJ...445..433A}
}

@ARTICLE{Mosser2012rot,
       author = {{Mosser}, B. and {Goupil}, M.~J. and {Belkacem}, K. and {Marques}, J.~P. and {Beck}, P.~G. and {Bloemen}, S. and {De Ridder}, J. and {Barban}, C. and {Deheuvels}, S. and {Elsworth}, Y. and {Hekker}, S. and {Kallinger}, T. and {Ouazzani}, R.~M. and {Pinsonneault}, M. and {Samadi}, R. and {Stello}, D. and {Garc{\'\i}a}, R.~A. and {Klaus}, T.~C. and {Li}, J. and {Mathur}, S. and {Morris}, R.~L.},
        title = "{Spin down of the core rotation in red giants}",
      journal = {\aap},
         year = 2012,
        month = dec,
       volume = {548},
          eid = {A10},
        pages = {A10},
          doi = {10.1051/0004-6361/201220106},
archivePrefix = {arXiv},
       eprint = {1209.3336},
 primaryClass = {astro-ph.SR},
       adsurl = {https://ui.adsabs.harvard.edu/abs/2012A&A...548A..10M}
}

@ARTICLE{Tayar2013,
       author = {{Tayar}, Jamie and {Pinsonneault}, Marc H.},
        title = "{Implications of Rapid Core Rotation in Red Giants for Internal Angular Momentum Transport in Stars}",
      journal = {\apjl},
         year = 2013,
        month = sep,
       volume = {775},
       number = {1},
          eid = {L1},
        pages = {L1},
          doi = {10.1088/2041-8205/775/1/L1},
archivePrefix = {arXiv},
       eprint = {1306.3986},
 primaryClass = {astro-ph.SR},
       adsurl = {https://ui.adsabs.harvard.edu/abs/2013ApJ...775L...1T}
}

@ARTICLE{Tayar2018,
       author = {{Tayar}, Jamie and {Pinsonneault}, Marc H.},
        title = "{Testing Angular Momentum Transport and Wind Loss in Intermediate-mass Core-helium Burning Stars}",
      journal = {\apj},
         year = 2018,
        month = dec,
       volume = {868},
       number = {2},
          eid = {150},
        pages = {150},
          doi = {10.3847/1538-4357/aae979},
archivePrefix = {arXiv},
       eprint = {1806.10649},
 primaryClass = {astro-ph.SR},
       adsurl = {https://ui.adsabs.harvard.edu/abs/2018ApJ...868..150T}
}

@ARTICLE{Kissin2015,
       author = {{Kissin}, Yevgeni and {Thompson}, Christopher},
        title = "{Rotation of Giant Stars}",
      journal = {\apj},
         year = 2015,
        month = jul,
       volume = {808},
       number = {1},
          eid = {35},
        pages = {35},
          doi = {10.1088/0004-637X/808/1/35},
archivePrefix = {arXiv},
       eprint = {1501.07217},
 primaryClass = {astro-ph.SR},
       adsurl = {https://ui.adsabs.harvard.edu/abs/2015ApJ...808...35K}
}

@ARTICLE{Fuller2019,
       author = {{Fuller}, Jim and {Piro}, Anthony L. and {Jermyn}, Adam S.},
        title = "{Slowing the spins of stellar cores}",
      journal = {\mnras},
         year = 2019,
        month = may,
       volume = {485},
       number = {3},
        pages = {3661-3680},
          doi = {10.1093/mnras/stz514},
archivePrefix = {arXiv},
       eprint = {1902.08227},
 primaryClass = {astro-ph.SR},
       adsurl = {https://ui.adsabs.harvard.edu/abs/2019MNRAS.485.3661F}
}

@ARTICLE{Li2025,
       author = {{Li}, Yaguang},
        title = "{Evidence that Mass Loss on the Red Giant Branch Decreases with Metallicity}",
      journal = {\apj},
         year = 2025,
        month = aug,
       volume = {988},
       number = {2},
          eid = {179},
        pages = {179},
          doi = {10.3847/1538-4357/ade3c7},
archivePrefix = {arXiv},
       eprint = {2505.12794},
 primaryClass = {astro-ph.SR},
       adsurl = {https://ui.adsabs.harvard.edu/abs/2025ApJ...988..179L}
}

@ARTICLE{Eggenberger2019,
       author = {{Eggenberger}, P. and {Deheuvels}, S. and {Miglio}, A. and {Ekstr{\"o}m}, S. and {Georgy}, C. and {Meynet}, G. and {Lagarde}, N. and {Salmon}, S. and {Buldgen}, G. and {Montalb{\'a}n}, J. and {Spada}, F. and {Ballot}, J.},
        title = "{Asteroseismology of evolved stars to constrain the internal transport of angular momentum. I. Efficiency of transport during the subgiant phase}",
      journal = {\aap},
         year = 2019,
        month = jan,
       volume = {621},
          eid = {A66},
        pages = {A66},
          doi = {10.1051/0004-6361/201833447},
archivePrefix = {arXiv},
       eprint = {1812.04995},
 primaryClass = {astro-ph.SR},
       adsurl = {https://ui.adsabs.harvard.edu/abs/2019A&A...621A..66E}
}

@ARTICLE{pinsonneault2025a,
       author = {{Pinsonneault}, Marc H. and {Zinn}, Joel C. and {Tayar}, Jamie and {Serenelli}, Aldo and {Garc{\'\i}a}, Rafael A. and {Mathur}, Savita and {Vrard}, Mathieu and {Elsworth}, Yvonne P. and {Mosser}, Benoit and {Stello}, Dennis and {Bell}, Keaton J. and {Bugnet}, Lisa and {Corsaro}, Enrico and {Gaulme}, Patrick and {Hekker}, Saskia and {Hon}, Marc and {Huber}, Daniel and {Kallinger}, Thomas and {Cao}, Kaili and {Johnson}, Jennifer A. and {Liagre}, Bastien and {Patton}, Rachel A. and {Santos}, {\^A}ngela R.~G. and {Basu}, Sarbani and {Beck}, Paul G. and {Beers}, Timothy C. and {Chaplin}, William J. and {Cunha}, Katia and {Frinchaboy}, Peter M. and {Girardi}, L{\'e}o and {Godoy-Rivera}, Diego and {Holtzman}, Jon A. and {J{\"o}nsson}, Henrik and {M{\'e}sz{\'a}ros}, Szabolcs and {Reyes}, Claudia and {Rix}, Hans-Walter and {Shetrone}, Matthew and {Smith}, Verne V. and {Spoo}, Taylor and {Stassun}, Keivan G. and {Wang}, Ji},
        title = "{APOKASC-3: The Third Joint Spectroscopic and Asteroseismic Catalog for Evolved Stars in the Kepler Fields}",
      journal = {\apjs},
         year = 2025,
        month = feb,
       volume = {276},
       number = {2},
          eid = {69},
        pages = {69},
          doi = {10.3847/1538-4365/ad9fef},
archivePrefix = {arXiv},
       eprint = {2410.00102},
 primaryClass = {astro-ph.SR},
       adsurl = {https://ui.adsabs.harvard.edu/abs/2025ApJS..276...69P}
}

@ARTICLE{cole1985,
       author = {{Cole}, P.~W. and {Demarque}, P. and {Deupree}, R.~G.},
        title = "{Convective heating of the inner core of red giants prior to the peak of the core helium flash}",
      journal = {\apj},
         year = 1985,
        month = apr,
       volume = {291},
        pages = {291-296},
          doi = {10.1086/163067},
       adsurl = {https://ui.adsabs.harvard.edu/abs/1985ApJ...291..291C}
}

@ARTICLE{2010ApJ...716.1269D,
       author = {{Denissenkov}, Pavel A. and {Pinsonneault}, Marc and {Terndrup}, Donald M. and {Newsham}, Grant},
        title = "{Angular Momentum Transport in Solar-type Stars: Testing the Timescale for Core-Envelope Coupling}",
      journal = {\apj},
         year = 2010,
        month = jun,
       volume = {716},
       number = {2},
        pages = {1269-1287},
          doi = {10.1088/0004-637X/716/2/1269},
archivePrefix = {arXiv},
       eprint = {0911.1121},
 primaryClass = {astro-ph.SR},
       adsurl = {https://ui.adsabs.harvard.edu/abs/2010ApJ...716.1269D}
}

@ARTICLE{Yorke2002,
       author = {{Yorke}, Harold W. and {Sonnhalter}, Cordula},
        title = "{On the Formation of Massive Stars}",
      journal = {\apj},
         year = 2002,
        month = apr,
       volume = {569},
       number = {2},
        pages = {846-862},
          doi = {10.1086/339264},
archivePrefix = {arXiv},
       eprint = {astro-ph/0201041},
 primaryClass = {astro-ph},
       adsurl = {https://ui.adsabs.harvard.edu/abs/2002ApJ...569..846Y}
}

@ARTICLE{Bildsten2012,
       author = {{Bildsten}, Lars and {Paxton}, Bill and {Moore}, Kevin and {Macias}, Phillip J.},
        title = "{Acoustic Signatures of the Helium Core Flash}",
      journal = {\apjl},
         year = 2012,
        month = jan,
       volume = {744},
       number = {1},
          eid = {L6},
        pages = {L6},
          doi = {10.1088/2041-8205/744/1/L6},
archivePrefix = {arXiv},
       eprint = {1111.6867},
 primaryClass = {astro-ph.SR},
       adsurl = {https://ui.adsabs.harvard.edu/abs/2012ApJ...744L...6B}
}

@ARTICLE{Basu2009,
       author = {{Basu}, Sarbani and {Chaplin}, William J. and {Elsworth}, Yvonne and {New}, Roger and {Serenelli}, Aldo M.},
        title = "{Fresh Insights on the Structure of the Solar Core}",
      journal = {\apj},
         year = 2009,
        month = jul,
       volume = {699},
       number = {2},
        pages = {1403-1417},
          doi = {10.1088/0004-637X/699/2/1403},
archivePrefix = {arXiv},
       eprint = {0905.0651},
 primaryClass = {astro-ph.SR},
       adsurl = {https://ui.adsabs.harvard.edu/abs/2009ApJ...699.1403B}
}

@ARTICLE{barnes2010,
       author = {{Barnes}, Sydney A.},
        title = "{A Simple Nonlinear Model for the Rotation of Main-sequence Cool Stars. I. Introduction, Implications for Gyrochronology, and Color-Period Diagrams}",
      journal = {\apj},
         year = 2010,
        month = oct,
       volume = {722},
       number = {1},
        pages = {222-234},
          doi = {10.1088/0004-637X/722/1/222},
       adsurl = {https://ui.adsabs.harvard.edu/abs/2010ApJ...722..222B}
}

@ARTICLE{bouvier1997,
       author = {{Bouvier}, J. and {Forestini}, M. and {Allain}, S.},
        title = "{The angular momentum evolution of low-mass stars.}",
      journal = {\aap},
         year = 1997,
        month = oct,
       volume = {326},
        pages = {1023-1043},
       adsurl = {https://ui.adsabs.harvard.edu/abs/1997A&A...326.1023B}
}

@ARTICLE{gallet2015,
       author = {{Gallet}, F. and {Bouvier}, J.},
        title = "{Improved angular momentum evolution model for solar-like stars. II. Exploring the mass dependence}",
      journal = {\aap},
         year = 2015,
        month = may,
       volume = {577},
          eid = {A98},
        pages = {A98},
          doi = {10.1051/0004-6361/201525660},
archivePrefix = {arXiv},
       eprint = {1502.05801},
 primaryClass = {astro-ph.SR},
       adsurl = {https://ui.adsabs.harvard.edu/abs/2015A&A...577A..98G}
}

@ARTICLE{denissenkov2010,
       author = {{Denissenkov}, Pavel A. and {Pinsonneault}, Marc and {Terndrup}, Donald M. and {Newsham}, Grant},
        title = "{Angular Momentum Transport in Solar-type Stars: Testing the Timescale for Core-Envelope Coupling}",
      journal = {\apj},
         year = 2010,
        month = jun,
       volume = {716},
       number = {2},
        pages = {1269-1287},
          doi = {10.1088/0004-637X/716/2/1269},
archivePrefix = {arXiv},
       eprint = {0911.1121},
 primaryClass = {astro-ph.SR},
       adsurl = {https://ui.adsabs.harvard.edu/abs/2010ApJ...716.1269D}
}

@ARTICLE{lanzafame2015,
       author = {{Lanzafame}, A.~C. and {Spada}, F.},
        title = "{Rotational evolution of slow-rotator sequence stars}",
      journal = {\aap},
         year = 2015,
        month = dec,
       volume = {584},
          eid = {A30},
        pages = {A30},
          doi = {10.1051/0004-6361/201526770},
archivePrefix = {arXiv},
       eprint = {1506.05298},
 primaryClass = {astro-ph.SR},
       adsurl = {https://ui.adsabs.harvard.edu/abs/2015A&A...584A..30L}
}

@ARTICLE{bouma2023,
       author = {{Bouma}, Luke G. and {Palumbo}, Elsa K. and {Hillenbrand}, Lynne A.},
        title = "{The Empirical Limits of Gyrochronology}",
      journal = {\apjl},
         year = 2023,
        month = apr,
       volume = {947},
       number = {1},
          eid = {L3},
        pages = {L3},
          doi = {10.3847/2041-8213/acc589},
archivePrefix = {arXiv},
       eprint = {2303.08830},
 primaryClass = {astro-ph.SR},
       adsurl = {https://ui.adsabs.harvard.edu/abs/2023ApJ...947L...3B}
}

@ARTICLE{cao2025,
       author = {{Cao}, Lyra and {Stassun}, Keivan G.},
        title = "{The Relationship of Stellar Radius Inflation to Rotation and Magnetic Starspots at 10─670 Myr}",
      journal = {\apjl},
         year = 2025,
        month = jul,
       volume = {988},
       number = {1},
          eid = {L1},
        pages = {L1},
          doi = {10.3847/2041-8213/ade875},
archivePrefix = {arXiv},
       eprint = {2506.18972},
 primaryClass = {astro-ph.SR},
       adsurl = {https://ui.adsabs.harvard.edu/abs/2025ApJ...988L...1C}
}

@ARTICLE{li2025b,
       author = {{Li}, Yaguang and {Huber}, Daniel and {Ong}, J.~M. Joel and {van Saders}, Jennifer and {Costa}, R.~R. and {Larsen}, Jens Reersted and {Basu}, Sarbani and {Bedding}, Timothy R. and {Dai}, Fei and {Chontos}, Ashley and {Carmichael}, Theron W. and {Hey}, Daniel and {Kjeldsen}, Hans and {Hon}, Marc and {Campante}, Tiago L. and {Monteiro}, M{\'a}rio J.~P.~F.~G. and {Lundkvist}, Mia Sloth and {Saunders}, Nicholas and {Isaacson}, Howard and {Howard}, Andrew W. and {Gibson}, Steven R. and {Halverson}, Samuel and {Rider}, Kodi and {Roy}, Arpita and {Baker}, Ashley D. and {Edelstein}, Jerry and {Smith}, Chris and {Fulton}, Benjamin J. and {Walawender}, Josh},
        title = "{K Dwarf Radius Inflation and a 10 Gyr Spin-down Clock Unveiled through Asteroseismology of HD 219134 from the Keck Planet Finder}",
      journal = {\apj},
         year = 2025,
        month = may,
       volume = {984},
       number = {2},
          eid = {125},
        pages = {125},
          doi = {10.3847/1538-4357/adc737},
archivePrefix = {arXiv},
       eprint = {2502.00971},
 primaryClass = {astro-ph.SR},
       adsurl = {https://ui.adsabs.harvard.edu/abs/2025ApJ...984..125L}
}

@ARTICLE{chiti2024,
       author = {{Chiti}, Federica and {van Saders}, Jennifer L. and {Heintz}, Tyler M. and {Hermes}, J.~J. and {Ong}, J.~M. Joel and {Hey}, Daniel R. and {Ramirez-Weinhouse}, Michele M. and {Dugas}, Alison},
        title = "{Rotation at the Fully Convective Boundary: Insights from Wide WD + MS Binary Systems}",
      journal = {\apj},
         year = 2024,
        month = dec,
       volume = {977},
       number = {1},
          eid = {15},
        pages = {15},
          doi = {10.3847/1538-4357/ad856c},
archivePrefix = {arXiv},
       eprint = {2403.12129},
 primaryClass = {astro-ph.SR},
       adsurl = {https://ui.adsabs.harvard.edu/abs/2024ApJ...977...15C}
}

@ARTICLE{saunders2024,
       author = {{Saunders}, Nicholas and {van Saders}, Jennifer L. and {Lyttle}, Alexander J. and {Metcalfe}, Travis S. and {Li}, Tanda and {Davies}, Guy R. and {Hall}, Oliver J. and {Ball}, Warrick H. and {Townsend}, Richard and {Creevey}, Orlagh and {Dodds}, Curt},
        title = "{Stellar Cruise Control: Weakened Magnetic Braking Leads to Sustained Rapid Rotation of Old Stars}",
      journal = {\apj},
         year = 2024,
        month = feb,
       volume = {962},
       number = {2},
          eid = {138},
        pages = {138},
          doi = {10.3847/1538-4357/ad1516},
archivePrefix = {arXiv},
       eprint = {2309.05666},
 primaryClass = {astro-ph.SR},
       adsurl = {https://ui.adsabs.harvard.edu/abs/2024ApJ...962..138S}
}

@ARTICLE{metcalfe2020,
       author = {{Metcalfe}, Travis S. and {van Saders}, Jennifer L. and {Basu}, Sarbani and {Buzasi}, Derek and {Chaplin}, William J. and {Egeland}, Ricky and {Garcia}, Rafael A. and {Gaulme}, Patrick and {Huber}, Daniel and {Reinhold}, Timo and {Schunker}, Hannah and {Stassun}, Keivan G. and {Appourchaux}, Thierry and {Ball}, Warrick H. and {Bedding}, Timothy R. and {Deheuvels}, S{\'e}bastien and {Gonz{\'a}lez-Cuesta}, Luc{\'\i}a and {Handberg}, Rasmus and {Jim{\'e}nez}, Antonio and {Kjeldsen}, Hans and {Li}, Tanda and {Lund}, Mikkel N. and {Mathur}, Savita and {Mosser}, Benoit and {Nielsen}, Martin B. and {Noll}, Anthony and {{\c{C}}elik Orhan}, Zeynep and {{\"O}rtel}, Sibel and {Santos}, {\^A}ngela R.~G. and {Yildiz}, Mutlu and {Baliunas}, Sallie and {Soon}, Willie},
        title = "{The Evolution of Rotation and Magnetic Activity in 94 Aqr Aa from Asteroseismology with TESS}",
      journal = {\apj},
         year = 2020,
        month = sep,
       volume = {900},
       number = {2},
          eid = {154},
        pages = {154},
          doi = {10.3847/1538-4357/aba963},
archivePrefix = {arXiv},
       eprint = {2007.12755},
 primaryClass = {astro-ph.SR},
       adsurl = {https://ui.adsabs.harvard.edu/abs/2020ApJ...900..154M}
}

@ARTICLE{vansaders2019,
       author = {{van Saders}, Jennifer L. and {Pinsonneault}, Marc H. and {Barbieri}, Mauro},
        title = "{Forward Modeling of the Kepler Stellar Rotation Period Distribution: Interpreting Periods from Mixed and Biased Stellar Populations}",
      journal = {\apj},
         year = 2019,
        month = feb,
       volume = {872},
       number = {2},
          eid = {128},
        pages = {128},
          doi = {10.3847/1538-4357/aafafe},
archivePrefix = {arXiv},
       eprint = {1803.04971},
 primaryClass = {astro-ph.SR},
       adsurl = {https://ui.adsabs.harvard.edu/abs/2019ApJ...872..128V}
}

@ARTICLE{vansaders2016,
       author = {{van Saders}, Jennifer L. and {Ceillier}, Tugdual and {Metcalfe}, Travis S. and {Silva Aguirre}, Victor and {Pinsonneault}, Marc H. and {Garc{\'\i}a}, Rafael A. and {Mathur}, Savita and {Davies}, Guy R.},
        title = "{Weakened magnetic braking as the origin of anomalously rapid rotation in old field stars}",
      journal = {\nat},
         year = 2016,
        month = jan,
       volume = {529},
       number = {7585},
        pages = {181-184},
          doi = {10.1038/nature16168},
archivePrefix = {arXiv},
       eprint = {1601.02631},
 primaryClass = {astro-ph.SR},
       adsurl = {https://ui.adsabs.harvard.edu/abs/2016Natur.529..181V}
}

@ARTICLE{mamajek2008,
       author = {{Mamajek}, Eric E. and {Hillenbrand}, Lynne A.},
        title = "{Improved Age Estimation for Solar-Type Dwarfs Using Activity-Rotation Diagnostics}",
      journal = {\apj},
         year = 2008,
        month = nov,
       volume = {687},
       number = {2},
        pages = {1264-1293},
          doi = {10.1086/591785},
archivePrefix = {arXiv},
       eprint = {0807.1686},
 primaryClass = {astro-ph},
       adsurl = {https://ui.adsabs.harvard.edu/abs/2008ApJ...687.1264M}
}

@ARTICLE{Roquette2021,
       author = {{Roquette}, J. and {Matt}, S.~P. and {Winter}, A.~J. and {Amard}, L. and {Stasevic}, S.},
        title = "{The influence of the environment on the spin evolution of low-mass stars - I. External photoevaporation of circumstellar discs}",
      journal = {\mnras},
         year = 2021,
        month = dec,
       volume = {508},
       number = {3},
        pages = {3710-3729},
          doi = {10.1093/mnras/stab2772},
archivePrefix = {arXiv},
       eprint = {2109.10296},
 primaryClass = {astro-ph.SR},
       adsurl = {https://ui.adsabs.harvard.edu/abs/2021MNRAS.508.3710R}
}

@ARTICLE{Amard2020,
       author = {{Amard}, Louis and {Matt}, Sean P.},
        title = "{The Impact of Metallicity on the Evolution of the Rotation and Magnetic Activity of Sun-like Stars}",
      journal = {\apj},
         year = 2020,
        month = feb,
       volume = {889},
       number = {2},
          eid = {108},
        pages = {108},
          doi = {10.3847/1538-4357/ab6173},
archivePrefix = {arXiv},
       eprint = {2001.10404},
 primaryClass = {astro-ph.SR},
       adsurl = {https://ui.adsabs.harvard.edu/abs/2020ApJ...889..108A}
}

@ARTICLE{spada2020,
       author = {{Spada}, F. and {Lanzafame}, A.~C.},
        title = "{Competing effect of wind braking and interior coupling in the rotational evolution of solar-like stars}",
      journal = {\aap},
         year = 2020,
        month = apr,
       volume = {636},
          eid = {A76},
        pages = {A76},
          doi = {10.1051/0004-6361/201936384},
archivePrefix = {arXiv},
       eprint = {1908.00345},
 primaryClass = {astro-ph.SR},
       adsurl = {https://ui.adsabs.harvard.edu/abs/2020A&A...636A..76S}
}

@ARTICLE{m67pleiadespaper,
       author = {{van Groeningen}, M.~G.~J. and {Castro-Ginard}, A. and {Brown}, A.~G.~A. and {Casamiquela}, L. and {Jordi}, C.},
        title = "{A machine-learning-based tool for open cluster membership determination in Gaia DR3}",
      journal = {\aap},
         year = 2023,
        month = jul,
       volume = {675},
          eid = {A68},
        pages = {A68},
          doi = {10.1051/0004-6361/202345952},
archivePrefix = {arXiv},
       eprint = {2303.08474},
 primaryClass = {astro-ph.GA},
       adsurl = {https://ui.adsabs.harvard.edu/abs/2023A&A...675A..68V}
}

@INPROCEEDINGS{Castelli2003,
       author = {{Castelli}, F. and {Kurucz}, R.~L.},
        title = "{New Grids of ATLAS9 Model Atmospheres}",
    booktitle = {Modelling of Stellar Atmospheres},
         year = 2003,
       editor = {{Piskunov}, N. and {Weiss}, W.~W. and {Gray}, D.~F.},
       series = {IAU Symposium},
       volume = {210},
        month = jan,
        pages = {A20},
          doi = {10.48550/arXiv.astro-ph/0405087},
archivePrefix = {arXiv},
       eprint = {astro-ph/0405087},
 primaryClass = {astro-ph},
       adsurl = {https://ui.adsabs.harvard.edu/abs/2003IAUS..210P.A20C}
}

@ARTICLE{Spruit1982,
       author = {{Spruit}, H.~C.},
        title = "{The flow of heat near a starspot}",
      journal = {\aap},
         year = 1982,
        month = apr,
       volume = {108},
       number = {2},
        pages = {356-360},
       adsurl = {https://ui.adsabs.harvard.edu/abs/1982A&A...108..356S}
}

@ARTICLE{Tayar2019b,
       author = {{Tayar}, Jamie and {Beck}, Paul G. and {Pinsonneault}, Marc H. and {Garc{\'\i}a}, Rafael A. and {Mathur}, Savita},
        title = "{Core-Envelope Coupling in Intermediate-mass Core-helium Burning Stars}",
      journal = {\apj},
         year = 2019,
        month = dec,
       volume = {887},
       number = {2},
          eid = {203},
        pages = {203},
          doi = {10.3847/1538-4357/ab558a},
archivePrefix = {arXiv},
       eprint = {1911.01443},
 primaryClass = {astro-ph.SR},
       adsurl = {https://ui.adsabs.harvard.edu/abs/2019ApJ...887..203T}
}

@ARTICLE{ybcpaper,
       author = {{Chen}, Yang and {Girardi}, L{\'e}o and {Fu}, Xiaoting and {Bressan}, Alessandro and {Aringer}, Bernhard and {Dal Tio}, Piero and {Pastorelli}, Giada and {Marigo}, Paola and {Costa}, Guglielmo and {Zhang}, Xing},
        title = "{YBC: a stellar bolometric corrections database with variable extinction coefficients. Application to PARSEC isochrones}",
      journal = {\aap},
         year = 2019,
        month = dec,
       volume = {632},
          eid = {A105},
        pages = {A105},
          doi = {10.1051/0004-6361/201936612},
archivePrefix = {arXiv},
       eprint = {1910.09037},
 primaryClass = {astro-ph.SR},
       adsurl = {https://ui.adsabs.harvard.edu/abs/2019A&A...632A.105C}
}

@ARTICLE{claudiam67,
       author = {{Reyes}, Claudia and {Stello}, Dennis and {Hon}, Marc and {Trampedach}, Regner and {Sandquist}, Eric and {Pinsonneault}, Marc H.},
        title = "{Isochrone fitting of the open cluster M67 in the era of Gaia and improved model physics}",
      journal = {\mnras},
         year = 2024,
        month = aug,
       volume = {532},
       number = {2},
        pages = {2860-2874},
          doi = {10.1093/mnras/stae1650},
archivePrefix = {arXiv},
       eprint = {2407.03526},
 primaryClass = {astro-ph.SR},
       adsurl = {https://ui.adsabs.harvard.edu/abs/2024MNRAS.532.2860R}
}

@ARTICLE{Mann2015,
       author = {{Mann}, Andrew W. and {Feiden}, Gregory A. and {Gaidos}, Eric and {Boyajian}, Tabetha and {von Braun}, Kaspar},
        title = "{How to Constrain Your M Dwarf: Measuring Effective Temperature, Bolometric Luminosity, Mass, and Radius}",
      journal = {\apj},
         year = 2015,
        month = may,
       volume = {804},
       number = {1},
          eid = {64},
        pages = {64},
          doi = {10.1088/0004-637X/804/1/64},
archivePrefix = {arXiv},
       eprint = {1501.01635},
 primaryClass = {astro-ph.SR},
       adsurl = {https://ui.adsabs.harvard.edu/abs/2015ApJ...804...64M}
}

@ARTICLE{Spruit1986,
       author = {{Spruit}, H.~C. and {Weiss}, A.},
        title = "{Colors and luminosities of stars with spots.}",
      journal = {\aap},
         year = 1986,
        month = sep,
       volume = {166},
        pages = {167-176},
       adsurl = {https://ui.adsabs.harvard.edu/abs/1986A&A...166..167S}
}

@ARTICLE{Cao2022b,
       author = {{Cao}, Lyra and {Pinsonneault}, Marc H.},
        title = "{Star-spots and magnetism: testing the activity paradigm in the Pleiades and M67}",
      journal = {\mnras},
         year = 2022,
        month = dec,
       volume = {517},
       number = {2},
        pages = {2165-2189},
          doi = {10.1093/mnras/stac2706},
archivePrefix = {arXiv},
       eprint = {2209.10549},
 primaryClass = {astro-ph.SR},
       adsurl = {https://ui.adsabs.harvard.edu/abs/2022MNRAS.517.2165C}
}

@ARTICLE{perezpaolino2025,
       author = {{P{\'e}rez Paolino}, Facundo and {Bary}, Jeffrey S. and {Hillenbrand}, Lynne A. and {Horner}, Benjamin and {Carvalho}, Adolfo},
        title = "{Spectral Biases, Starspot Morphology, and Dynamo Transitions on the Pre-main Sequence: Insights from the X-Shooter WTTS Library}",
      journal = {\apj},
         year = 2025,
        month = sep,
       volume = {990},
       number = {2},
          eid = {205},
        pages = {205},
          doi = {10.3847/1538-4357/adf6ad},
archivePrefix = {arXiv},
       eprint = {2505.10837},
 primaryClass = {astro-ph.SR},
       adsurl = {https://ui.adsabs.harvard.edu/abs/2025ApJ...990..205P}
}

@ARTICLE{Jeffries2021,
       author = {{Jeffries}, R.~D. and {Jackson}, R.~J. and {Sun}, Qinghui and {Deliyannis}, Constantine P.},
        title = "{The effects of rotation on the lithium depletion of G- and K-dwarfs in Messier 35}",
      journal = {\mnras},
         year = 2021,
        month = jan,
       volume = {500},
       number = {1},
        pages = {1158-1177},
          doi = {10.1093/mnras/staa3141},
archivePrefix = {arXiv},
       eprint = {2010.04217},
 primaryClass = {astro-ph.SR},
       adsurl = {https://ui.adsabs.harvard.edu/abs/2021MNRAS.500.1158J}
}

@ARTICLE{Jeffries2023,
       author = {{Jeffries}, R.~D. and {Jackson}, R.~J. and {Wright}, Nicholas J. and {Weaver}, G. and {Gilmore}, G. and {Randich}, S. and {Bragaglia}, A. and {Korn}, A.~J. and {Smiljanic}, R. and {Biazzo}, K. and {Casey}, A.~R. and {Frasca}, A. and {Gonneau}, A. and {Guiglion}, G. and {Morbidelli}, L. and {Prisinzano}, L. and {Sacco}, G.~G. and {Tautvai{\v{s}}ien{\.{e}}}, G. and {Worley}, C.~C. and {Zaggia}, S.},
        title = "{The Gaia-ESO Survey: empirical estimates of stellar ages from lithium equivalent widths (EAGLES)}",
      journal = {\mnras},
         year = 2023,
        month = jul,
       volume = {523},
       number = {1},
        pages = {802-824},
          doi = {10.1093/mnras/stad1293},
archivePrefix = {arXiv},
       eprint = {2304.12197},
 primaryClass = {astro-ph.SR},
       adsurl = {https://ui.adsabs.harvard.edu/abs/2023MNRAS.523..802J}
}

@ARTICLE{Cao2026a,
       author = {{Cao}, Lyra and {Pinsonneault}, Marc H. and {Sharifi}, Kayvon},
        title = "{A magnetic origin for the Pleiades lithium spread}",
        journal = {submitted \apj},
        year = 2026
}

@ARTICLE{Campbell1984,
       author = {{Campbell}, B.},
        title = "{Color anomalies and starspots in Hyades dwarfs.}",
      journal = {\apj},
         year = 1984,
        month = aug,
       volume = {283},
        pages = {209-217},
          doi = {10.1086/162295},
       adsurl = {https://ui.adsabs.harvard.edu/abs/1984ApJ...283..209C}
}

@ARTICLE{Stauffer2003,
       author = {{Stauffer}, John R. and {Jones}, Burton F. and {Backman}, Dana and {Hartmann}, Lee W. and {Barrado y Navascu{\'e}s}, David and {Pinsonneault}, Marc H. and {Terndrup}, Donald M. and {Muench}, August A.},
        title = "{Why Are the K Dwarfs in the Pleiades So Blue?}",
      journal = {\aj},
         year = 2003,
        month = aug,
       volume = {126},
       number = {2},
        pages = {833-847},
          doi = {10.1086/376739},
archivePrefix = {arXiv},
       eprint = {astro-ph/0306127},
 primaryClass = {astro-ph},
       adsurl = {https://ui.adsabs.harvard.edu/abs/2003AJ....126..833S}
}

@ARTICLE{Soderblom1993,
       author = {{Soderblom}, David R. and {Jones}, Burton F. and {Balachandran}, Suchitra and {Stauffer}, John R. and {Duncan}, Douglas K. and {Fedele}, Stephen B. and {Hudon}, J.~D.},
        title = "{The Evolution of the Lithium Abundances of Solar-Type Stars. III. The Pleiades}",
      journal = {\aj},
         year = 1993,
        month = sep,
       volume = {106},
        pages = {1059},
          doi = {10.1086/116704},
       adsurl = {https://ui.adsabs.harvard.edu/abs/1993AJ....106.1059S}
}

@ARTICLE{GullySantiago2017,
       author = {{Gully-Santiago}, Michael A. and {Herczeg}, Gregory J. and {Czekala}, Ian and {Somers}, Garrett and {Grankin}, Konstantin and {Covey}, Kevin R. and {Donati}, J.~F. and {Alencar}, Silvia H.~P. and {Hussain}, Gaitee A.~J. and {Shappee}, Benjamin J. and {Mace}, Gregory N. and {Lee}, Jae-Joon and {Holoien}, T.~W.-S. and {Jose}, Jessy and {Liu}, Chun-Fan},
        title = "{Placing the Spotted T Tauri Star LkCa 4 on an HR Diagram}",
      journal = {\apj},
         year = 2017,
        month = feb,
       volume = {836},
       number = {2},
          eid = {200},
        pages = {200},
          doi = {10.3847/1538-4357/836/2/200},
archivePrefix = {arXiv},
       eprint = {1701.06703},
 primaryClass = {astro-ph.SR},
       adsurl = {https://ui.adsabs.harvard.edu/abs/2017ApJ...836..200G}
}

@ARTICLE{Cao2022a,
       author = {{Cao}, Lyra and {Pinsonneault}, Marc H. and {Hillenbrand}, Lynne A. and {Kuhn}, Michael A.},
        title = "{Age Spreads and Systematics in {\ensuremath{\lambda}} Orionis with Gaia DR2 and the SPOTS Tracks}",
      journal = {\apj},
         year = 2022,
        month = jan,
       volume = {924},
       number = {2},
          eid = {84},
        pages = {84},
          doi = {10.3847/1538-4357/ac307f},
archivePrefix = {arXiv},
       eprint = {2110.11363},
 primaryClass = {astro-ph.SR},
       adsurl = {https://ui.adsabs.harvard.edu/abs/2022ApJ...924...84C}
}

@ARTICLE{David2019,
       author = {{David}, Trevor J. and {Hillenbrand}, Lynne A. and {Gillen}, Edward and {Cody}, Ann Marie and {Howell}, Steve B. and {Isaacson}, Howard T. and {Livingston}, John H.},
        title = "{Age Determination in Upper Scorpius with Eclipsing Binaries}",
      journal = {\apj},
         year = 2019,
        month = feb,
       volume = {872},
       number = {2},
          eid = {161},
        pages = {161},
          doi = {10.3847/1538-4357/aafe09},
archivePrefix = {arXiv},
       eprint = {1901.05532},
 primaryClass = {astro-ph.SR},
       adsurl = {https://ui.adsabs.harvard.edu/abs/2019ApJ...872..161D}
}

@ARTICLE{Jackson2018,
       author = {{Jackson}, R.~J. and {Deliyannis}, Constantine P. and {Jeffries}, R.~D.},
        title = "{The inflated radii of M dwarfs in the Pleiades}",
      journal = {\mnras},
         year = 2018,
        month = may,
       volume = {476},
       number = {3},
        pages = {3245-3262},
          doi = {10.1093/mnras/sty374},
archivePrefix = {arXiv},
       eprint = {1802.04288},
 primaryClass = {astro-ph.SR},
       adsurl = {https://ui.adsabs.harvard.edu/abs/2018MNRAS.476.3245J}
}

@ARTICLE{Cao2023,
       author = {{Cao}, Lyra and {Pinsonneault}, Marc H. and {van Saders}, Jennifer L.},
        title = "{Core-envelope Decoupling Drives Radial Shear Dynamos in Cool Stars}",
      journal = {\apjl},
         year = 2023,
        month = jul,
       volume = {951},
       number = {2},
          eid = {L49},
        pages = {L49},
          doi = {10.3847/2041-8213/acd780},
archivePrefix = {arXiv},
       eprint = {2301.07716},
 primaryClass = {astro-ph.SR},
       adsurl = {https://ui.adsabs.harvard.edu/abs/2023ApJ...951L..49C}
}

@ARTICLE{yreclab,
       author = {{Cao}, Lyra and {Pinsonneault}, Marc H.},
        title = "{yreclab: a public scriptable stellar evolution code targeting WebAssembly}",
        journal = {submitted \apj},
         year = 2026
}

@inproceedings{emscripten,
author = {Zakai, Alon},
title = {Emscripten: an LLVM-to-JavaScript compiler},
year = {2011},
isbn = {9781450309424},
publisher = {Association for Computing Machinery},
address = {New York, NY, USA},
url = {https://doi.org/10.1145/2048147.2048224},
doi = {10.1145/2048147.2048224},
booktitle = {Proceedings of the ACM International Conference Companion on Object Oriented Programming Systems Languages and Applications Companion},
pages = {301–312},
numpages = {12},
location = {Portland, Oregon, USA},
series = {OOPSLA '11}
}

@ARTICLE{Fields2023,
       author = {{Fields}, Carl E. and {Townsend}, Richard H.~D. and {Dotter}, A.~L. and {Zingale}, Michael and {Timmes}, F.~X.},
        title = "{MESA-Web: A cloud resource for stellar evolution in astronomy curricula}",
      journal = {arXiv e-prints},
         year = 2023,
        month = sep,
          eid = {arXiv:2309.15930},
        pages = {arXiv:2309.15930},
          doi = {10.48550/arXiv.2309.15930},
archivePrefix = {arXiv},
       eprint = {2309.15930},
 primaryClass = {astro-ph.SR},
       adsurl = {https://ui.adsabs.harvard.edu/abs/2023arXiv230915930F}
}

@ARTICLE{Demarque1964,
       author = {{Demarque}, P.~R. and {Percy}, J.~R.},
        title = "{A Series of Solar Models.}",
      journal = {\apj},
         year = 1964,
        month = aug,
       volume = {140},
        pages = {541},
          doi = {10.1086/147947},
       adsurl = {https://ui.adsabs.harvard.edu/abs/1964ApJ...140..541D}
}

@ARTICLE{Bahcall2005,
       author = {{Bahcall}, John N. and {Basu}, Sarbani and {Pinsonneault}, Marc and {Serenelli}, Aldo M.},
        title = "{Helioseismological Implications of Recent Solar Abundance Determinations}",
      journal = {\apj},
         year = 2005,
        month = jan,
       volume = {618},
       number = {2},
        pages = {1049-1056},
          doi = {10.1086/426070},
archivePrefix = {arXiv},
       eprint = {astro-ph/0407060},
 primaryClass = {astro-ph},
       adsurl = {https://ui.adsabs.harvard.edu/abs/2005ApJ...618.1049B}
}

@ARTICLE{Delahaye2006,
       author = {{Delahaye}, Franck and {Pinsonneault}, M.~H.},
        title = "{The Solar Heavy-Element Abundances. I. Constraints from Stellar Interiors}",
      journal = {\apj},
         year = 2006,
        month = sep,
       volume = {649},
       number = {1},
        pages = {529-540},
          doi = {10.1086/505260},
archivePrefix = {arXiv},
       eprint = {astro-ph/0511779},
 primaryClass = {astro-ph},
       adsurl = {https://ui.adsabs.harvard.edu/abs/2006ApJ...649..529D}
}

@ARTICLE{Villante2014,
       author = {{Villante}, Francesco L. and {Serenelli}, Aldo M. and {Delahaye}, Franck and {Pinsonneault}, Marc H.},
        title = "{The Chemical Composition of the Sun from Helioseismic and Solar Neutrino Data}",
      journal = {\apj},
         year = 2014,
        month = may,
       volume = {787},
       number = {1},
          eid = {13},
        pages = {13},
          doi = {10.1088/0004-637X/787/1/13},
archivePrefix = {arXiv},
       eprint = {1312.3885},
 primaryClass = {astro-ph.SR},
       adsurl = {https://ui.adsabs.harvard.edu/abs/2014ApJ...787...13V}
}

@ARTICLE{2013A&A...558A..33A,
       author = {{Astropy Collaboration} and {Robitaille}, Thomas P. and {Tollerud}, Erik J. and {Greenfield}, Perry and {Droettboom}, Michael and {Bray}, Erik and {Aldcroft}, Tom and {Davis}, Matt and {Ginsburg}, Adam and {Price-Whelan}, Adrian M. and {Kerzendorf}, Wolfgang E. and {Conley}, Alexander and {Crighton}, Neil and {Barbary}, Kyle and {Muna}, Demitri and {Ferguson}, Henry and {Grollier}, Fr{\'e}d{\'e}ric and {Parikh}, Madhura M. and {Nair}, Prasanth H. and {Unther}, Hans M. and {Deil}, Christoph and {Woillez}, Julien and {Conseil}, Simon and {Kramer}, Roban and {Turner}, James E.~H. and {Singer}, Leo and {Fox}, Ryan and {Weaver}, Benjamin A. and {Zabalza}, Victor and {Edwards}, Zachary I. and {Azalee Bostroem}, K. and {Burke}, D.~J. and {Casey}, Andrew R. and {Crawford}, Steven M. and {Dencheva}, Nadia and {Ely}, Justin and {Jenness}, Tim and {Labrie}, Kathleen and {Lim}, Pey Lian and {Pierfederici}, Francesco and {Pontzen}, Andrew and {Ptak}, Andy and {Refsdal}, Brian and {Servillat}, Mathieu and {Streicher}, Ole},
        title = "{Astropy: A community Python package for astronomy}",
      journal = {\aap},
         year = 2013,
        month = oct,
       volume = {558},
          eid = {A33},
        pages = {A33},
          doi = {10.1051/0004-6361/201322068},
archivePrefix = {arXiv},
       eprint = {1307.6212},
 primaryClass = {astro-ph.IM},
       adsurl = {https://ui.adsabs.harvard.edu/abs/2013A&A...558A..33A}
}

@software{Eggleton2011,
       author = {{Eggleton}, P.~P. and {Tout}, Christopher and {Pols}, Onno and {Izzard}, Rob and {Eldridge}, John and {Lesaffre}, Pierre and {Stancliffe}, Richard and {Church}, Ross and {Lau}, Herbert},
        title = "{STARS: A Stellar Evolution Code}",
 howpublished = {Astrophysics Source Code Library, record ascl:1107.008},
         year = 2011,
        month = jul,
          eid = {ascl:1107.008},
archivePrefix = {ascl},
       eprint = {1107.008},
       adsurl = {https://ui.adsabs.harvard.edu/abs/2011ascl.soft07008E}
}

@ARTICLE{Eggleton1972,
       author = {{Eggleton}, Peter P.},
        title = "{Composition changes during stellar evolution}",
      journal = {\mnras},
         year = 1972,
        month = jan,
       volume = {156},
        pages = {361},
          doi = {10.1093/mnras/156.3.361},
       adsurl = {https://ui.adsabs.harvard.edu/abs/1972MNRAS.156..361E}
}

@ARTICLE{Paxton2004,
       author = {{Paxton}, Bill},
        title = "{EZ to Evolve ZAMS Stars: A Program Derived from Eggleton's Stellar Evolution Code}",
      journal = {\pasp},
         year = 2004,
        month = jul,
       volume = {116},
       number = {821},
        pages = {699-701},
          doi = {10.1086/422345},
archivePrefix = {arXiv},
       eprint = {astro-ph/0405130},
 primaryClass = {astro-ph},
       adsurl = {https://ui.adsabs.harvard.edu/abs/2004PASP..116..699P}
}

@ARTICLE{CD1993,
       author = {{Christensen-Dalsgaard}, J. and {Proffitt}, C.~R. and {Thompson}, M.~J.},
        title = "{Effects of Diffusion on Solar Models and Their Oscillation Frequencies}",
      journal = {\apjl},
         year = 1993,
        month = feb,
       volume = {403},
        pages = {L75},
          doi = {10.1086/186725},
       adsurl = {https://ui.adsabs.harvard.edu/abs/1993ApJ...403L..75C}
}

@ARTICLE{Espinosa2013,
       author = {{Espinosa Lara}, F. and {Rieutord}, M.},
        title = "{Self-consistent 2D models of fast-rotating
early-type stars}",
      journal = {\aap},
         year = 2013,
        month = apr,
       volume = {552},
          eid = {A35},
        pages = {A35},
          doi = {10.1051/0004-6361/201220844},
archivePrefix = {arXiv},
       eprint = {1212.0778},
 primaryClass = {astro-ph.SR},
       adsurl =
{https://urldefense.com/v3/__https://ui.adsabs.harvard.edu/abs/2013A&A...552A..35E__;!!KGKeukY!zb0l0JFmdUd0TmmuE_wRBqokBHxE0J55VZZhN3J9y9bTxxYjYYLzLvqag61ioS6AtVJchu6Oe8xDxQH82Cfs$ }
}

@ARTICLE{Rieutord2016J,
       author = {{Rieutord}, Michel and {Espinosa Lara}, Francisco and
{Putigny}, Bertrand},
        title = "{An algorithm for computing the 2D structure of fast
rotating stars}",
      journal = {Journal of Computational Physics},
         year = 2016,
        month = aug,
       volume = {318},
        pages = {277-304},
          doi = {10.1016/j.jcp.2016.05.011},
archivePrefix = {arXiv},
       eprint = {1605.02359},
 primaryClass = {astro-ph.SR},
       adsurl =
{https://urldefense.com/v3/__https://ui.adsabs.harvard.edu/abs/2016JCoPh.318..277R__;!!KGKeukY!zb0l0JFmdUd0TmmuE_wRBqokBHxE0J55VZZhN3J9y9bTxxYjYYLzLvqag61ioS6AtVJchu6Oe8xDxWqUGtkZ$ }
}
\bibliographystyle{aasjournalv7}

\end{document}